\documentclass{aa}

\usepackage[fleqn]{amsmath}
\usepackage{amssymb}
\usepackage{natbib}
\bibpunct{(}{)}{;}{a}{}{,}

\usepackage{graphicx}
\graphicspath{{M2735fig/}}

\usepackage[bf]{subfigure}
\def\thesubfigure{\alph{subfigure}.}

\newcommand{\idiff}[2][]{\,\mathrm{d}#2^{{#1}}}
\newcommand{\diff}[2][]{\,\mathrm{d}#2^{{#1}}}
\newcommand{\deriv}[3][]{\frac{\mathrm{d}^{{#1}}{#2}}{\mathrm{d}{#3}^{{#1}}}}

\newcommand{\pderiv}[3][]{\frac{\partial^{{#1}}{#2}}{\partial {#3}^{{#1}}}}
\newcommand{\lt}{\ensuremath{<}}

\def\lambdaint#1{\ensuremath{\int_0^{+\infty} #1 \idiff\lambda}}

\def\ten#1{\ensuremath{10^{#1}}}
\def\alb{\ensuremath{{\cal A}}}

\def\TR{\ensuremath{T_{\rm r}}}
\def\TV{\ensuremath{T_{\rm v}}}
\def\TISM{\ensuremath{T_{\rm a}}}
\def\Trad{\ensuremath{T_{\rm rad}}}
\def\kB{\ensuremath{k_{\rm B}}}
\def\mH{\ensuremath{m_{\rm H}}}

\def\Teff{\ensuremath{T_\mathrm{eff}}}
\def\omegak{\ensuremath{\Omega_{\rm K}}}
\def\Teq{\ensuremath{T_{\rm eq}}}
\def\ci{\ensuremath{c_{\rm i}}}

\def\Hdisk{\ensuremath{{\cal H}}}
\def\Ti{\ensuremath{T_{\rm i}}}
\def\Te{\ensuremath{T_{\rm e}}}
\def\Tstar{\ensuremath{T_\star}}

\def\Tgrav{\ensuremath{T_{\rm g}}}
\def\Fout{\ensuremath{F_{\rm out}}}

\def\deg{\ensuremath{^\mathrm{o}}}
\def\taui{\ensuremath{\tau_{\rm i}}}
\def\dtaui{\ensuremath{\Delta\tau_{\rm i}}}
\def\taue{\ensuremath{\tau_{\rm e}}}
\def\expo#1{\mathrm{e}^{#1}}
\def\kappae{\ensuremath{\kappa_\mathrm{e}}}
\def\kappar{\ensuremath{\kappa_\mathrm{R}}}
\def\kappap{\ensuremath{\kappa_\mathrm{P}}}
\def\invtauR{\frac{4}{3\xi\tauR}}
\def\Sigmae{\ensuremath{\Sigma_\mathrm{e}}}
\def\rhoi{\ensuremath{\rho_{\rm i}}}
\def\rhoe{\ensuremath{\rho_{\rm e}}}
\def\hi{\ensuremath{h_{\rm i}}}
\def\he{\ensuremath{h_{\rm e}}}
\def\nui{\ensuremath{\nu_{\rm i}}}
\def\mui{\ensuremath{\mu_{\rm i}}}
\def\phif{\ensuremath{\phi_{\rm f}}}
\def\phis{\ensuremath{\phi_*}}

\def\mum{\ensuremath{\mu\mathrm{m}}}

\def\epsilon{\varepsilon}
\def\theta{\vartheta}
\def\latin#1{\textit{#1}}
\def\Msun{\ensuremath{M_\odot}}
\def\Rsun{\ensuremath{R_\odot}}
\def\Mdot{\ensuremath{\dot M}}
\def\yr{\ensuremath{\mathrm{yr}}}

\def\Md{\ensuremath{M_{\rm d}}}
\def\Fstar{\ensuremath{F_*}}
\def\Fdisk{\ensuremath{F_\mathrm{disc}}}
\def\sci#1#2{\ensuremath{#1\!\times\! 10^{#2}}}
\def\tauR{\ensuremath{\tau_\mathrm{R}}}
\def\tauP{\ensuremath{\tau_\mathrm{P}}}
\def\tauPe{\ensuremath{\tau^e_\mathrm{P}}}
\def\tauPstar{\ensuremath{\tau^*_\mathrm{P}}}
\def\dtaua{\eta(2\tauP)-\eta\left(\invtauR\right)}
\def\dtaub{\eta(2\tauP)-\eta(2\tauPe)}
\def\plm{{\cal P}(l, m)}
\def\qlm{{\cal Q}(l, m)}
\def\Ftot{\ensuremath{F_\lambda}}
\def\Fi{\ensuremath{F^\mathrm{i}_\lambda}}
\def\Fe{\ensuremath{F^\mathrm{e}_\lambda}}
\def\Fs{\ensuremath{F^\mathrm{s}_\lambda}}
\def\Tphot{\ensuremath{T^\mathrm{phot}_\lambda(i)}}
\def\tauil{\ensuremath{\tau^\mathrm{i}_\lambda}}
\def\tauel{\ensuremath{\tau^\mathrm{e}_\lambda}}

\usepackage{ulem}
\def\correct#1{#1}

\title{The vertical structure of T~Tauri accretion discs}
\subtitle{III. Consistent interpretation of spectra and visibilities with a two-layer model}
\author{R\'egis Lachaume\inst{1} \and Fabien Malbet\inst{1} \and Jean-Louis Monin\inst{1,2}}
\offprints{R.~Lachaume, \email{\\Regis.Lachaume@obs.ujf-grenoble.fr}}
\institute{%
    Laboratoire d'Astrophysique UMR UJF-CNRS 5571, 
    Observatoire de Grenoble, 
    Universit\'e Joseph Fourier, BP 53, 
    F-38041 Grenoble cedex 9,
    France
\and
    Institut Universitaire de France
}
\date{Received 27 May 2002 / Accepted 17 June 2002}
\authorrunning{Lachaume, Malbet \& Monin}
\titlerunning{The vertical structure of T~Tauri discs. III}

\abstract{%
We present a two-layer accretion disc model developed to simultaneously fit
optical long baseline visibilities and spectral energy distributions of
T~Tauri accretion discs.  This model allows us to access easily the physical
conditions in the disc as the mid-plane or the surface temperature.

Our model includes viscous heating, absorption of stellar irradiation, and
thermalisation with the surrounding medium.  The disc is modelled with
concentric cylinders for which the vertical radiation transfer is computed
using two layers with vertically averaged temperatures: the outer layer is
heated by the stellar irradiation and by the inner layer, and the inner layer
by viscous dissipation and by the outer layer.  We investigate three
prescriptions for the geometrical thickness of the disc: it is either
proportional the scale height (model~1), given ad hoc (model~2), or zero
(model~3).  We then derive the disc structure in the case of the $\alpha$ and
$\beta$ viscosity prescriptions, as well as for various optical thickness
regimes of the disc.

This analytical model allows us to disentangle regions where the mid-plane
temperature and the effective temperature are dominated by accretion from
regions dominated by reprocessing of stellar light.  In the case of
$\alpha$-prescription, we find that the structure of model~2 gives
predictions very close to those of numerical simulations from previous authors.

From the disc structure, we derive the spectral energy distributions, images
and interferometric visibilities.  We analyse the influence of the disc
parameters on the resulting structure and on the observable outputs.  We apply
our model to interpret consistently the spectral energy distributions and
visibilities of SU~Aur and FU~Ori for which interferometric data are available,
and that are not known to be part of a multiple system.  We were not
able to derive a consistent fit for T~Tau North, which might come from caveats
in the flux correction from its South component, but were able to separately
derive fits for its spectrum and its visibilities.

We find that even a single interferometric measurement at one infrared
wavelength can bring a very strong constraint on disc models. We predict that
future massive interferometric observations of accretion discs will provide a
breakthrough in the understanding of accretion disc physics.

}

\begin{document}


\maketitle


\section{Introduction}

Since the initial models of viscous accretion discs by \citet{Shakura73} and
\citet{LyndenBell74}, the physics of the close environment of T~Tauri stars
(TTS) has been extensively studied in order to interpret their spectral energy
distribution (SED).  For the sake of simplicity, models traditionally
separated discs into two categories, sometimes called active discs, on one
hand, in which viscous dissipation is predominant, and passive discs, on the
other hand, for which irradiation by the central star is the main heating
process.  Early models used quasi-Keplerian steady accretion discs, assumed to
be geometrically thin for a wide range of accretion rate; they predicted a
fixed slope for the infrared spectrum: $\lambda F_\lambda \sim \lambda^{-4/3}$.
However many TTS present flatter SEDs, and disc flaring was among the first
attempts to explore disc vertical structure as an explanation for such SED
flattening \citep{Adams87,Kenyon87}: in a flared disc, the surface of the
remote parts is tilted toward the star and gets more stellar light than
forecast by the standard model, resulting in a warmer disc further away from
the star. Since then, several models have been proposed in order to explain
both standard SEDs and flatter ones.

An analytical study of the radiative transfer in the vertical structure of
discs was first carried out by \citet{Hubeny90} for active discs, then by
\citet[hereafter paper~I]{paper1} for passive discs.  In the latter model, the
topmost layers of the disc, illuminated by the star, are hotter than the disc
photosphere, resulting in excess continuum and line emission.  Later on,
\citet{Chiang97,Chiang99} used a simplified two-layer passive disc model based
on the same super-heating mechanism and derived SEDs, confirming conclusions of
paper~I and producing results consistent with observations.  More recently,
\citet[hereafter paper~II]{paper2} generalised the analytical study of paper~I
to discs heated by several processes, and used its formalism to derive the the
vertical structure of active discs.

On the other hand, numerical integration of the equations of radiative transfer
was carried out by various authors in order to derive the vertical structure of
accretion discs with fewer \latin{a priori} approximations.
\citet{Bell94,Bell97} developed an active disc model in order to explain
FU~Orionis outbursts; \citet{DAlessio98,DAlessio99} dealt with the more general
case of a disc heated both by viscous heating and stellar irradiation.

From a general point of view, all these studies predict spectra consistent with
observations, but they have rarely been checked consistently against the
spatial information revealed by recent optical and infrared high angular
resolution imaging.  Recently, the advent of optical interferometry has set
newer constraints on disc models.  \citet{Malbet98,Akeson00,Akeson02,Malbet01p}
obtained the first visibility measurements of TTS and FU~Orionis stars.
However, they failed to consistently fit both SEDs and visibilities with a
standard disc model: most of the time, the disc parameters derived from the SED
data are in disagreement with those derived from visibility data.  No other
attempt to compare self-consistent disc models and interferometric measurements
has been carried out so far for low mass pre-main sequence stars.

In this paper, we tackle the issue of analytically describing a disc in
presence of the two main heating processes, viscous heating and stellar
irradiation, the latter requiring a correct description of the flaring.  This
model suits the TTS and FU~Ori-type stars for which viscous heating cannot be
ignored; for more massive stars (Herbig Ae/Be) it has been shown
\citep{Dullemond01} that viscosity, as a heating mechanism, can be ignored, so
that our model is not relevant.  In Sect.~\ref{sec:model}, we
present a two-layer version of the model developed in paper~I and carry out an 
analytical determination of the structure of the disc. We derive a
set of equations giving the mid-plane temperature and the flaring index, from
which the whole structure can be determined.  In Sect.~\ref{sec:num}, we
briefly present the numerical approach.  In Sect.~\ref{sec:ana}, we compare the
results of the model with other models and analyse the influence of some disc
parameters on the observables.  In Sect.~\ref{sec:res}, we apply our model to
the few low-mass young stellar objects (YSOs) observed in optical interferometry.


\section{Model description}
\label{sec:model}

\subsection{The two-layer structure}

The standard accretion disc model by \citet{Shakura73} determines the disc
emission with its effective temperature and does not take the vertical
temperature profile into account. However, this approximation fails when the
disc surface is super-heated by stellar irradiation (paper I): the illumination
by the star is predominant in a optically thin outer layer of the disc while
the optically thick part of the disc is governed by the radiative transfer of
thermal radiation.  This phenomenon results in emission at different
temperatures and has a strong incidence on the SED.

To further explore the resulting structure of the disc, we use a two-layer
vertical structure.  This approach was first proposed by \citet{Chiang97} in
the case of passive discs, and we apply it here in a more general case where
viscous dissipation, stellar irradiation, and thermalization with the
surrounding medium are taken into account. In the case of a passive disc the
disc physical properties variations with distance to the star follow a unique
power law \citep{Chiang97}.  In an active disc, where the temperature is
dominated by viscous dissipation, this is no longer the case
\correct{(paper~II, see Fig.~3)}, due to the sensitivity of the vertical
structure to the disc material opacity.

\begin{figure}[t]
     \includegraphics[width=0.95\hsize]{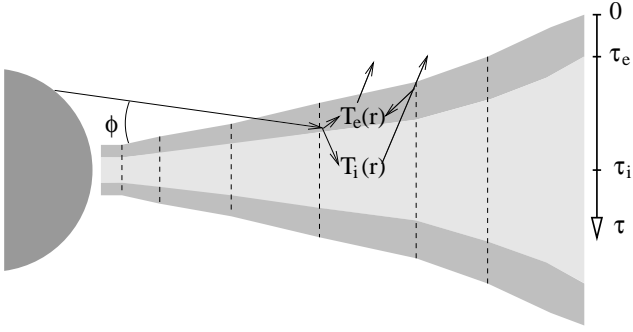}
     \caption{Radial slice of a two-layer disc.}
     \label{fig:tld}
\end{figure}

In this paper, we present a model of disc that can be seen as a simplification
of the formalism of paper~II. In this model, we consider an optically thick
disc heated both by stellar irradiation and viscous dissipation.  An optically
and geometrically thin outer layer is directly heated by the star and by the
inner layer; the optically thick inner layer is heated by viscous dissipation
and by the outer layer.  In the outer layer we use a vertically averaged
temperature to solve the radiative transfer, and in the inner layer we use the
mid-plane temperature.  A radial slice of such a disc is represented in
Fig.~\ref{fig:tld}.  The inner and outer layers have the optical thicknesses
{\taui} and {\taue}, and the temperatures {\Ti} and {\Te} respectively.

The incidence of stellar radiation onto the disc is given by the angle $\phi$,
which is an average of this incidence over the surface of the star.  The height
of the outer layer is {\Hdisk}.

In this section, we focus our attention on two quantities: the mid-plane
temperature and the flaring angle.  The structure of the disc can be derived
from these two quantities; for instance, the mid-plane temperature governs the
scale height and the flaring angle the surface temperature.  In order to obtain
{\Ti} and $\gamma$ we proceed in four steps: in Sect.~\ref{sec:heating}, we
derive the effective temperatures {\TV}, {\TR} and {\TISM} corresponding to
viscous heating, stellar light reprocessing and thermalization with the disc
surrounding medium.  In Sect.~\ref{sec:transfer}, equations of transfer are
obtained, giving the temperatures {\Ti} and {\Te} as a function of the optical
depths {\taui} and {\taue} and the effective temperatures {\TV}, {\TR} and
{\TISM}.  In Sect.~\ref{sec:struct}, the structure of the disc (scale height,
column density, optical depth, etc.) is connected to the temperatures {\Ti} and
{\Te}.  In Sect.~\ref{sec:equation}, the results from the first three steps are
combined to derive the mid-plane temperature {\Ti} and the flaring (variation
of $\Hdisk/r$ with $r$) in a set of coupled equations.

\subsection{Heating sources}
\label{sec:heating}

\citet{LyndenBell74} showed that the effective temperature of a geometrically
thin active disc is
\begin{equation}
     \TV = \frac38 \frac{\omegak^2 \Mdot f}{\sigma\pi},
     \label{eq:TV}
\end{equation}
where $f$ is given by
\begin{equation}
     f      = 1-\sqrt{\frac {r_*}r}.
     \label{eq:f}
\end{equation}
$\dot M$, $\omegak$ and $r_*$ are respectively the accretion rate, the
Keplerian angular velocity at radius $r$, and the radius of the star.

We suppose that the surface of the disc presents a mean albedo {\alb} constant
over the whole disc. The effective temperature associated with stellar heating
then reads
\begin{equation}
     \TR^4 \approx \phi \Teq^4.
\end{equation}
{\Teq} is the Local Thermal Equilibrium temperature of a body with an albedo
{\alb} at a distance $r$ of the star.  It is given by
\begin{equation}
     \Teq^4 = (1-\alb) \frac{T_*^4}{2} \left( \frac {r_*}{r} \right)^2,
\end{equation}
where $T_*$ is the effective temperature of the star.

\subsection{Equations of transfer}
\label{sec:transfer}

Since the layers are assumed to be isothermal, the radiative transfer has
a simple form, depending only on the optical depths of the layers and their
temperatures.

If viscous dissipation only occurs, we expect a surface temperature of
\begin{subequations}
\label{eq:t-irr}
\begin{equation}
     \Te^4 = \frac{\TV^4}{2}
\end{equation}
as in a stellar atmosphere within the Eddington approximation.
Concerning the temperature in the mid-plane of the disc, as showed in paper~II,
the stellar atmosphere formalism cannot be used as is.  A corrective term 
{\dtaui} must be applied to the optical thickness {\taui}:
\begin{equation}
     \Ti^4 = \frac{3}{4} \TV^4 \left( \taui - \dtaui \right).
     \label{eq:Tivstaui}
\end{equation}
\end{subequations}
For the sake of simplicity we ignored the usual constant term $2/3$ before the
large optical depth.  

If reprocessing only occurs, the outer layer is super-heated because it is
optically thin:  the stellar irradiation is dissipated along a slanted path of
optical depth unity in the visible, \latin{i.e.} much less than one for the
reprocessed IR radiation in the vertical direction.  We follow
\citet{Chiang97} and state that both layers present a vertically averaged
temperature.  The outer layer catches the flux $\TR^4$ and emits
half of it upward, while being optically thin, so that
\begin{subequations}
\label{eq:t-visc}
\begin{align}
   \Te^4 &= \frac{\TR^4}{2\taue},\\
\intertext{and emits the other half down to the optically thick inner layer, hence}
   \Ti^4 &= \frac{\TR^4}{2}.
\end{align}
\end{subequations}

In the case of thermalization with the disc surrounding medium, the
temperatures are
\begin{subequations}
\label{eq:t-ism}
\begin{align}
     \Ti^4 = \TISM^4,\\
     \Te^4 = \TISM^4.
\end{align}
\end{subequations}

In the general case, the three heating contributions considered above occur
together.  The superposition principle can be applied to the specific
radiative intensities; that is the radiation at a given location of the disc
is the sum of the radiations corresponding to different heating processes
(see paper~II). Summing over Eqs.~(\ref{eq:t-irr}, \ref{eq:t-visc},
\ref{eq:t-ism}) we derive
\begin{subequations}
\begin{align}
     \Ti^4 &= \TISM^4 + \frac{\TR^4}{2}
                  + \frac34 \TV^4 \left(\taui-\dtaui \right),
     \label{eq:Ti}\\
     \Te^4 &= \TISM^4 + \frac{\TR^4}{2\taue}
                  + \frac{\TV^4}{2}.
     \label{eq:Te}
\end{align}
\end{subequations}
This result can also be obtained from the equations of transfer 28--31
of paper~II using the following approximations: (i) the surface is optically 
thin, (ii) both layers are isothermal, (iii) viscosity only occurs in the inner
layer, (iv) diffusion is ignored, (v) no stellar radiation penetrates the
inner layer, and (vi) the vertical optical thickness of the disc is larger
than unity.

\subsection{Structure of the disc}
\label{sec:struct}

If we ignore self-gravity, the vertical gravitational field is proportional
to the distance $z$ from the mid-plane.  Since the inner layer is isothermal,
the density reads
\begin{equation}
     \rho(z) = \rhoi \exp{\left(-\frac{z^2}{2 \hi^2} \right)},
     \label{eq:rhoz}
\end{equation}
where the density scale height {\hi} is linked to the sound speed {\ci} and the
gravitational field $g_z$ by
\begin{equation}
     \hi = \ci \left( \pderiv{g_z}{z} \right)^{-1/2}.
     \label{eq:hi}
\end{equation}
We consider the gravitational temperature {\Tgrav} at which the thermal
energy of a particle balances the gravitational energy, given by
\begin{equation}
     \kB \Tgrav = \frac{G M_* \mH}{r},
\end{equation}
where {\mH} is the atomic mass of hydrogen.  Then,
\begin{equation}
     \hi = r \sqrt{\frac{\Ti}{\mui\Tgrav}},
     \label{eq:hivsTi}
\end{equation}
where {\mui} is the mean molecular mass in the mid-plane of the disc. 

In the standard disc model by \citet{LyndenBell74}, the mass column 
$\Sigma$ is linked to the uniform accretion rate {\Mdot} by
\begin{equation}
     \Sigma = \frac{\Mdot}{3\pi\nui} f,
     \label{eq:sigma}
\end{equation}
where {\nui} is the vertically averaged viscosity.

In order to go further, we need a prescription for the kinematic viscosity.
\citet{Shakura73} use the so-called $\alpha$-prescription:
\begin{equation}
     \nui = \alpha \ci \hi,
     \label{eq:nui}
\end{equation}
where $\alpha$ is a dimensionless factor ranging from \ten{-4} to \ten{-1}.
If we ignore the thin outer layer, the viscosity is uniform over the entire
height because of the isothermal assumption. So, Eq.~(\ref{eq:sigma}), together
with Eq.~(\ref{eq:nui}), becomes
\begin{equation}
     \Sigma = \Sigma_0 \left( \frac{\mui\Tgrav}{\Ti} \right), \quad
     \Sigma_0 = \frac{\Mdot f}{3 \omegak\alpha \,\pi r^2}.
\end{equation}

The mass column also meets
\begin{equation}
     \Sigma = 2 \int_0^{+\infty} \rho(z) \diff{z}.
\end{equation}
so that, with Eqs.~(\ref{eq:sigma}, \ref{eq:rhoz})
\begin{equation}
     \rhoi = \rho_0 \left( \frac{\mui\Tgrav}{\Ti} \right)^{\frac32}, \quad
     \rho_0 = \sqrt{\frac 2\pi} \frac{\Mdot f}{6\omegak\alpha\,\pi r^3}.
     \label{eq:rhoivsTi}
\end{equation}

In order to keep an analytical description, we assume that the opacity $\kappa$
locally follows a power law:
\begin{equation}
     \kappa(\rho, T) = k \rho^l T^m.
     \label{eq:kappa}
\end{equation}

We now assume that the outer layer has an optical depth much smaller than that
of the inner layer, that is $\taue \ll \taui$.  This is indeed true in
optically thick regions of the disc, where $\taui \gg 1$ and $\taue \ll 1$.
In optically thin regions, the validity of this approximation must be
self-consistently checked after computation.  With such an approximation, the
optical depth of the disc can be written as if the outer layer were absent:
\begin{equation}
     \taui = \int_0^{+\infty} \rho(z) \kappa(z) \diff z,
\end{equation}
so that
\begin{equation}
     \taui = \sqrt{\frac{\pi}{2(l+1)}} k \rhoi^{l+1} \Ti^m \hi.
\end{equation}
The corrective term {\dtaui} depends on the distribution $\theta$ of the
viscous heating. Paper~I gives
\begin{align}
     \theta(z) &= \frac {
        \int_z^{+\infty} \rho(\zeta) \nu(\zeta) \diff\zeta
     }{
        \int_0^{+\infty} \rho(\zeta) \nu(\zeta) \diff\zeta
     },\\
     \dtaui   &= \int_0^{+\infty} \rho(z) \kappa(z) \theta(z) \diff z.
\end{align}
After some calculations we finally obtain
\begin{equation}
     \taui-\dtaui = \sqrt{\frac \pi2} \xi_l \rhoi^{l+1} \Ti^m \hi
     \label{eq:dtaui}
\end{equation}
with
\begin{equation}
     \xi_l = \frac{1}{\sqrt{l+1}} - \frac{1}{\sqrt{l+2}}
     \label{eq:xi}
\end{equation}

We call $\omega$ the ratio of opacity for radiation reprocessed by the
outer layer to opacity for stellar radiation.  $\omega$ is smaller
than unity.  With such a definition, the optical depth of the outer layer
is
\begin{equation}
     \taue \approx \omega\phi.
\end{equation}
\begin{figure}[t]
   \includegraphics[width=0.7\hsize]{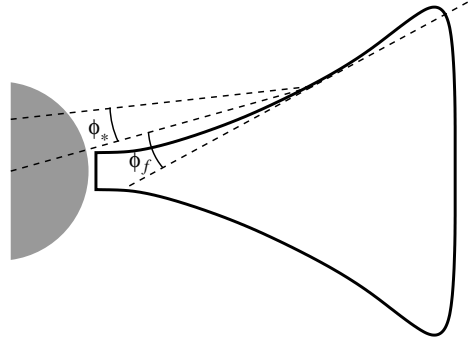}
   \caption{Incidence angle of stellar radiation onto the disc}
   \label{fig:phi}
\end{figure}

The incidence angle of stellar radiation onto the disc is related to the
flaring of the disc.  Yet, in a flat disc, this incidence is non-zero because
of the vertical extent of the star. Figure~\ref{fig:phi} shows that $\phi$ is
the sum of two contributions: the mean extent of the star {\phis} and the tilt
of the disc surface toward the star \phif.  Since the star is not a point-like
source, we need to integrate over the stellar surface to find {\phis}. We 
introduce the flaring angle as the logarithmic variation of the disc thickness 
$\Hdisk/r$ with respect to the radial location:
\begin{equation}
     \gamma = \pderiv{\ln{\Hdisk}}{\ln{r}} - 1.
     \label{eq:gamma}
\end{equation}
$\gamma$ represents the variation of the opening angle with the distance to
star, \latin{i.e.} $\Hdisk/r \sim r^\gamma$.  If the opening angle of the disc
is constant, \latin{e.g.} in a flat disc, $\gamma$ is zero. With such a
definition, the incidence angle is
\begin{subequations}
\begin{align}
     \phi  &=       \phif + \phis       \\
\intertext{with}
     \phif &=       \gamma \frac \Hdisk r, \\
     \phis &\approx \frac{4}{3\pi} \frac {r_*}{r},
\end{align}
\end{subequations}
The expression of {\phis} is an asymptotic value at large $r$, as used in 
\citet{Chiang97}.
Then, we need to connect {\Hdisk} to the previously determined scale height
{\hi} and use the closure relation
\begin{equation}
     \Hdisk = Q \hi \label{eq:Q}.
\end{equation}
\citet{Chiang97} notice that $Q$ is a slowly varying quantity close to 4 and
imposed $Q = 4$ over the disc.  \citet{DAlessio99} consistently determine the 
location of the outer layer, and we used their results for the fiducial model 
to derive that $Q$ ranges from 3 to 5 (see Fig.~\ref{fig:compareQ}).

Finally,
\begin{equation}
     \taue =
           \omega \left( \phis
                        + \gamma Q \sqrt{\frac{\Tgrav}{\mui\Ti}}
                  \right).
\end{equation}

\subsection{Implicit equations governing the structure}
\label{sec:equation}

First, we use the previous developments to derive the mid-plane temperature
{\Ti} as a function of the disc parameters.  Second, we derive the flaring 
index on which the temperature also depends.  Last, we determine a
boundary condition on {\Hdisk} that allows to derive it over the
while disc using $\Hdisk/r \propto r^\gamma$.

\subsubsection{Mid-plane temperature}

All quantities of Eq.~(\ref{eq:Ti}) have been expressed in terms of {\Ti} and
\mui, of the physical constants of the disc, and of the quantities $\gamma$ and
$Q$. Therefore, we derive an equation on {\Ti}
\begin{subequations}
\label{eq:implTi}
\begin{align}
     \Ti^4         &= \sum_k t_k(r, \Ti(r))^4 \label{eq:tcontrib}\\
\intertext{with}
     t_1(r, \Ti)^4 &= \frac38 \xi_l\TV^4\Tgrav^{1+\frac32 l}\Sigma_0\rhoi^l
                      \Ti^{-\frac32l+m-1},\\
     t_2(r, \Ti)^4 &= \frac12 \Teq^4 \Tgrav^{-\frac12}\gamma Q\mui^{-\frac12}
                      \Ti^{\frac12},\\
     t_3(r, \Ti)^4 &= \frac12 \Teq^4\phis,\\
     t_4(r, \Ti)^4 &= \TISM^4.
\end{align}
\end{subequations}

Equation~(\ref{eq:tcontrib}) states that the mid-plane temperature is the sum
of different heating processes. $t_k$ represents the thermal contribution of
the $k$-th process.  $t_1(r)$ is the mid-plane temperature for an optically
thick disc dominated by viscous heating; $t_2(r)$ is the mid-plane temperature
for a flared disc dominated by stellar irradiation, if the geometrical extent
of the star is ignored due to the distance; $t_3(r)$ is the mid-plane
temperature for a flat disc dominated by stellar irradiation; $t_4(r)$ is the
mid-plane temperature in a disc thermalised with the ambient medium.  The
radial and thermal dependencies of the contributions $t_k$ are
\begin{subequations}
\label{eq:tk:dep}
\begin{align}
     t_1(r, \Ti)^4 &\sim r^{-3l-9/2} f(r)^{l+2} \Ti^{-3/2l+m-1},\\
     t_2(r, \Ti)^4 &\sim r^{-3/2} \Ti^{1/2},\\
     t_3(r, \Ti)^4 &\sim r^{-3},\\
     t_4(r, \Ti)^4 &\sim 1.
\end{align}
\end{subequations}

\subsubsection{Flaring index} 
\label{sec:flaring}

The case of the flaring is stlightly more complicated.  In \citet{Chiang97},
the disc thickness was both a power law of the distance to the star ($\Hdisk
\propto r^{9/7}$ at large radii) and proportional to the scale height $\hi$.
In presence of viscous dissipation, it is no longer possible to ensure
that both properties are met, since the scale height is not a simple
power law.  We investigate three possibilities:
\begin{enumerate}
   \item The thickness is proportional to the scale height (model~1).
   \item The thickness is given by the same law as in the passive
   disc (model~2).
   \item The disc is flat, which occurs if the dust grains settle in the
   mid-plane (model~3).
\end{enumerate}

The flaring index of model~1 can be expressed as the average of the specific
flaring indices $\gamma_k$ imposed by the various heating processes weighted by
their characteristic temperatures $t_k$ (see Appendix~\ref{sec:gamma} for a
demonstration).
\begin{subequations}
\label{eq:implgamma}
\begin{align}
   \gamma   &= \frac{\displaystyle\sum_k\gamma_k t_k^4}{\displaystyle\sum_k t_k^4}\\
\intertext{with}
   \gamma_1 &= \frac{1-3l-2m}{20+6l-4m} + \frac{l+2}{20+6l-4m} g,\\
   \gamma_2 &= \frac27,\\
   \gamma_3 &= \frac18,\\
   \gamma_4 &= \frac12,
\end{align}
\end{subequations}
where
\begin{equation}
   g = \frac{1-f}{f}
\end{equation}
arises from the departure of viscous heating from the radial power law in
$r^{-3}$. This effect occurs close to the star and is characterised by $f$
(Eqs.~\ref{eq:TV}, \ref{eq:f}).  $\gamma_1$ is the flaring index for an
optically thick active disc, therefore depending on the opacity law; $\gamma_2$
the flaring index in a passive disc, if the geometrical extent of the star is
ignored; $\gamma_3$ the flaring index in a flat passive disc; $\gamma_4$ the
flaring index in a disc thermalised with the ambient medium.

In \citet{Chiang97}, the thickness meets $\Hdisk \propto r^\gamma$, with
$\gamma = 1/8$ close to the star and $\gamma = 2/7$ at large radii, depending
whether the dominant effect in stellar light reprocessing is the extent of 
the star ($t_3$) or the disc flaring ($t_2$).  We have assumed in
model~2 that, even in the presence of viscosity, we can still write
\begin{equation}
   \gamma = \frac{\gamma_2 t_2^4 + \gamma_3 t_3^4}{t_2^4 + t_3^4},
   \label{eq:flaring2}
\end{equation}
which means the disc thickness {\Hdisk} is only governed by the irradiation
terms ($t_2$ and $t_3$) and varies from $1/8$ to $2/7$ from the central
parts to the outer ones.

In model~3, the disc is flat, that is $\gamma = 0$.

\subsubsection{Disc thickness on the outer edge}
\label{sec:Q}

In model~1 we have assumed, like in \citet{Chiang97}, that the closure relation
$\Hdisk = Q\hi$ applies, with $Q$ constant over the disc.  The authors took $Q
= 4$, which is seemingly a good approximation for low-mass discs ($\Mdot
\approx 10^{-8}\,\Msun/\yr$) but which might no longer be correct over the wide
range of masses that we shall consider ($\Mdot \approx
10^{-8}$--$10^{-4}\,\Msun/\yr$).  For this reason we prefer to determine the
value of $Q$, and chose to carry out the calculation at the outer edge of the
disc.

In model~2, the thickness is no longer proportional to the scale height,
so that we have to determine $\Hdisk$ with the relation 
$\deriv{\ln\Hdisk}{\ln r} = \gamma+1$, that requires a boundary condition,
and therefore determine $\Hdisk$ at the other edge of the disc.  We
perform this calculation by computing $Q = \Hdisk/\hi$.

So, we need to determine the value of $Q$ at the outer boundary of the disc in
both models, which leads us to approximately solving the vertical hydrostatic
equilibrium and the radiative transfer of the incoming radiation.  We obtain
\begin{align}
                Q &= \sqrt{\frac{2}{\delta_\rho(Q)} \ln \left( \frac{\kappae\rhoi r}{\gamma Q\omega} \sqrt{\frac\Te\Ti \delta_h(Q)}\right) }, \label{eq:scQ}\\
   \intertext{%
      where $\delta_h$ and $\delta_\rho$ are corrections due to the
      deviation of the gravity field from the linear law $g_z \propto z$
      when the elevation becomes of the order of $r$:%
   }
   \delta_\rho(Q) &= 2 \left(\frac{r}{Q\hi}\right)^2 \left[ 1 - \left(1+(Q\hi/r)^2\right)^{-1/2} \right], \label{eq:deltarho}\\
   \delta_h  (Q)  &= \left((1+(Q\hi/r^2)\right)^{3/2}.\label{eq:deltah}
\end{align}
A demonstration can be found in Appendix~\ref{sec:Qedge}.



\section{Numerical approach}
\label{sec:num}

Though analytical, the expressions of {\Ti} and $\gamma$ do not allow direct
determination, because the equations giving $\gamma$ and $\Ti$ are coupled and
the opacities do not follow a power law over the whole disc.  This section 
first describes the numerical method we used for the derivation of the
structure.  Then, we briefly explain who the observables (SEDs and
visibilities) have been derived.

\subsection{Opacities}

Because the heating and the flaring of the disc is dependent on opacities, 
their adequate description is mandatory.  The computation of the SED and
image requires monochromatic opacities that we took from two sources:
\begin{itemize}
   \item At low temperatures ($T \lesssim 1800$\,K) the opacity is dominated
   by the dust.  We use opacities computed by \citet{Henning96} for a mixture 
   of inhomogeneous aggregates; they include continuum emission and the
   silicate feature in the mid-infrared.
   \item At higher temperatures ($T \gtrsim 1800$\,K) the opacity comes from
   to the gas.  We could not find such opacities;  instead, we used 
   piecewise power laws $\kappa_\lambda(\rho, T) = \kappa_0(\rho, T) 
   \lambda^{-p}$, and chose $(\rho, T)$ and $q$, so that they approximately
   match Rosseland opacities by \citet{Bell94} and Planck opacities by 
   \citet{Henning96}.  Despite of this poor description, the influence
   on the observables is small because it only concerns the inner disc, where
   both optical depth ($\tau \gg 1$) and thickness ($\Hdisk \ll r$) are not
   critical.
\end{itemize}
These opacities were smoothed in order to avoid numerical instabilities between
the different domains;  it is justified since we use a vertically averaged
structure in both layers. 

The structure requires the Planck and Rosseland opacities, as well as
the determination of the opacity ratio $\omega$.  They have been determined
from the monochromatic opacities as explained in Appendix~\ref{sec:thin}.  In
order to keep the power law opacities formalism, we determined the local 
equivalent $\kappar \propto \rho^l T^m$ with
\begin{subequations}
   \begin{align}
      l &= \pderiv{\log \kappar}{\log \rho},\\
      m &= \pderiv{\log \kappar}{\log T}.
   \end{align}
   \label{eq:lm}
\end{subequations}

\subsection{Physical conditions in the disc}

The disc is divided in concentric cylinders and the two-layer formalism
is applied for each one:  it consists in computing $t_k$, refining
the values for $\Ti$ and $\gamma$ (Eqs.~\ref{eq:implTi},\ref{eq:implgamma}), 
computing the corresponding opacities with the associated $l$ and $m$ 
(Eq.~\ref{eq:lm}), deriving $\Hdisk$, and iterating until convergence.
In model~1, the computation is carried out in ascending order of radius
Cylinders shadowed by inner ones do not receive stellar light and are determined 
with the same equations, but using $\Tstar = 0$\,K.  $\Hdisk$ is determined
as $\Hdisk = Qh$ with $Q$ constant; the self-consistent determination of
$Q$ at the outer edge has not been implemented so far. In model~2, $Q$ is 
determined at the outer edge (Eq.~\ref{eq:Q}), then the computation is 
performed from outer to inner cylinders.  For each cylinder, $\Hdisk$ is given 
by the thickness of the enclosing cylinder $\Hdisk^\mathrm{out}$ using
\begin{equation}
  \Hdisk = \Hdisk^\mathrm{out} \left(\frac{r}{r^\mathrm{out}}\right)^{1+\gamma}.
\end{equation}

The validity of the derived structure was checked with two tests.  On one hand,
when setting $\gamma = 0$ and $\taue = 1$, the results are consistent with
those of the standard disc as determined in \citet{Bertout88}.  On the other
hand, the model locally determines the flaring index (Eq.~\ref{eq:implgamma})
without using the radial structure, and we checked that the derived radial
structure fulfilled $\deriv\Hdisk r = \gamma$.

\subsection{Spectral energy distributions}

In order to compute a synthetic image or a SED, we divide each disc annulus in
angular sectors.  For each sector we determine the emergent flux in the
observer's direction as the sum of three contributions: the contributions of
the outer layers {\Fe}, of the inner layer {\Fi}, and of simple isotropic
scattering of stellar radiation by the surface of the disc {\Fs}.  These fluxes
depend on the wavelength-dependent optical thicknesses {\tauil} and {\tauel} of
the inner and outer layers.  The incidence of the line of sight onto the
visible surface of the disc is given by its angle $i$, and its incidence
onto the opposite surface by $i'$.
\begin{subequations}
\begin{align}
   \Ftot &= \Fi + \Fe + \Fs \\
\intertext{with}
   \Fi   &= \left(1 - \expo{-\frac{2\tauil}{\cos i}} \right) 
           \expo{-\frac{\tauel}{\cos i}} 
           B_\lambda(\Tphot),
           \label{eq:Fi}\\
   \Fe   &= \left( \left(1-\expo{-\frac{\tauel}{\cos i}}\right) 
              + \left( 1-\expo{-\frac{\tauel}{\cos i'}} \right) 
                   \expo{-\frac{2\tauil}{\cos i}} 
           \right) 
           B_\lambda(\Te),
           \label{eq:Fe}\\
   \Fs   &= \frac{\alb \sin(\phi)}{2} \left(\frac {r_*}{r}\right)^2 
           B_\lambda(T_*),
           \label{eq:Fs}
\end{align}
\end{subequations}
where $B_\lambda$ is the black-body function. {\Tphot} is the temperature of
the inner layer at optical depth $\cos i$ at wavelength $\lambda$ if $\tauil >
\cos i$, and its mid-plane temperature in the other case.  

This determination allows a fast computation but is only an approximation.  It
is only valid when the outer parts of the disc shadows neither the star nor
other parts of the disc; therefore a more adequate radiative transfer
determination must be carried out for edge-on discs, which is beyond the scope
of this paper.

As a test, we have checked that the bolometric flux predicted by the SED is
consistent with the one predicted by the effective temperatures of the
heating processes.  We noticed discrepancies of 2--3\% in most cases, up to 
8\% in a few ones; they are connected to the use of the mean opacities 
{\kappar} and {\kappap} in the structure calculation.

\subsection{Visibilities}

The image and visibilities are determined using the same method as for the SED.
However, optical visibilities at long baselines can be both sensitive to
large-scale structures up to 1'' (\latin{e.g.} Palomar Testbed Interferometer
field of view) and to small scale structures down to 0.1~mas.  As the
visibility function is the Fourier transform of the image, a straightforward
determination would require a huge number of image pixels ($\approx
10^4\times10^4$).

As one of the goals of the present model is rapidity, we avoid this
determination.  Our image $I(x, y)$ contains the central structure with about
$10^3\times10^3$ pixels.  All flux falling outside this image is integrated as
a single value {\Fout}. Then, the non-normalised visibility $\cal V$ is given
by
\begin{equation} 
   {\cal V} (u, v) = 
   \begin{cases} 
      \tilde{I}(0, 0) + \Fout           &\text{if $u = v = 0$}\\ 
      \tilde{I}(u/\lambda, v/\lambda)   &\text{otherwise} 
   \end{cases} 
   \label{eq:calV} 
\end{equation}
where $\tilde{I}$ is the Fourier transform of $I$. This equation states that
the {\Fout} contribution cancels as soon as the baseline is non-zero, which
means that the outer parts of the disc are fully resolved for any arbitrarily
small baseline and would produce a visibility of 0 if they were alone. Two
approximations are made: (i) the outer parts of the disc do not contain
small-scale structures (ii) for too small baselines Eq.~(\ref{eq:calV}) is not
valid. Our model does include neither irregularities nor steep variations in
the outer parts of the disc so that approximation (i) is valid.  The smallest
baseline for which Eq.~(\ref{eq:calV}) is valid is determined by the field of
view of the image\footnote{The visibility is the discretised Fourier transform
of the image and the smallest non-zero baselines obtained are given by $B =
\lambda/\theta$, where $\theta$ is the field of the image.  As a consequence
Eq.~(\ref{eq:calV}) is only valid for baselines $\gtrsim B$.} and we
choose this field in order to produce correct results for the smallest
baselines used in long-base optical interferometry ($\approx 10$~m).


\section{Model analysis}
\label{sec:ana}

In this section study the influence of disc flaring hypotheses, compare our
model with other authors' models, analyse the influence of different heating
processes, show the influence of viscosity prescription and examine the
influence of some disc parameters on both the structure and the observables.
For the sake of comparison, authors compute a fiducial model, typical
of a T~Tauri star; its parameters are displayed in Table~\ref{tab:param1}.

\begin{table}[t]
   \caption{Parameters used for the fiducial model}
   \label{tab:param1}
   \begin{tabular}{ll}
      \hline\hline
      Parameters        & fiducial  model  \\
      \hline
      $r_*$ ($\Rsun$)            & $2.0$\\
      $r_{\mathrm{min}}$ (\Rsun) &  6.0 \\
      $r_{\mathrm{max}}$ (AU)    &  100 \\
      $M_*$ (\Msun)              &  1.0 \\
      {\Mdot} (\Msun/\yr)        & \sci{1.0}{-8} \\
      $T_*$ (K)                  & $4000$        \\
      $i$ (\deg)                 & $0$           \\
      $A_V$ (mag)                & $0.0$         \\
      \alb                       & $0.2$         \\
      $\alpha$                   & \sci{1}{-2}   \\
      \hline
   \end{tabular}
\end{table}

From now on, we use model~2 in our study, unless specified otherwise 
(sect.~\ref{sec:flhyp} and appendices).

\subsection{Importance of disc flaring hypotheses}
\label{sec:flhyp}

\begin{figure*}[t!]
   \subfigure[$\Ti$]{\includegraphics[width=0.31\hsize]{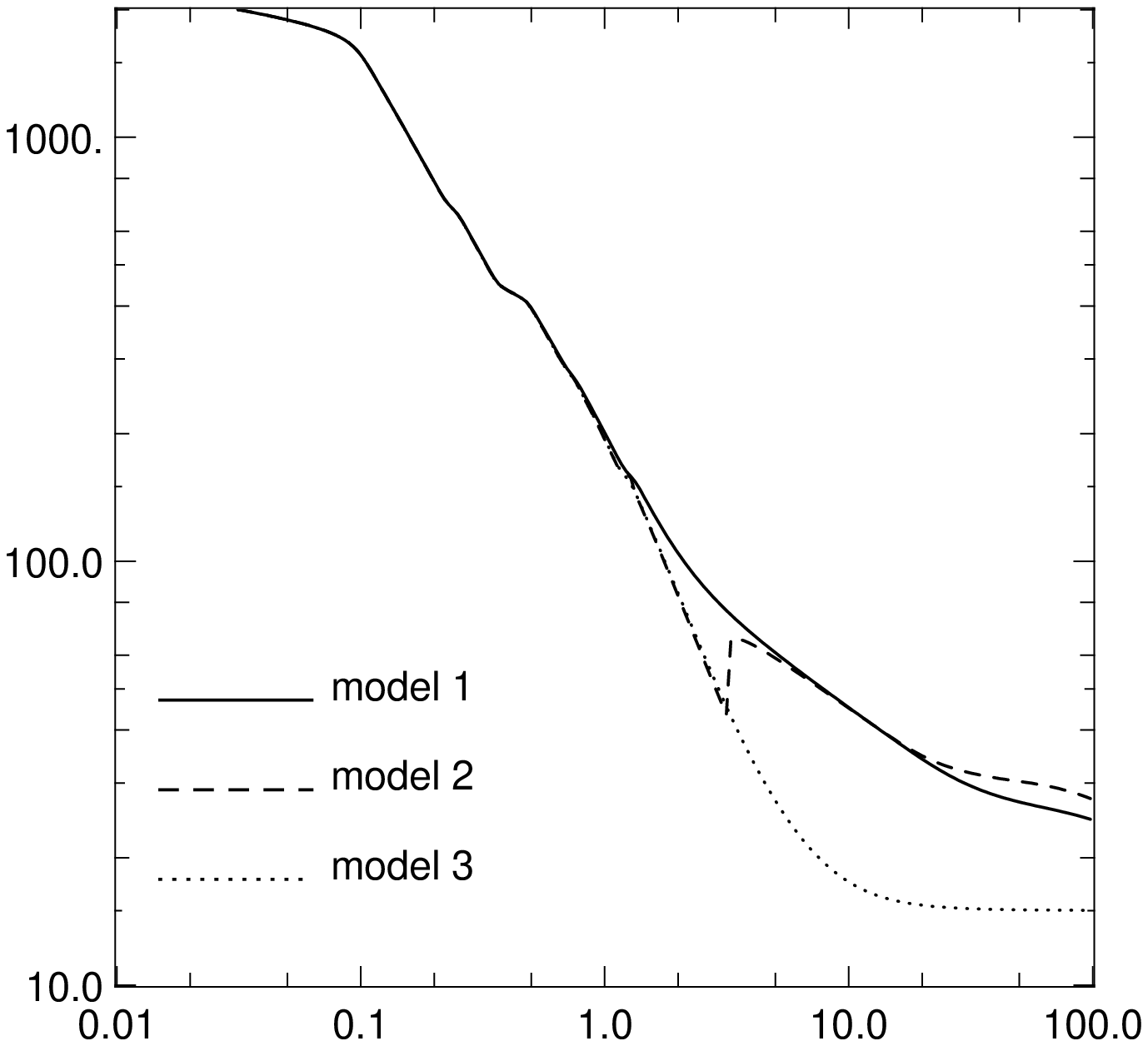}}\hfill
   \subfigure[$\Te$]{\includegraphics[width=0.31\hsize]{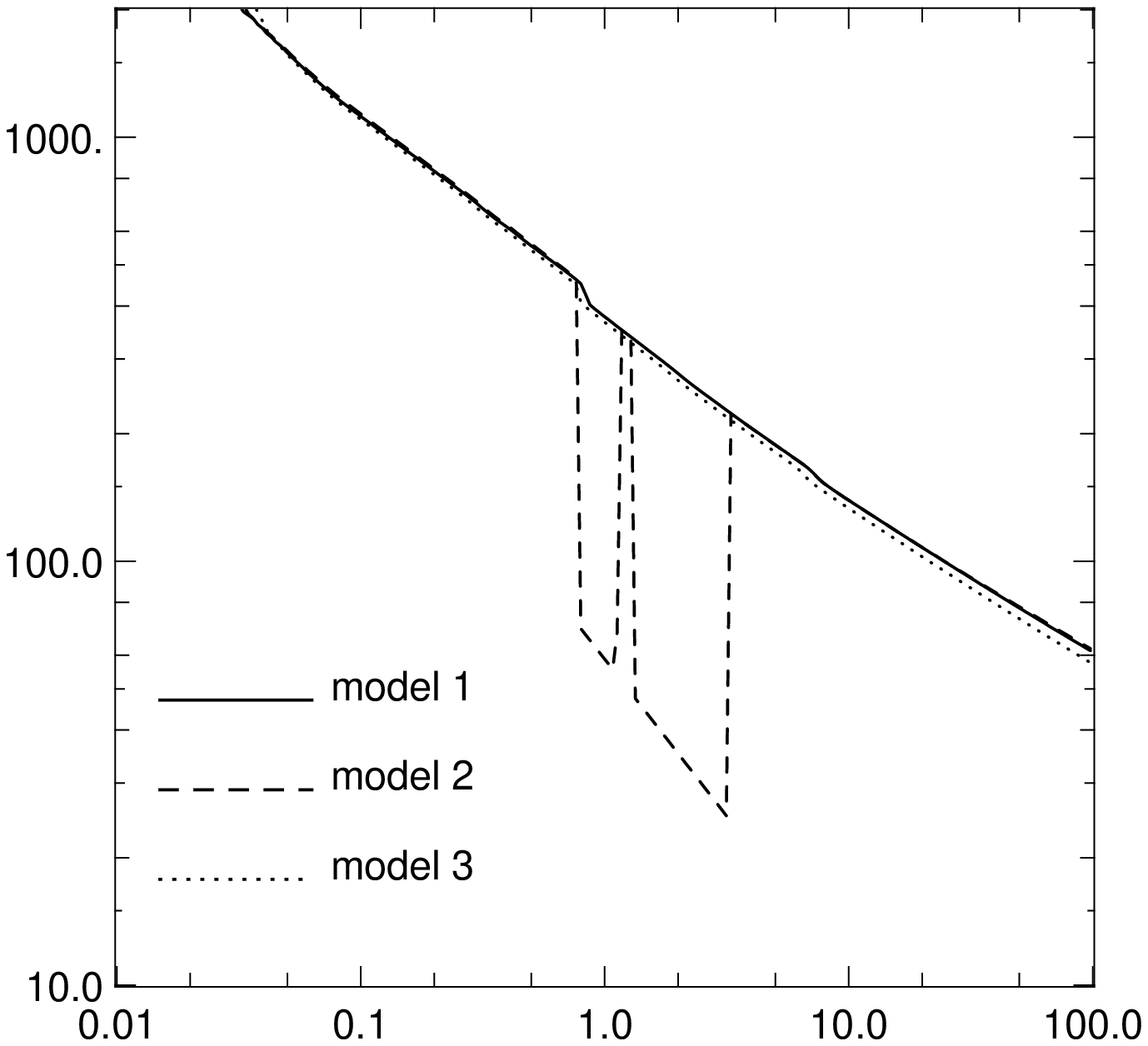}}\hfill
   \subfigure[$h/r$]{\includegraphics[width=0.31\hsize]{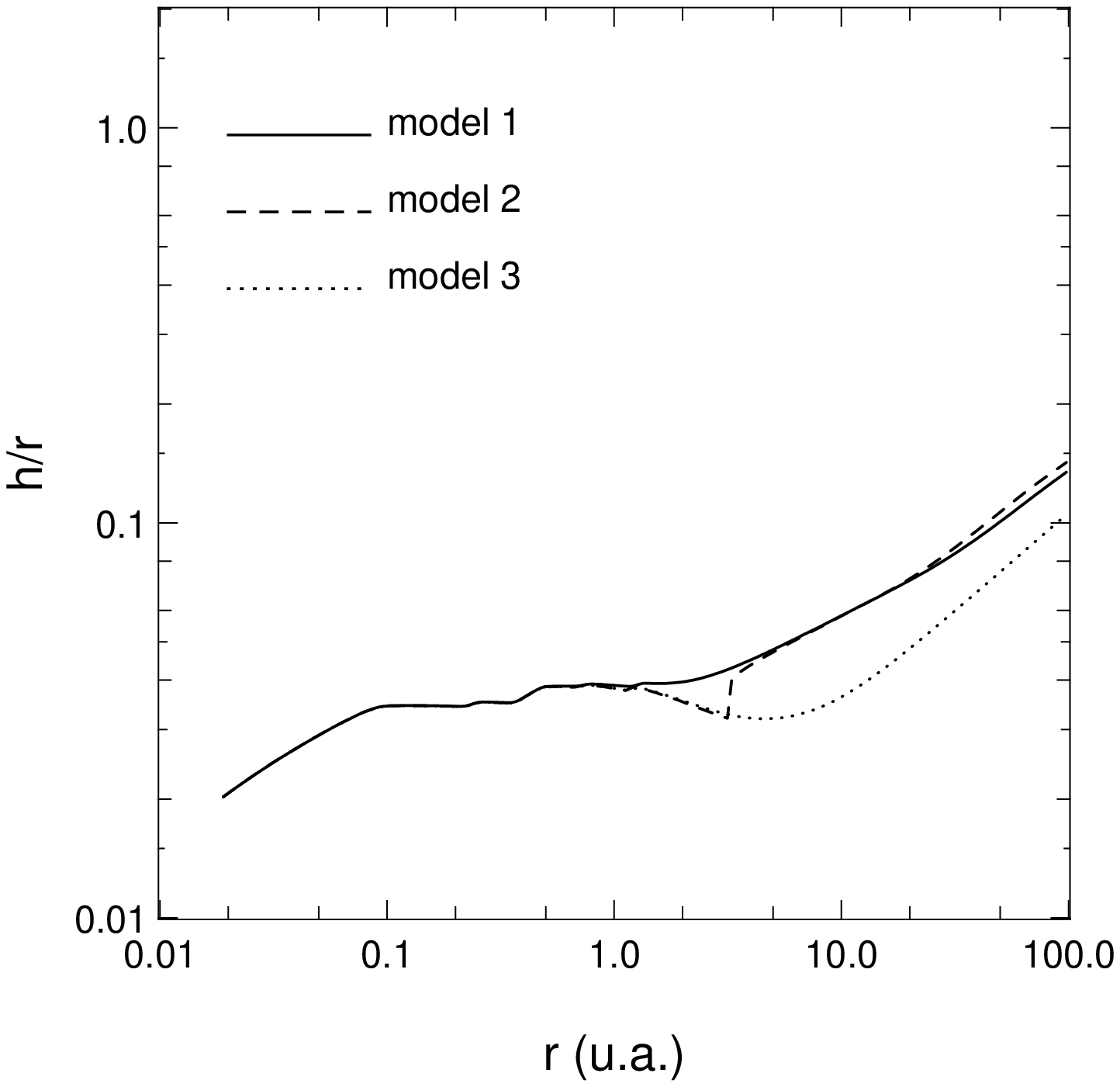}}\hfill
   \vglue -10pt%
   \caption{%
      Influence of disc flaring hypotheses on the structure. Model~1:
      thickness proportional to the scale height, model~2: self-similar
      thickness, model~3: flat disc.  Left: mid-plane temperature, middle:
      surface temperature, right: relative scale height.
   }
   \label{fig:hyp}
\end{figure*}

Figure~\ref{fig:hyp} compares the structure of the fiducial T~Tauri disc in
three cases:  the disc thickness is proportional to the scale height (model~1),
the disc thickness is given by the reprocessing terms only (model~2), and the
disc is flat (model~3).

The major difference between models 1 and 2 is the presence of self-shadowing
in the first one: at a radius of a few AUs, the surface temperature of model 1
drops because the disc is not directly illuminated by the star (see middle
panel of Fig.~\ref{fig:hyp}).  The cause is a decrease of the relative scale
height $h/r$ (see right panel on the same figure), hence a similar behaviour of
the disc surface because of the proportionality $\Hdisk \propto h$.  \latin{A
contrario}, model~2 assumes a thickness increasing faster than $r$  ($\Hdisk
\propto r^{1+\gamma}$), so the surface is always illuminated.

Model~3 is a flat disc, therefore it catches much less light from the central
star:  the outer parts are much cooler (Figure~\ref{fig:hyp}, $\Ti$, left
panel).  If the surface has the same temperature (cf. $\Te$, middle panel),  it
produces a much smaller flux because the outer layer happens to be extremely
thin.   Such a disc has a lesser scale height because it is cooler (cf. $h/r$,
right panel).

As we shall see in Sect.~\ref{subsec:comp-models},  numerical simulations tend
to prove that there is no self-shadowing effect in T~Tauri discs.
\citet{DAlessio99} establishes that even with a scale height flaring inwards,
there is still enough material at large $z$ to catch stellar light, so that all
parts of the disc are illuminated.  Moreover,  we noticed that model~2, with
the \latin{ad hoc} flaring index, compares much better with other simulations
than model~1.  So, we shall continue our study with model~2.

\subsection{Viscosity prescription}

\begin{figure*}[t!]
   \subfigure[$\Ti$]{\includegraphics[width=0.31\hsize]{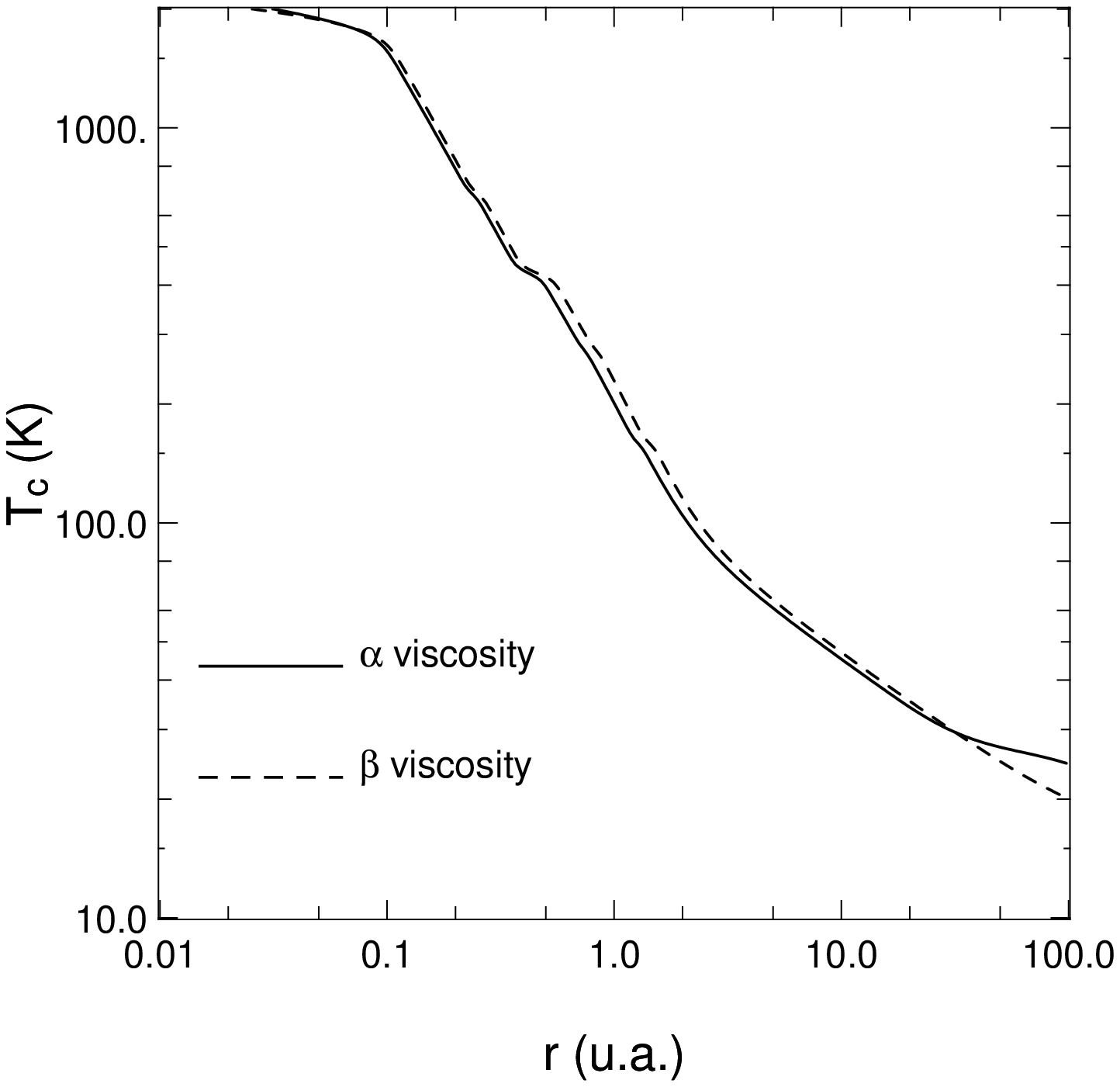}}\hfill
   \subfigure[$\Md$]{\includegraphics[width=0.31\hsize]{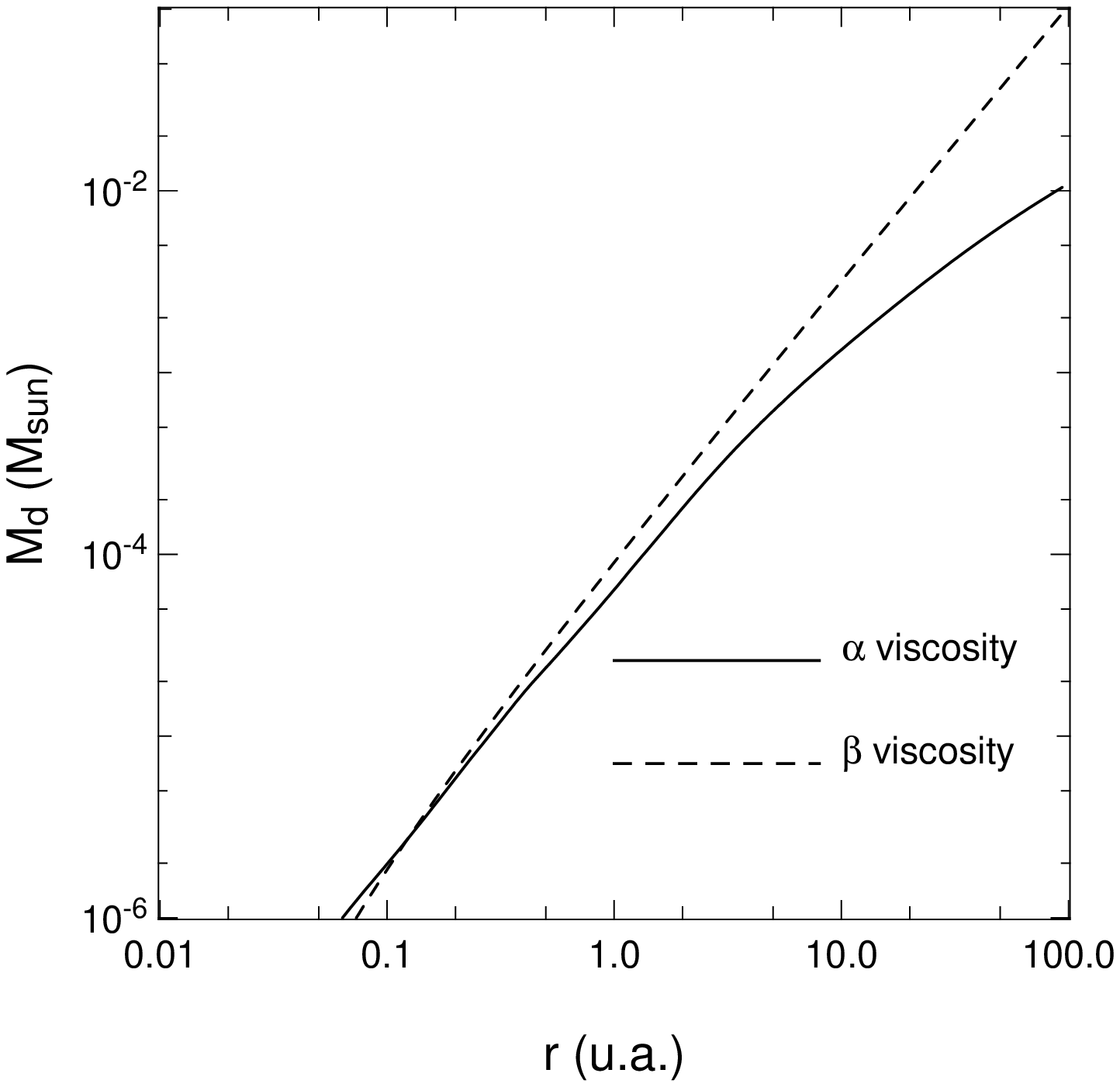}}\hfill
   \subfigure[SED]{\includegraphics[width=0.31\hsize]{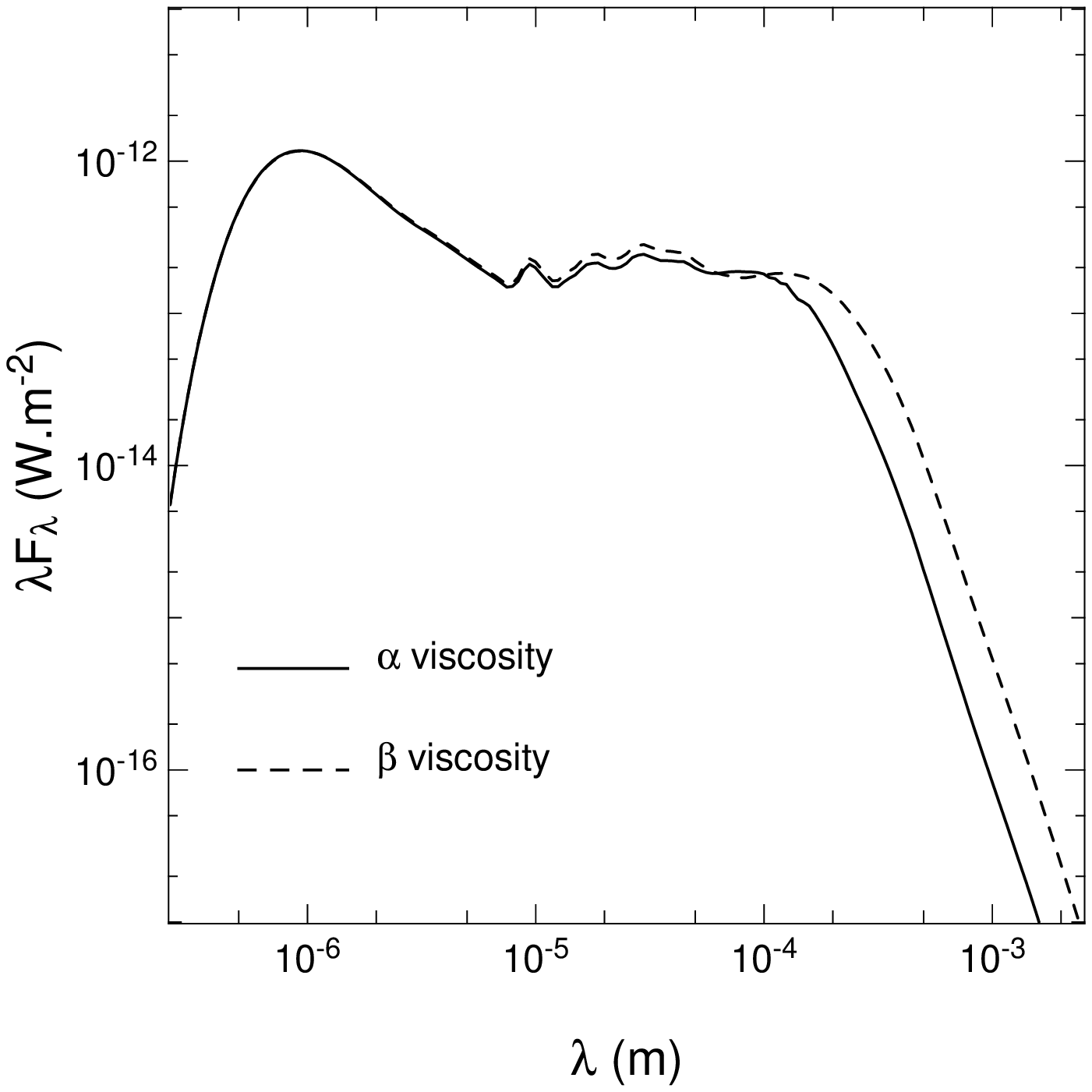}}\hfill
   \vglue -10pt%
   \caption{%
      Influence of the viscosity prescription.  Solid lines:
      $\alpha$-prescription by \citet{Shakura73}; dashed lines:
      $\beta$-prescription by \citet{Hure01}.  Right panel: mid-plane
      temperature vs. radius; middle panel: mid-plane temperature vs. radius;
      right panel: SED.
   }
   \label{fig:visc}
\end{figure*}

We compare the standard $\alpha$ viscosity prescription with the $\beta$
prescription derived from laboratory experiments by \citet{Hure01}.  $\beta$
is defined as
\begin{equation}
     \nui = \beta \omegak r^2
\end{equation}
and is a slowly varying factor ranging from \ten{-6} to \ten{-2}.  We
keep it constant over the disc.

We chose $\beta = 10^{-5}$ so that the viscosity {\nui} has the same order of
magnitude as with the $\alpha$-model in the central parts of the disc. This
prescription gives much higher column density and disc mass in the outer parts
of the disc, as displayed on Fig.~\ref{fig:visc}.  Note that the impact of
viscosity on the observables is small, since both viscous effective temperature
and reprocessing flux do not directly depend on it.  In the mid-IR, the SED is
only slightly affected, via a subtle modification of the disc thickness
{\Hdisk}, with an impact of the reprocessing temperature.  Larger wavelength
observations ($> 100$\mum) probe the outer parts of the disc, and should become
more sensitive to the influence of the viscosity law, because the flux emerging
from these optically thin parts is proportional to $\kappa \Sigma \Ti$.  

\subsection{Comparison with other models}
\label{subsec:comp-models}

\begin{figure*}[p]
   \null\hfill
   \subfigure[{\Ti}]{\includegraphics[width=0.32\hsize]{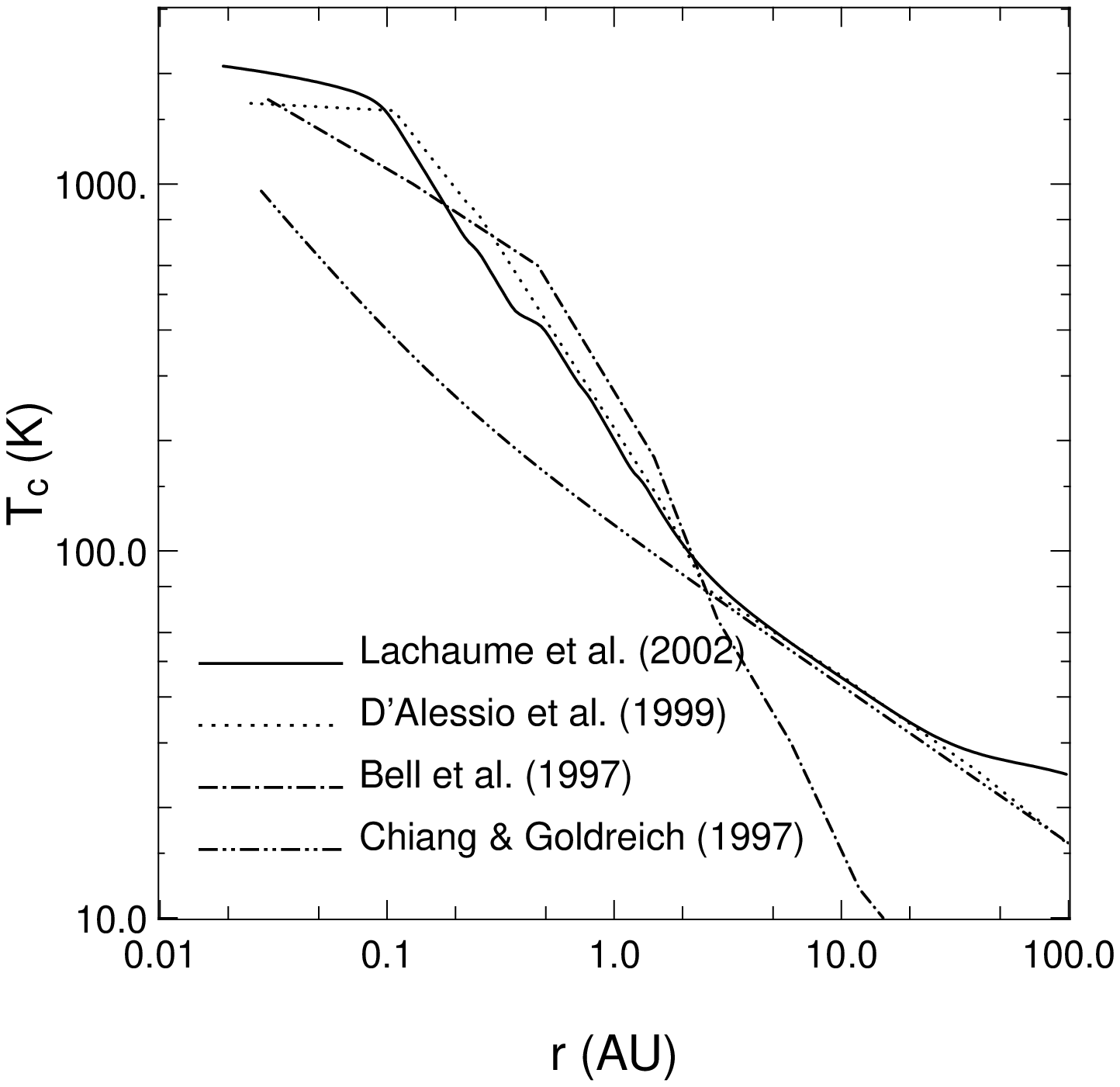}\label{fig:compareTc}}\hfill
   \subfigure[{\TR}]{\includegraphics[width=0.32\hsize]{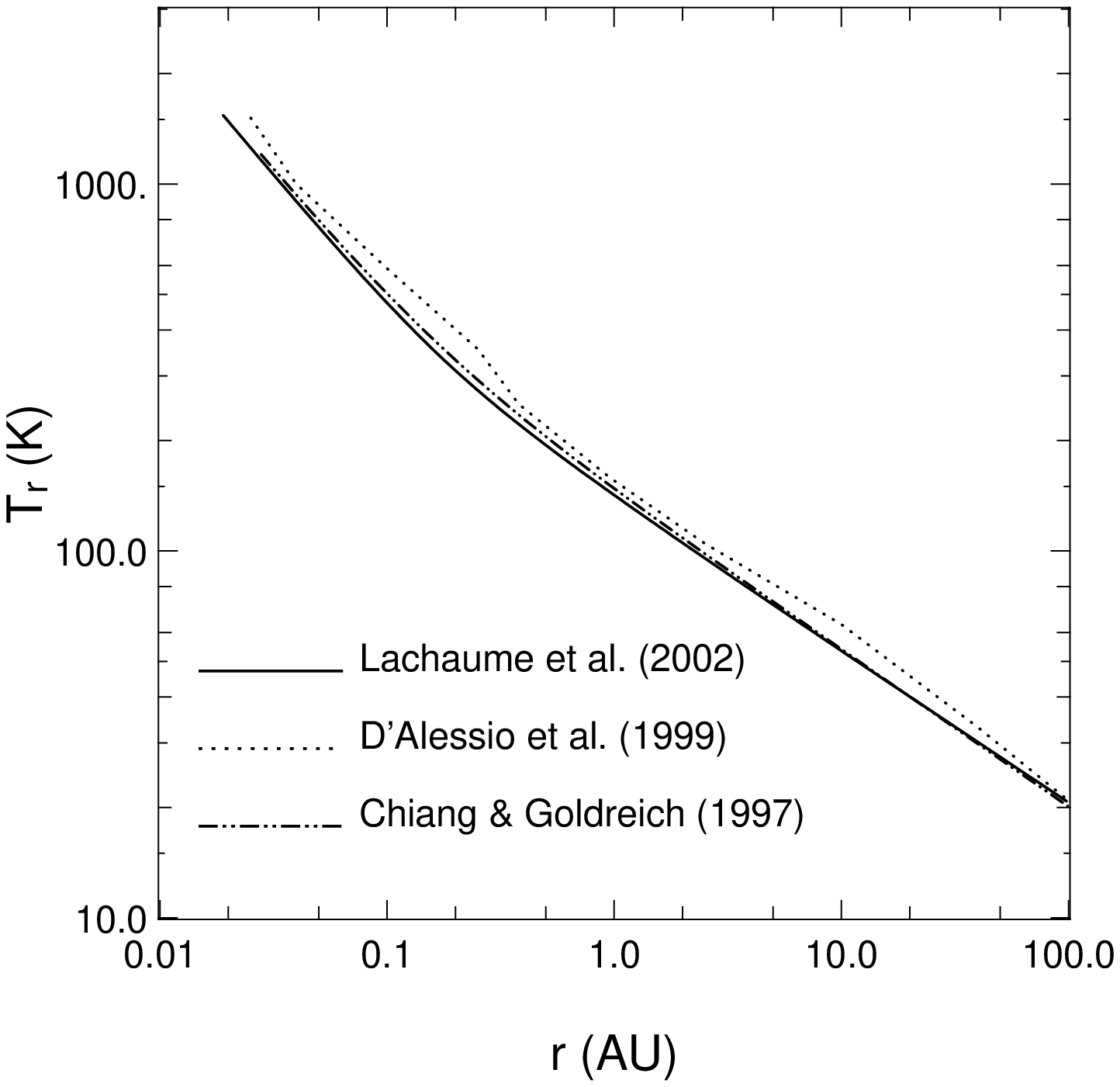}\label{fig:compareTr}}\hfill
   \subfigure[{\Te}]{\includegraphics[width=0.32\hsize]{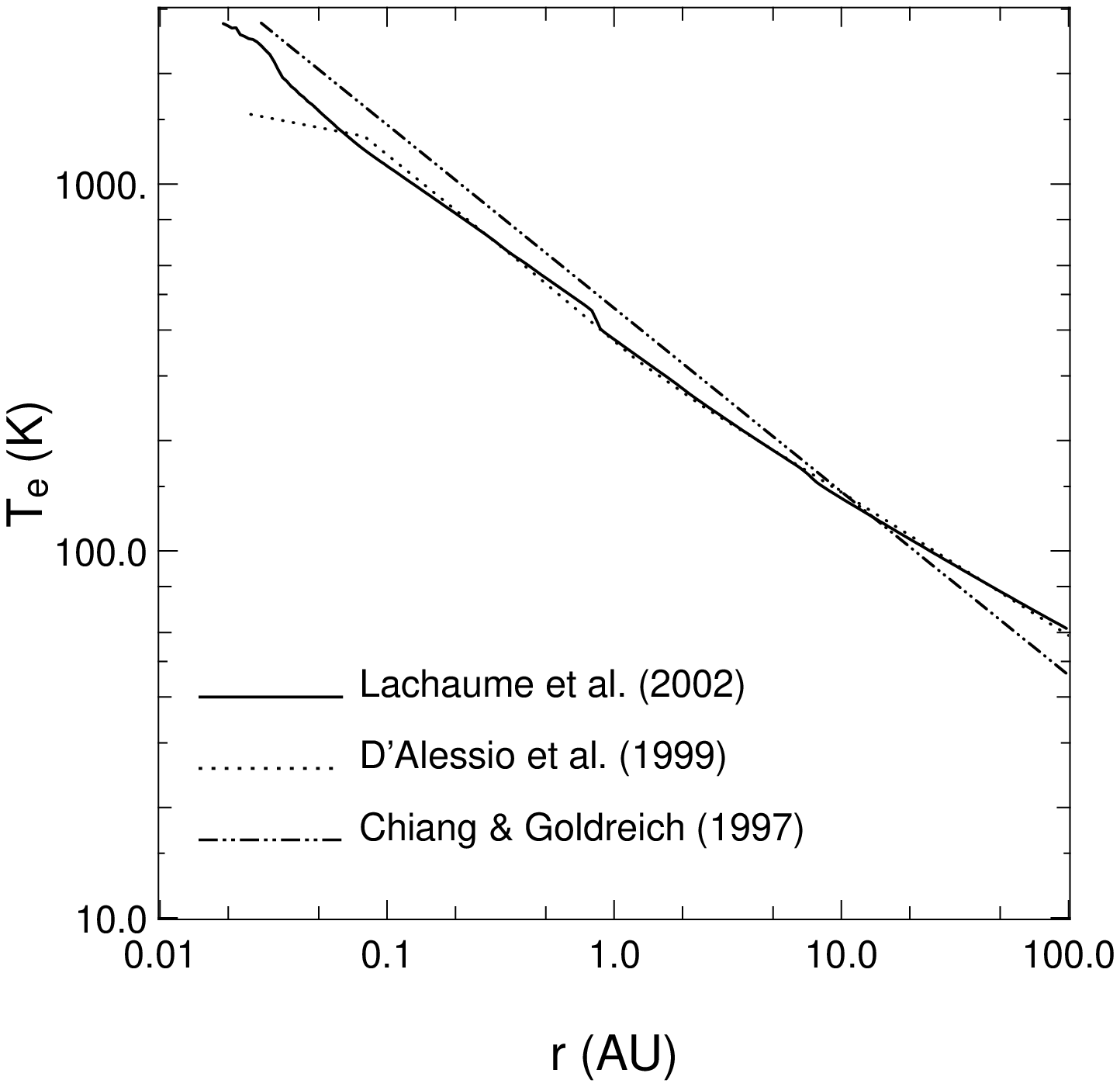}\label{fig:compareTe}}\hfill
   \hfill\null\\
   \null\hfill
   \subfigure[$\hi/r$]{\includegraphics[width=0.32\hsize]{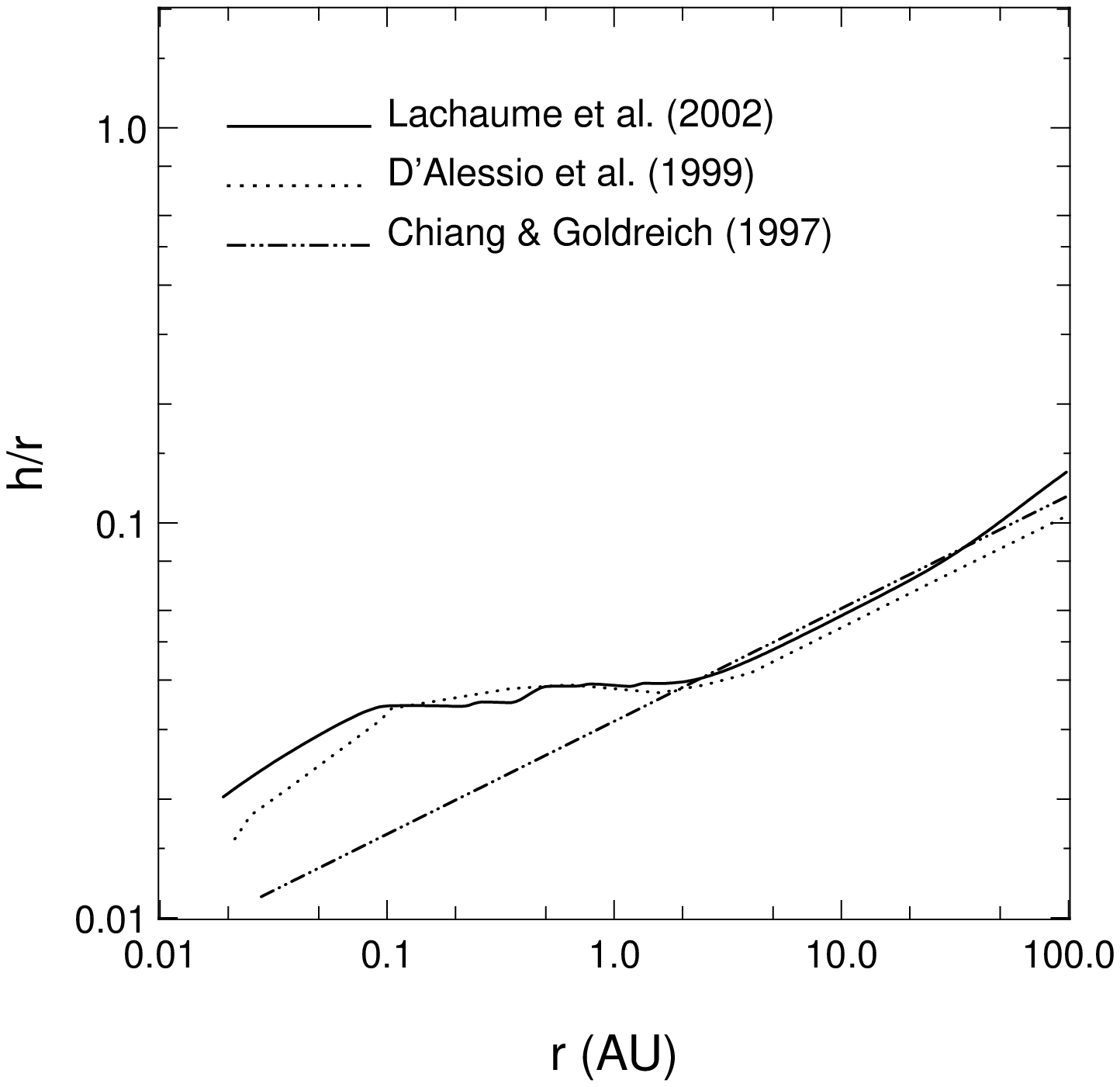}\label{fig:compareh}}\hfill
   \subfigure[$\Hdisk/r$]{\includegraphics[width=0.32\hsize]{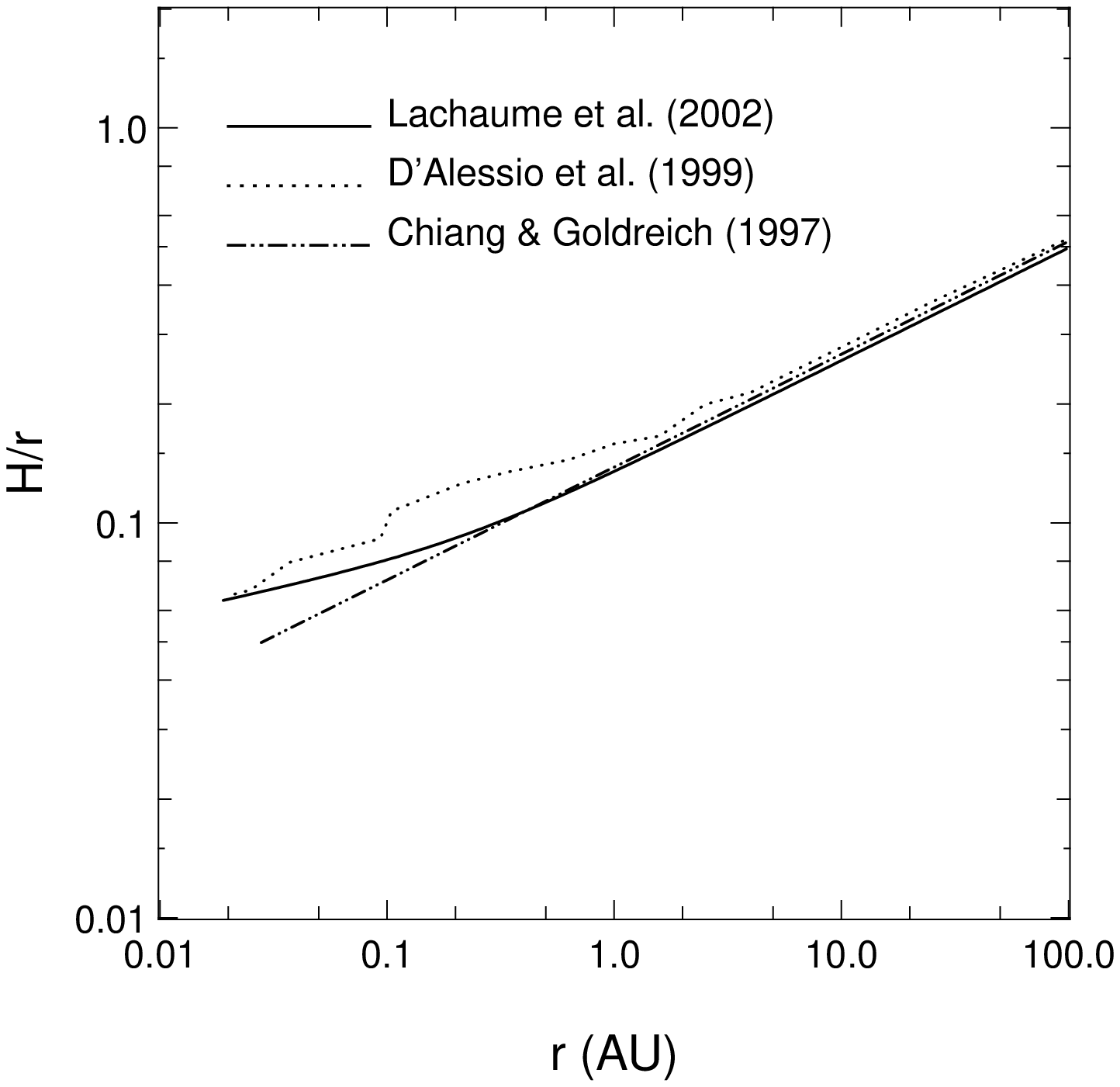}\label{fig:compareH}}\hfill
   \subfigure[$\Sigma$]{\includegraphics[width=0.32\hsize]{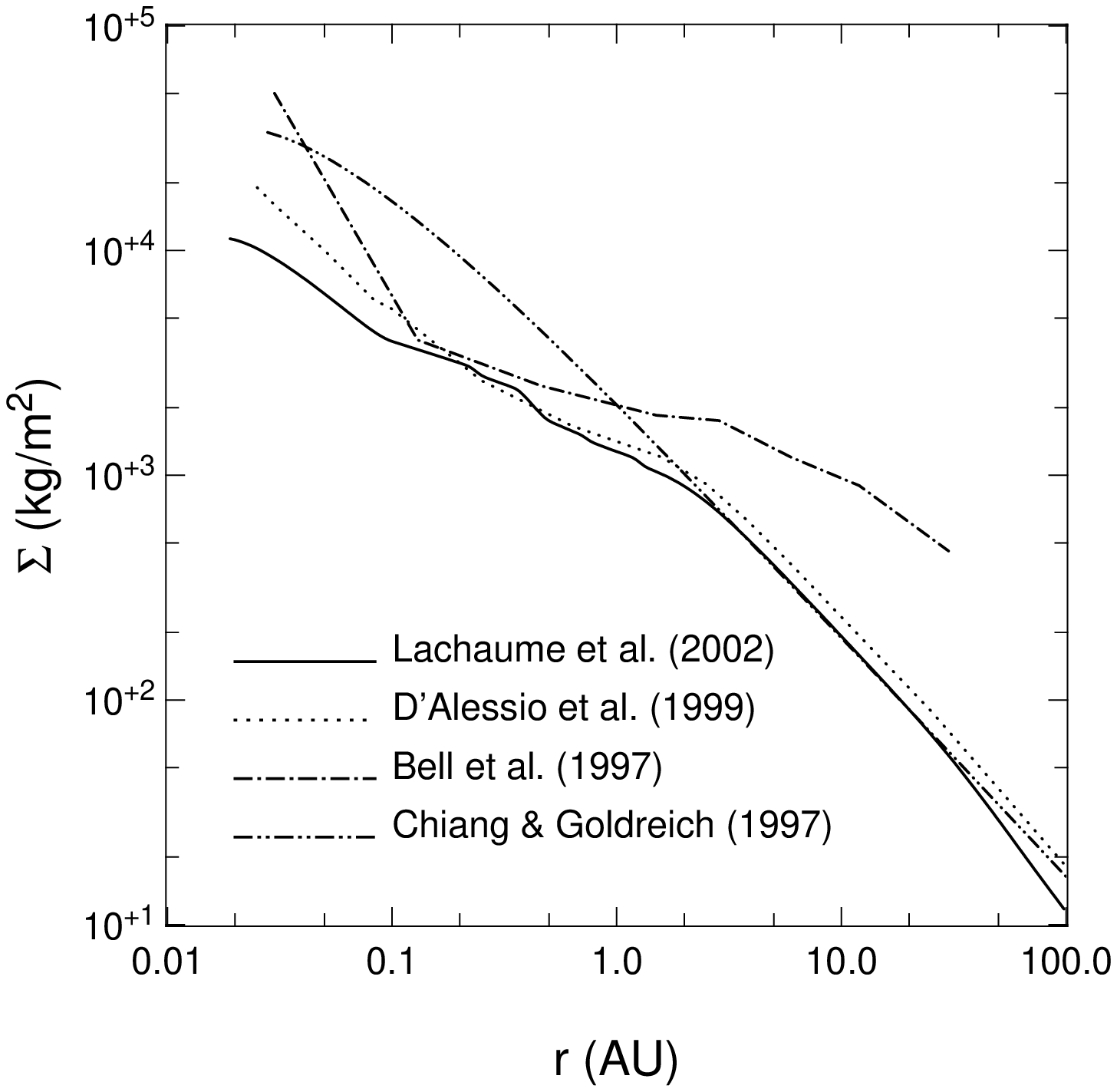}\label{fig:comparesigma}}%
   \hfill\null\\
   \null\hfill
   \subfigure[{\Md}]{\includegraphics[width=0.32\hsize]{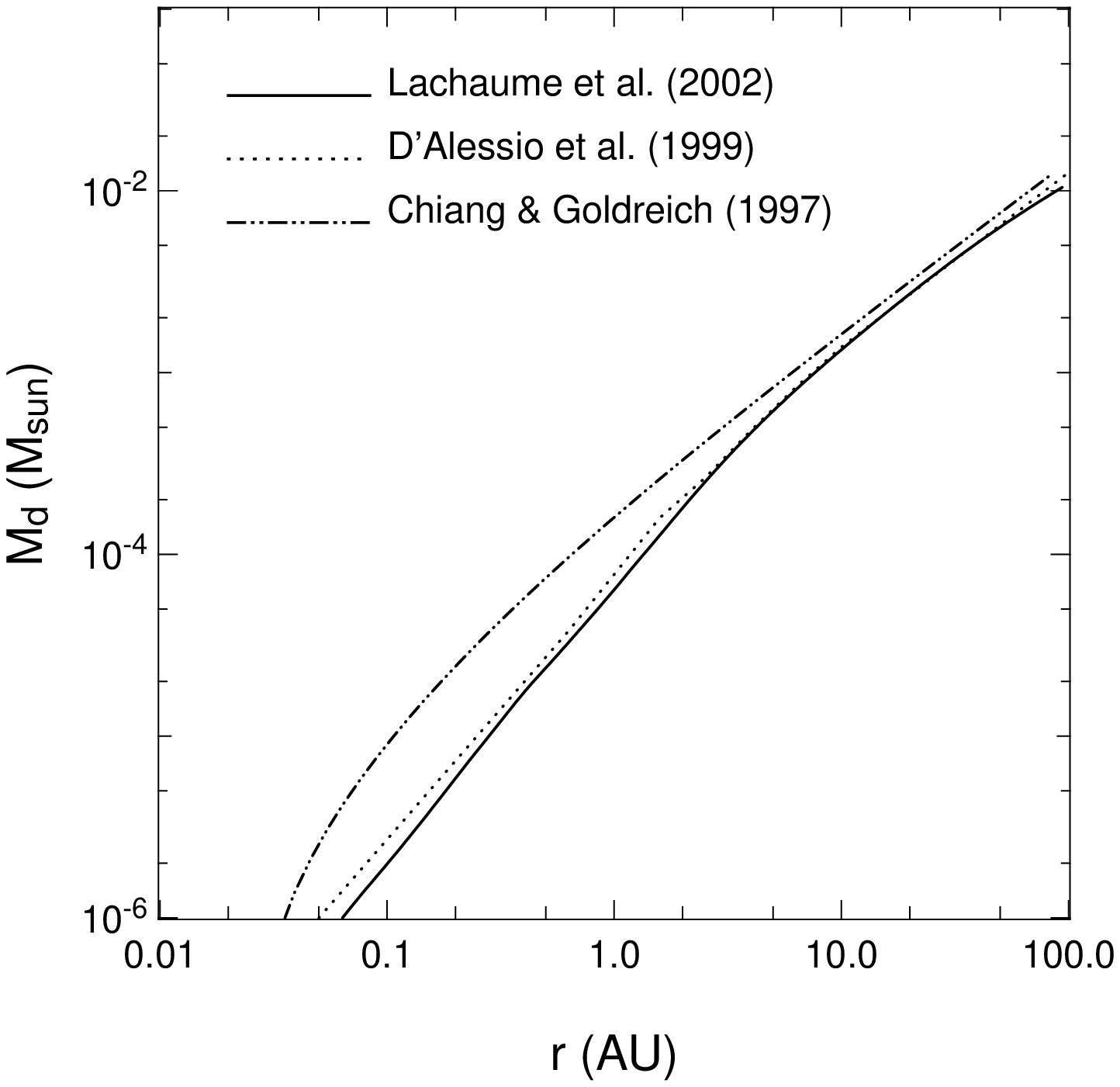}\label{fig:compareMd}}\hfill
   \subfigure[$Q$]{\includegraphics[width=0.32\hsize]{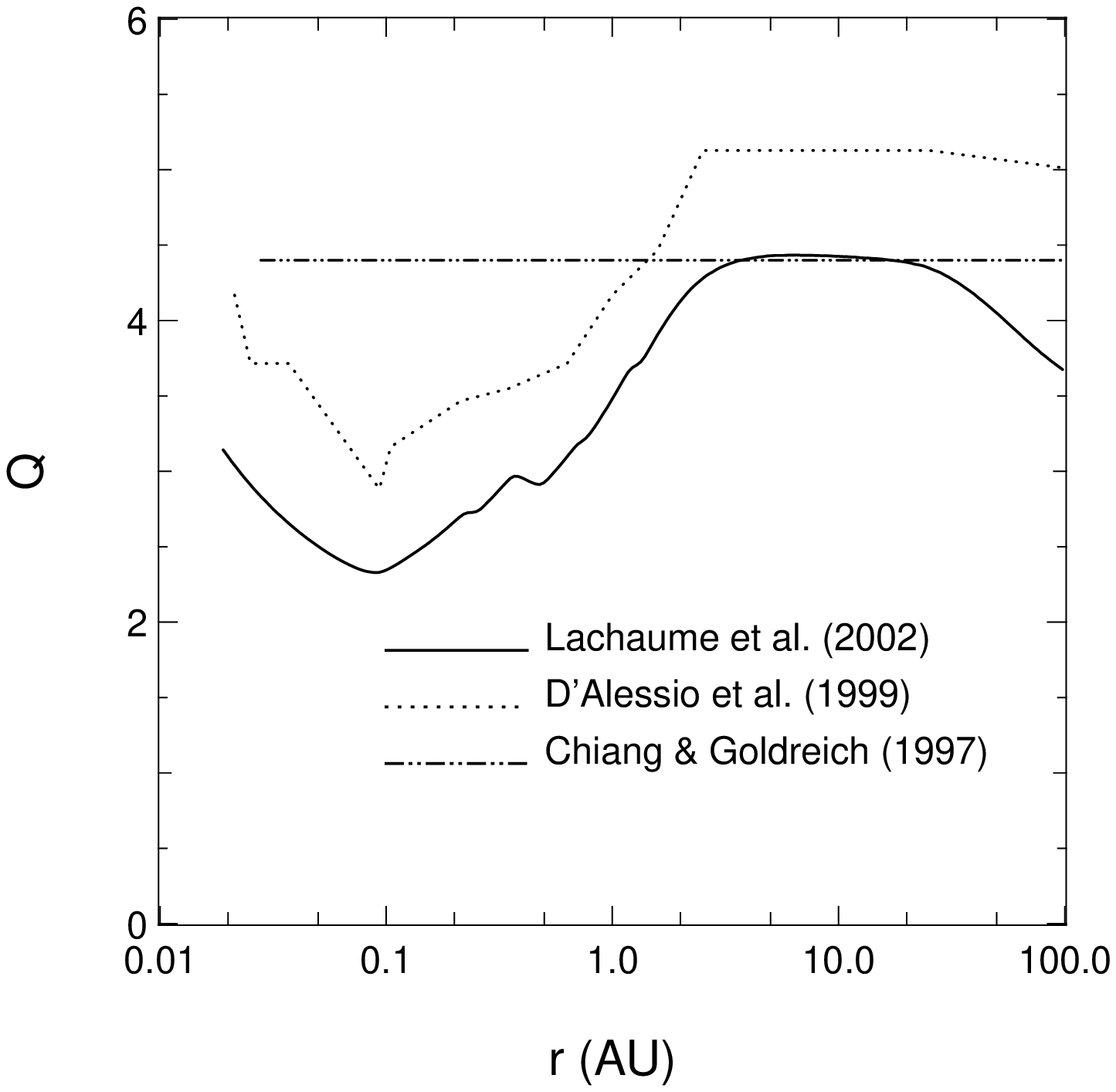}\label{fig:compareQ}}\hfill
   \subfigure[SED]{\includegraphics[width=0.32\hsize]{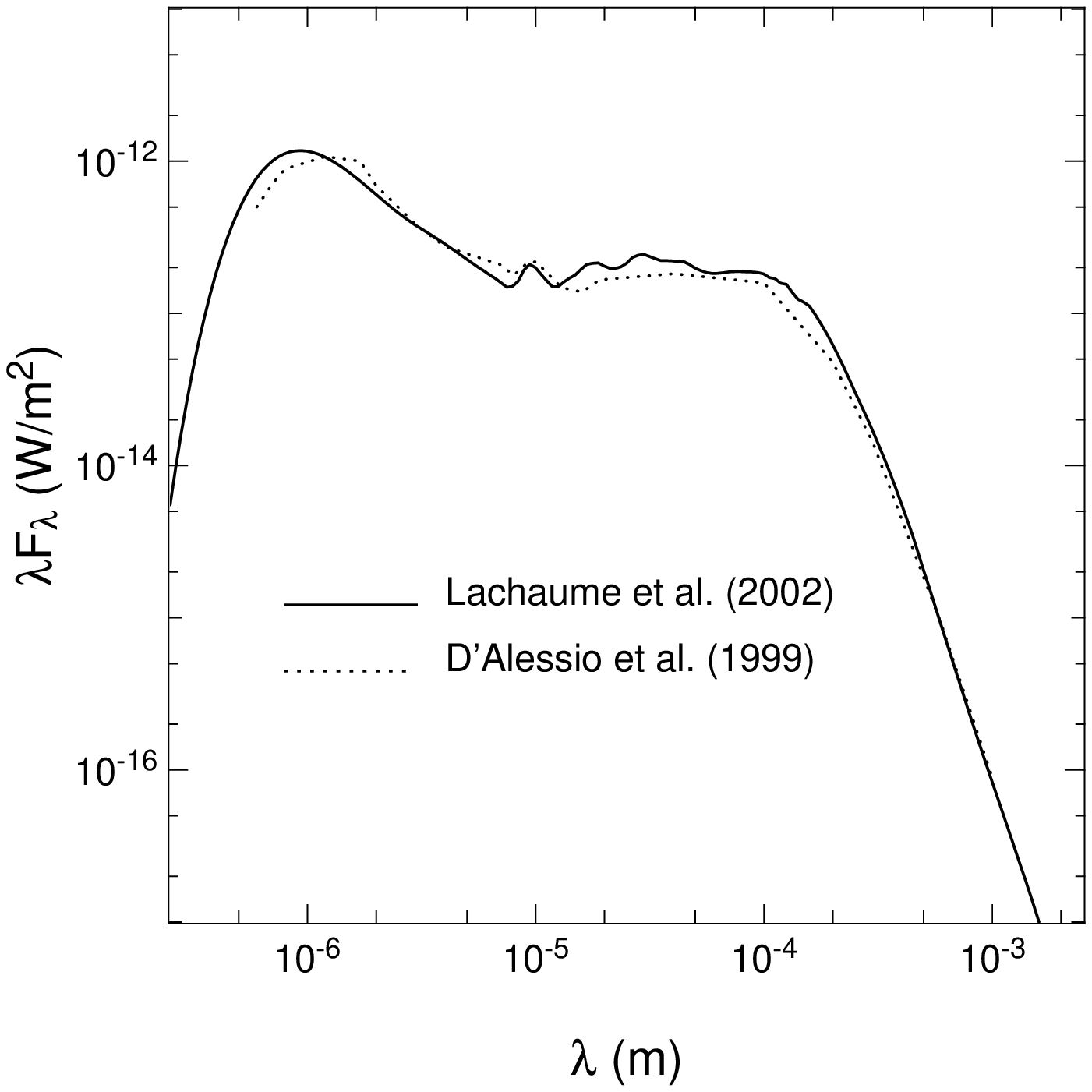}\label{fig:comparesed}}%
   \hfill\null
   \caption{%
      Comparison between disc models: present work with $\alpha$ and
      $\beta$ prescription, \citet{DAlessio99}, \citet{Bell97} and
      \citet{Chiang97}.  The panels are: (a) mid-plane temperature (b)
      effective temperature associated to stellar irradiation (c) surface
      temperature (d) scale height (e) mean height of reprocessing (f) column
      density (g) cumulative disc mass (h) relative geometrical thickness of
      the disc (k) spectral energy distribution of the pole-on disc.%
   }
   \label{fig:compare} 
\end{figure*}

We computed the two-layer structure of the fiducial model
(Table~\ref{tab:param1}) and compare it model with others listed in
Table~\ref{tab:models} together with their main characteristics.
Figure~\ref{fig:compare} displays some physical conditions describing the
radial structure of the fiducial disc, as forecast by these models.  As a
general result, despite of the approximation made, our model is in good 
agreement with previous ones.  

\begin{table}[t]
   \def\albeta{$\alpha,\beta$}
   \caption{%
      Disc model characteristics compared in section~\ref{subsec:comp-models}.
      We mention whether stellar light reprocessing, viscous heating and
      convective transport are taken into account, whether the vertical
      transfer is solved or vertically averaged, and the viscosity prescription
      used.
   }
   \label{tab:models}
   \begin{tabular}{llllll}
      \hline\hline
                              &       a &       b &        c &        d &        e\\
      \hline
      reprocessing            &     yes &      no &      yes &      yes &       no\\
      viscosity               &     yes &     yes &      yes &       no &      yes\\
      convection              &      no &      no &       no &      yes &       no\\
      vertical transfer       &      no &      1D &       1D &       no &       1D\\
      viscosity prescription  & \albeta & \albeta & $\alpha$ & 
      $-$ & $\alpha$\\
      \hline
      \multicolumn{5}{l}{a: present model}\\
      \multicolumn{5}{l}{b: paper~II}\\
      \multicolumn{5}{l}{c: \citet{DAlessio99}}\\
      \multicolumn{5}{l}{d: \citet{Chiang97}}\\
      \multicolumn{5}{l}{e: \citet{Bell97}}\\
      \hline
   \end{tabular}
\end{table}

Some discrepancies with the \citet{Bell97} and \citet{Chiang97} models come
from different heating hypotheses.  \citet{Chiang97} obtain lower values for
temperatures in the inner parts of the disc, and higher mass columns since
$\Sigma \nu$ is constant (see Fig.~\ref{fig:compareTc}~\& \ref{fig:compareMd}).
The reason is that they do not include viscous heating, predominant in the
first AUs of T~Tauri discs (see Fig.~\ref{fig:contrib} for $r \lesssim
1$~AU).  In the outer parts of the disc, dominated by reprocessing, the
\citet{Bell97} model differs significantly because it does not take this
process into account.

Our model and that of \citet{DAlessio99} present very close predictions
in terms of central temperature, surface temperature or scale height.
\citet{Hure01b} already noticed that the flat vertically average disc
model is a good approximate of an active disc;  we have now demonstrated
that a two-layer model is a faithful description of discs both illuminated
by the central star and heated by viscosity.  However there are still
some discrepancies:
\begin{itemize}
   \item  Our prediction of the disc thickness $\Hdisk$ is an underestimation
   in the range 0.1--1\,AU, so is the effective temperature of reprocessing
   $\TR$.
   \item Despite of a smaller effective temperature than in \citet{DAlessio99},
   our SED is a slight overestimation in the range 20--100\,$\mu$m. 
   There is a likely reason: the two-layer structure tends to over-estimate the
   temperature of emission by the outer layer of the disc, hence the 
   far-infrared excess.
   \item  We find important discrepancies in the region $r \lesssim 0.1$\,AU,
   because our opacities at intermediate temperatures ($T \gtrsim 1800$\,K) are
   different from those of \citet{DAlessio99}.
\end{itemize}
These discrepancies do not affect much the results concerning SED and
long-baseline visibility model fitting.

\subsection{Location of heating processes}

\begin{figure*}[t!]
   \subfigure[$\Ti$]{\includegraphics[width=0.31\hsize]{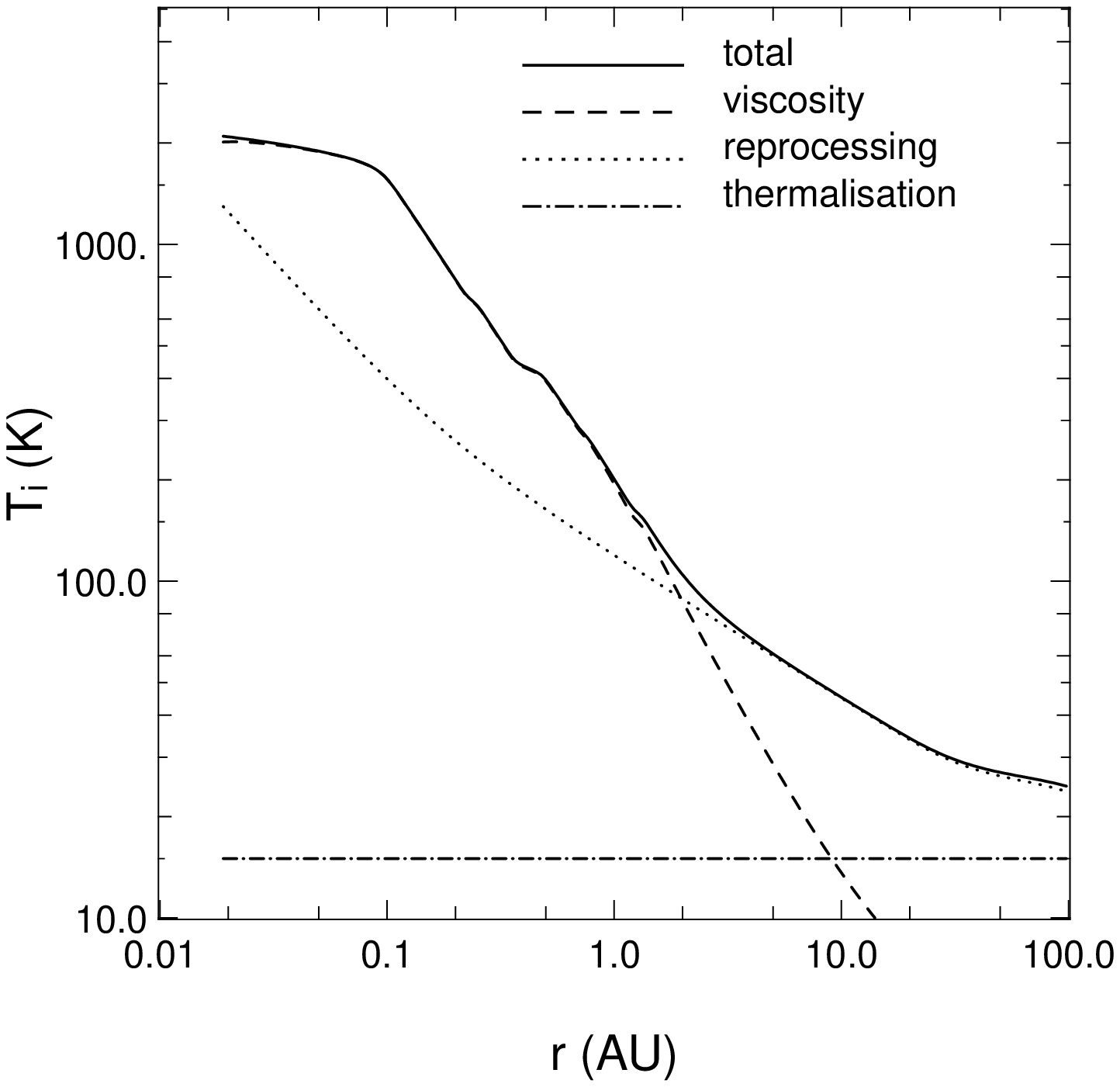}}\hfill
   \subfigure[$\Teff$]{\includegraphics[width=0.31\hsize]{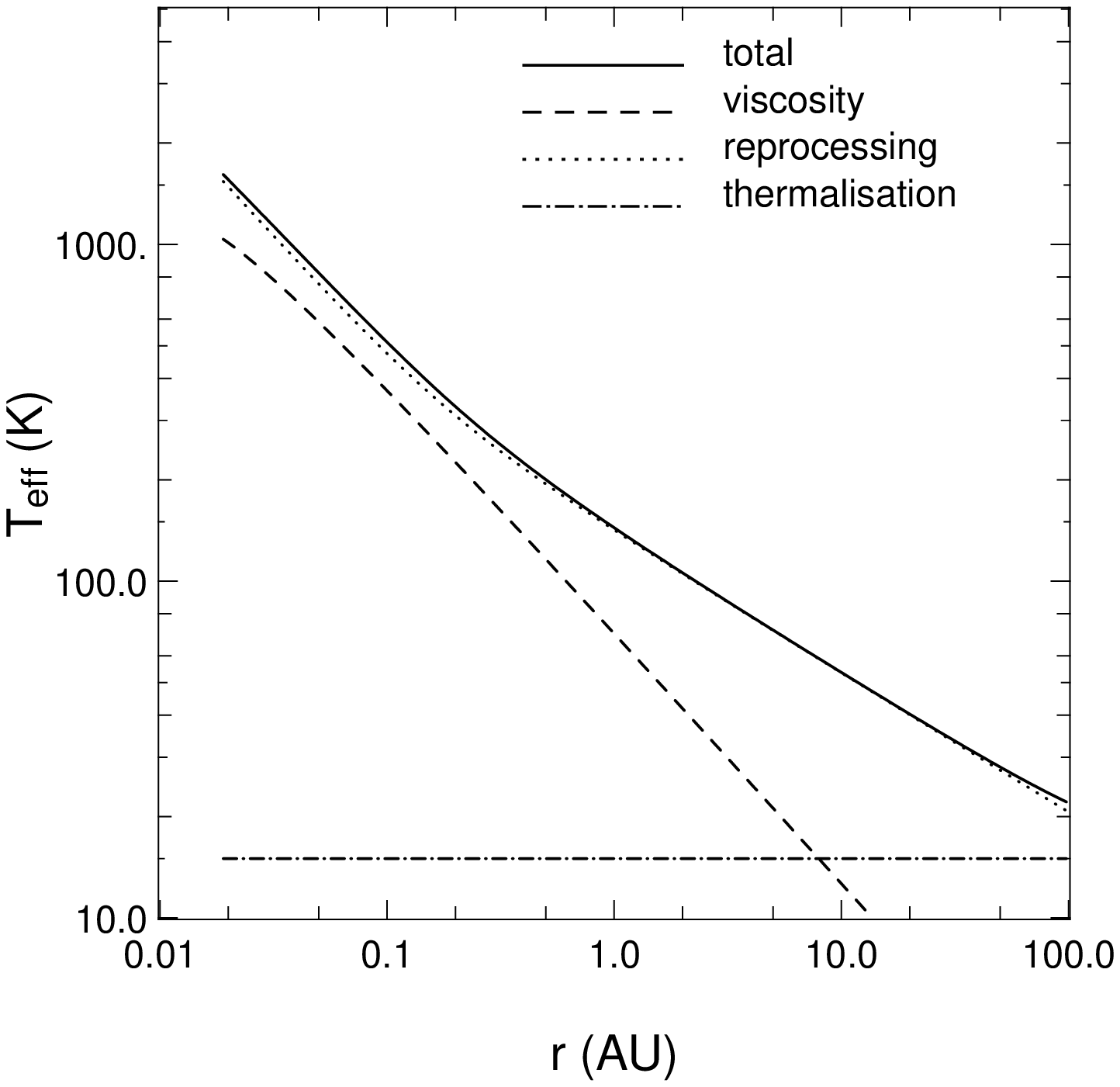}}\hfill
   \subfigure[$\Te$]{\includegraphics[width=0.31\hsize]{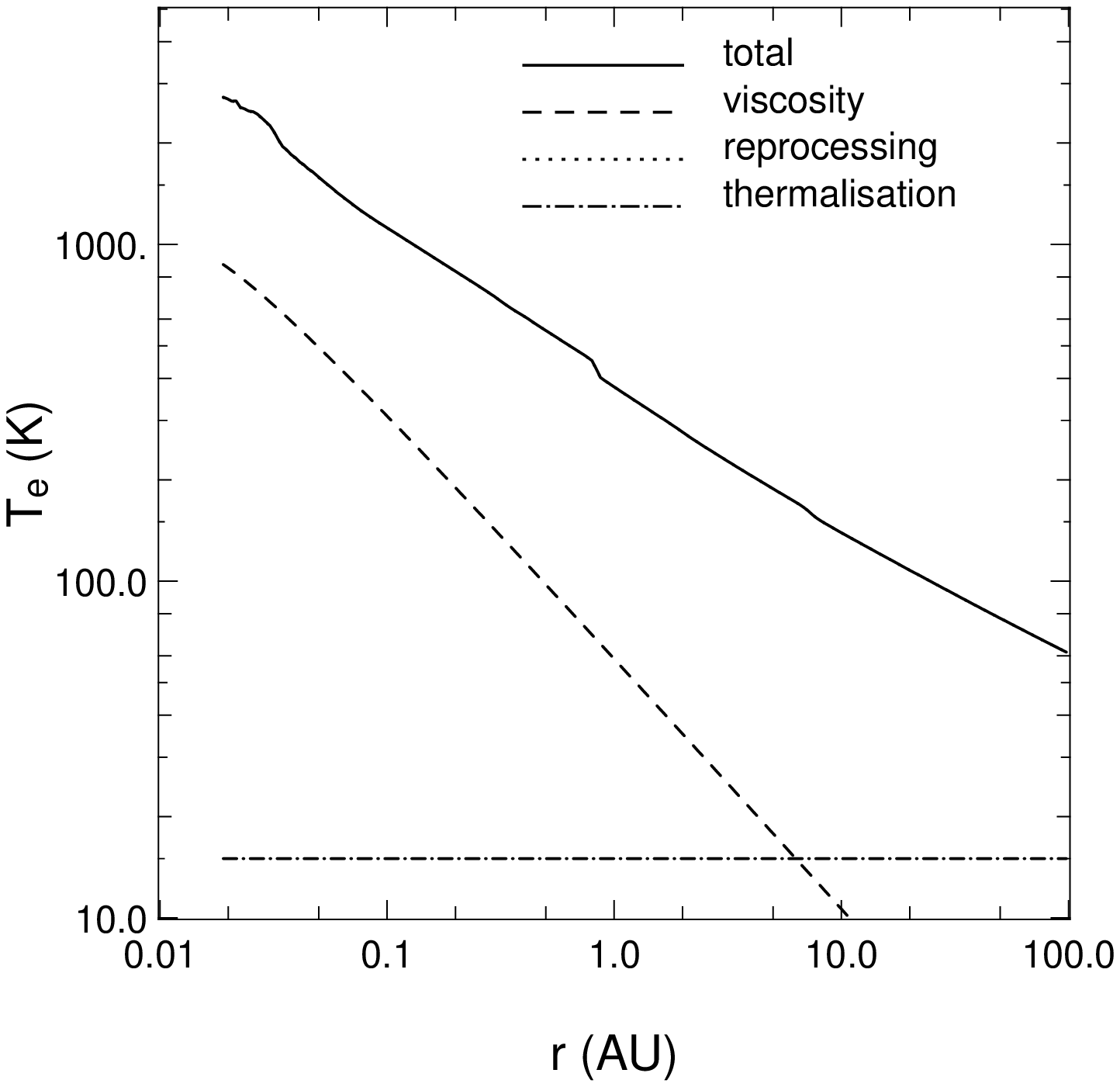}}
   \caption{%
      Contribution of heating processes: radial profile of these contributions
      to the temperature of the disc (model~2).  Left: mid-plane temperature, 
      middle: surface temperature, right: surface temperature.
   }
   \label{fig:contrib}
\end{figure*}

Figure~\ref{fig:contrib} displays the radial profile of contributions of each 
heating process to the temperature of a typical T~Tauri disc.  The mid-plane
temperature of the fiducial model is dominated by reprocessing for 
$r \gtrsim 2$\,AU and by viscous dissipation for $r \lesssim 2$\,AU.  However,
in terms of effective temperature, the disc is dominated by reprocessing
at any radial location, with a marginal contribution of viscosity for
$r \lesssim 0.2$\,AU.   The outer layer is completely dominated by stellar
light reprocessing at any distance from the star.  One important conclusion to 
draw is that the notion of a disc dominated by an heating process depends on 
what quantity we are interested in.

\subsection{Thermal flux vs. scattered light}

\begin{figure*}[t!]
   \subfigure[J band]{\includegraphics[width=0.31\hsize]{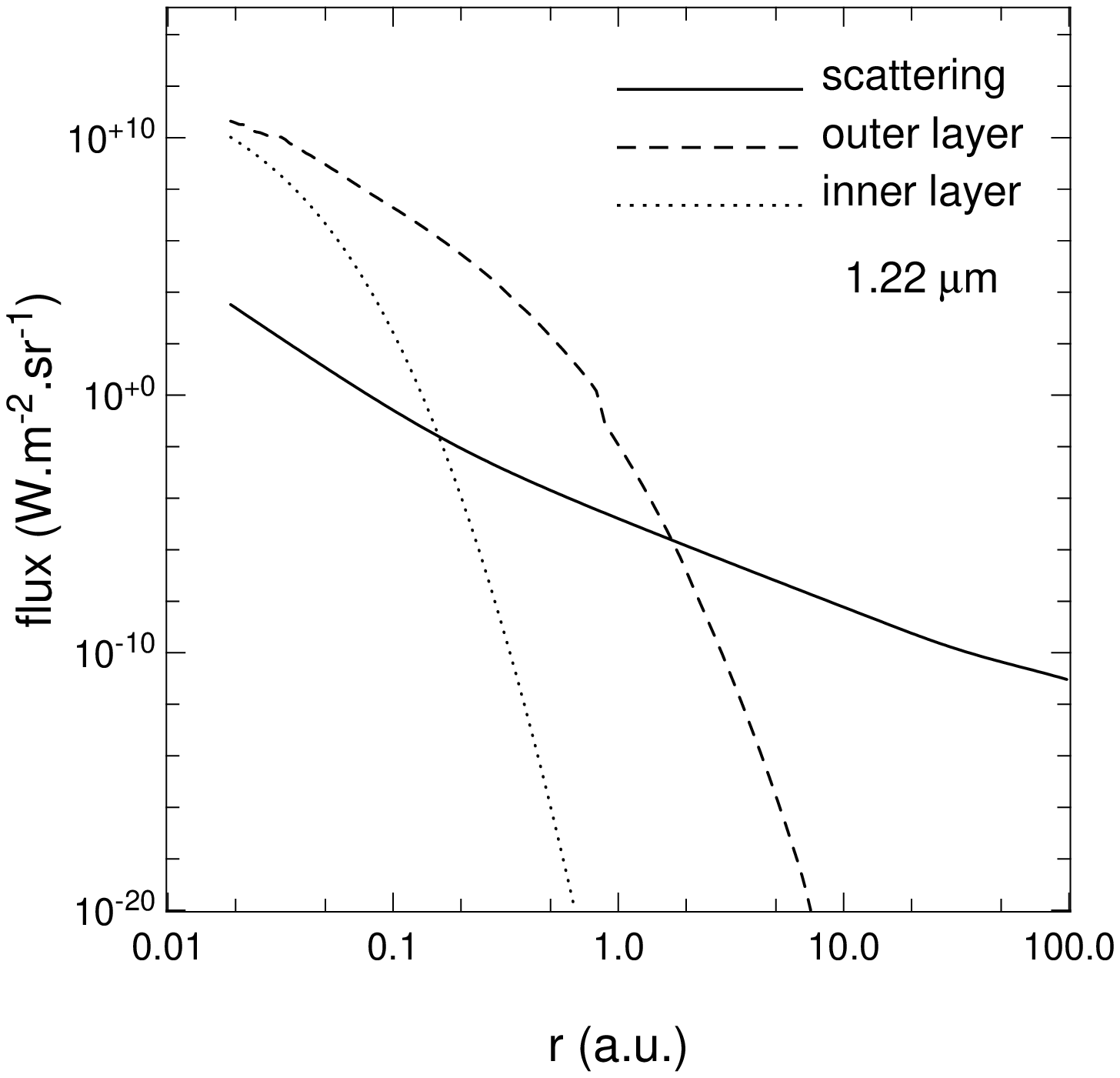}}\hfill
   \subfigure[K band]{\includegraphics[width=0.31\hsize]{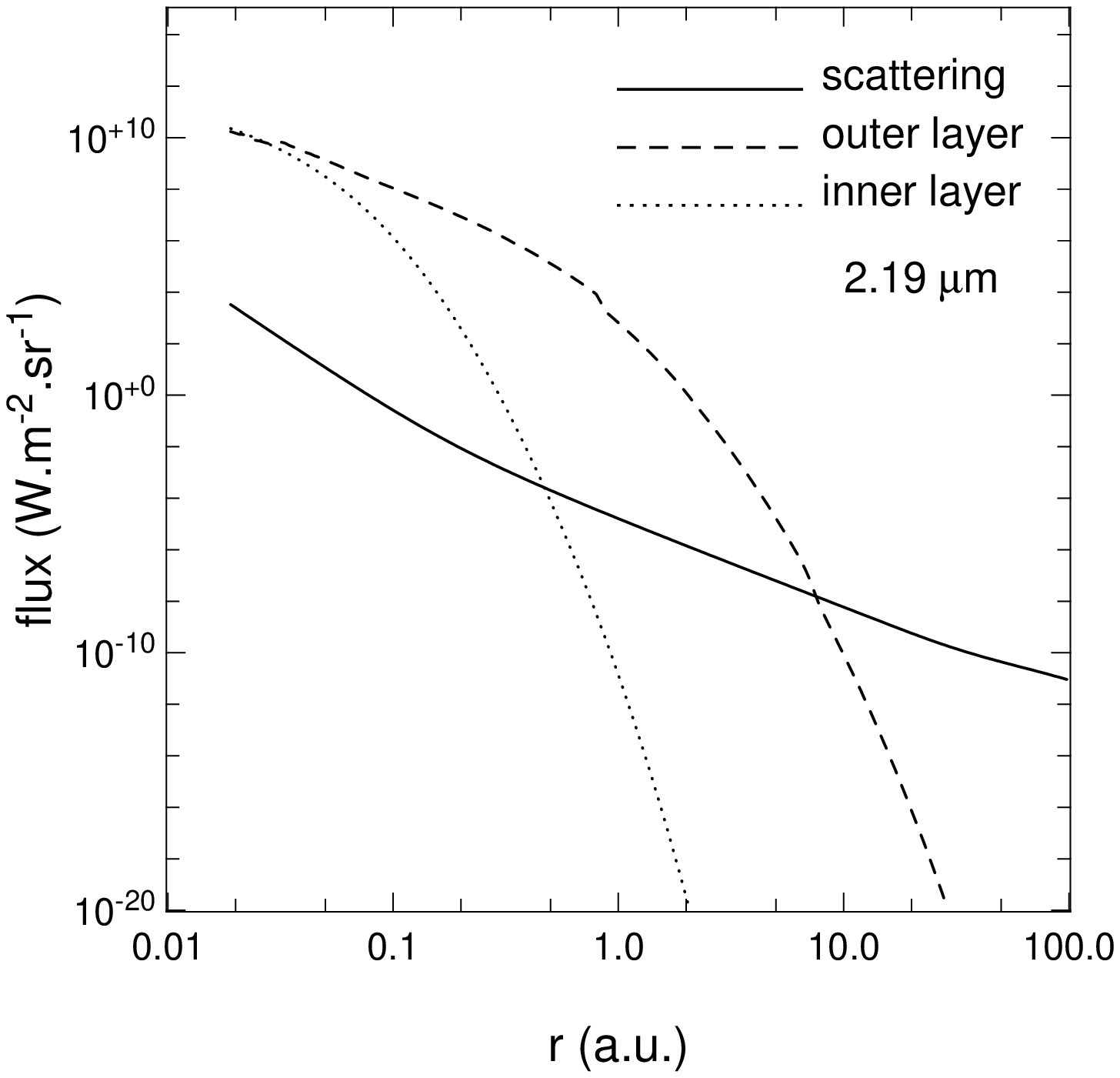}}\hfill
   \subfigure[N band]{\includegraphics[width=0.31\hsize]{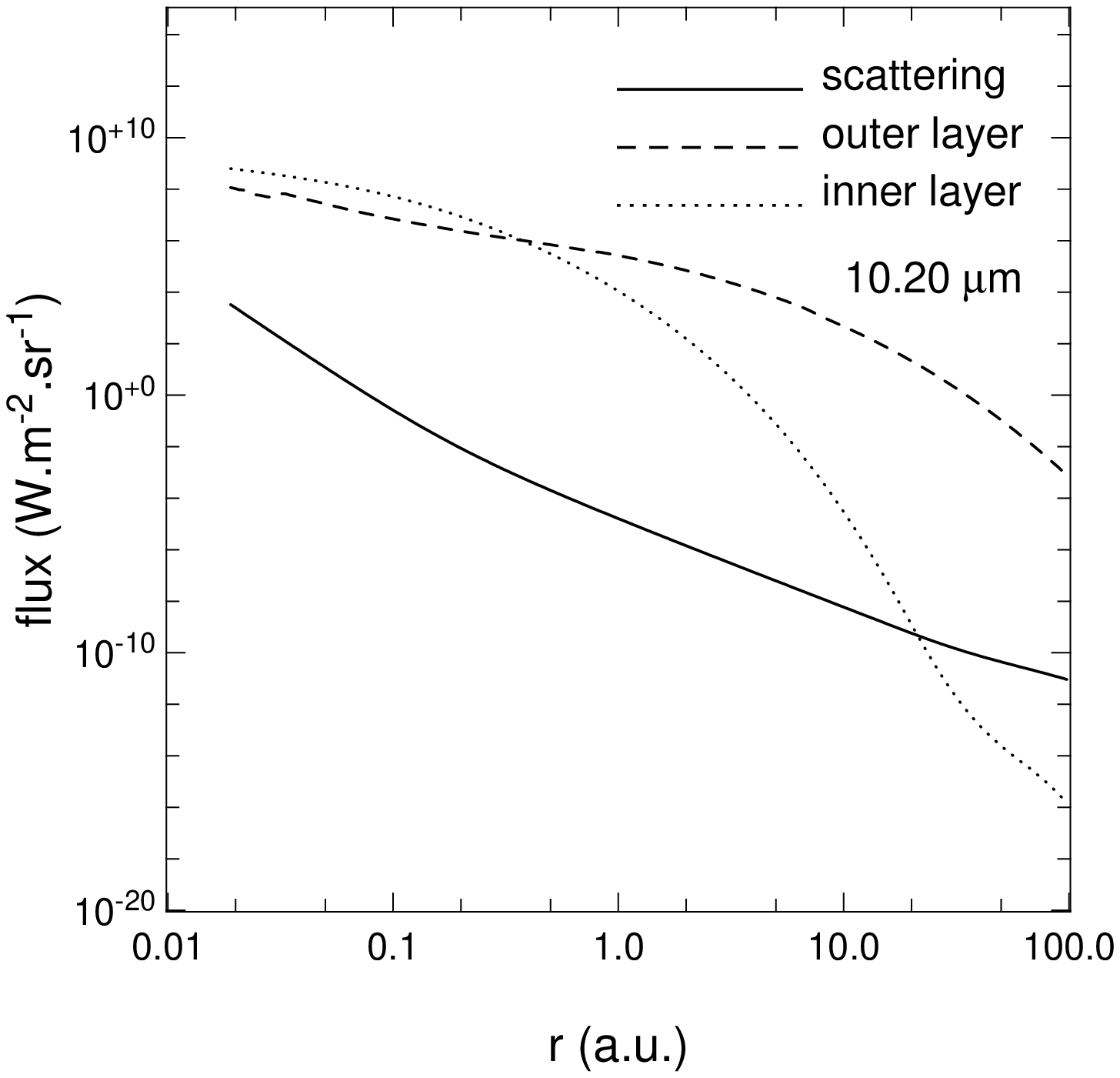}}
   \caption{%
      Scattered light and thermal flux from a typical T~Tauri disc (model~2)
      vs. the distance to the star. Left: J-band; middle: K-band; right: N-band.  Solid lines:
      scattered light; dashed lines: thermal emission of the outer layer;
      dotted lines: thermal emission of the inner layer.
   }
   \label{fig:scat}
\end{figure*}

Simple disc models often ignore scattering when they focus on the inner parts,
but at large radius scattered light is dominant.  A reason for this is that the
thermal flux decreases exponentially with $r$, since it is proportional to
$B_\lambda(T(r))$, while reprocessed flux is roughly in proportionnal to
$r^{-2}$ --- more exactly $\propto r^{-2} \phi$.  Figure~\ref{fig:scat}
represents the contribution of thermal and scattered light to the emergent flux
from a typical T~Tauri disc, in different spectral bands.  

Let us first notice that the flux emerging from a standard ``one-layer'' disc
is dominated by scattered light for $r > 0.15$\,AU in J, $r > 0.5$\,AU in K,
and $r > 20$\,AU in N (contribution of the inner layer, Fig.~\ref{fig:scat}).
Ignoring scattering in AU-scale predictions with the standard model proves out
to be a bad approach in the near-IR.  

In a two-layer disc, the outer layer is hotter and produces a larger thermal
flux (contribution of the outer layer, same figure).  Scattering is therefore
dominant only at large radii:  $r > 2$\,AU in J, $r > 8$\,AU in K, and $r
\gtrsim 500$\,AU in N, so that ignoring it as a first approach can be an
interesting simplification, when dealing with the AU-scale observations
provided by IR interferometry, provided that the field of view of the 
interferometer is limited (see Sect.~\ref{sec:albedo}).  In other words,
the thermal emission is more extended in a two-layer disc, which implies
for instance lower optical visibilities.

\subsection{Influence of disc parameters}

Figure~\ref{fig:paramsp} displays the influence of four parameters on
the disc structure and the observables ---the SED and the normalised 
visibility.

The visibility curves are built using a point-like source, the star, with a
flux {\Fstar}, over-imposed on an extended source {\Fdisk}.  At large
baselines, the visibility amplitude reaches a plateau yielding information on
the flux ratio $|V|^2 \approx 1/(1+\Fstar/\Fdisk)^2$.  The higher the plateau,
the smaller the ratio $\Fdisk/\Fstar$.  The baseline $B$ at which the plateau
is reached gives  the disc extent $\theta \approx \lambda/B$.  The higher the
baseline $B$, the smaller the disc.

At first sight the SED and visibilities present general features: the presence
of a silicate feature in emission, due to the super-heated outer layer, and a
steep drop of visibility when the wavelengths approaches the N band, also
linked to the silicates:  at 10\,$\mu$m the outer layer becomes suddenly
optically thicker and contributes more to the total flux; since it is hotter
and therefore presents a larger extent than the inner layer, the disc appears
much larger.

\subsubsection{Albedo}
\label{sec:albedo}

When the albedo varies, the outer parts of the structure and the SED
are altered, because reprocessing is lowered.  The effect on the structure
of the inner parts is not visible, because reprocessing is not important
there.  The visible SED is only slightly changed because stellar light
dominates scattered light by a factor $> 10$.

While one usually ignores scattering in simple disc models, its incidence on
the visibility curve can be important.  The scattered flux represents about
$\alb \Hdisk/r$ of the stellar flux for a disc with a relative outer
geometrical thickness $\Hdisk/r$, \latin{i.e.} $\approx 5$ to 10\% in most
discs.  At near-infrared wavelengths and for moderate accretion rates, direct
stellar flux and disc reprocessed flux are comparable, so scattered light
cannot be ignored.  Since reprocessing occurs at a large scale (like
scattering, see Fig.~\ref{fig:contrib}), an extended structure is resolved at
very small baselines, hence the visibility drop near the origin. This effect
might not be seen if the interferometer field of view is not large enough,
hence a possible discrepancy up to 0.1 in normalised visibility. 

An important scattering also alters the variation of the visibility with
wavelength by a differential effect: scattering is much more important at short
wavelengths, where it induces a larger drop in visibility.

\subsubsection{Inner disc radius}

As the inner truncation of the disc increases, the near-infrared part of the
SED is depleted, because this part of the spectrum is mostly produced by the
hot inner parts of the disc.

The very long baseline ($B > 400$~m) visibility amplitudes are higher, because
the flux of the disc is lower compared to the flux of the star.  The visibility
plateau is reached for smaller baselines ($\approx 50$--$100$~m) because the
disc has a larger mean square dimension.  For small baselines, the visibility
is lower because the resolved outer parts of the disc have a greater
contribution compared to its still unresolved parts.  

The inner truncation also affects the variation of visibility with wavelength,
because of a resolution effect.  At small wavelengths, the resolution is high
and visibilities follow the very long baseline trend and the visibility is
higher for a large inner radius.  At large wavelengths, the resolution is low
and the visibility is lower for a large inner truncation.

\subsubsection{Viscosity}

As $\alpha$ increases, the viscosity becomes more efficient.  The amount of
material needed to produce a given accretion rate decreases.  Therefore the
optical thickness of the disc becomes lower and the mid-plane temperature
follows the same trend.  The discrepancy in the IR SED and visibility
amplitudes are small, because the effective temperature of viscosity does not
directly depend on $\alpha$.  However, when $\alpha$ increases, $\Sigma$
decreases, so that the thickness $\Hdisk$ also decreases and less stellar light
is caught by the disc.  Therefore, the mid IR SED is depleted and the
visibility is slightly higher.  The SED is much more affected in the
submillimetre, because it probes the optically thin outer parts of the disc, so
that the flux is proportional to $\Sigma \propto \alpha^{-1}$.

\subsubsection{Accretion rate}

As the accretion rate increases, the mid-plane temperature is higher in the
regions dominated by viscous heating ($r < 1$--$10$~AU), but remains unchanged
in the outer regions dominated by reprocessing.  The near- and mid-infrared
parts of the SED drastically change with accretion rate, since the viscous
effective temperature is proportional to the accretion rate.  At small
accretion rates, the temperature inversion at the surface produces a
silicate feature in emission at 10\,$\mu$m; when $\Mdot$ increases,
the temperature inversion disappears, because the reprocessing becomes
secondary, and the feature disappears.  The far-infrared
SED does not change much for moderate accretion rates
($10^{-9}$--$10^{-7}$~{\Msun}/{\yr}), because the regions that significantly
contribute at large wavelength are dominated by stellar light reprocessing
(almost) independent of the accretion.  The slight decrease in the far
infrared, when accretion increases, is due to the change of $\Hdisk$
with the amount of material.

The visibility presents a lower plateau for larger accretion rates, because the
flux from the disc becomes higher compared to the flux from the star.  For
large accretion rates ($>10^{-7}$~{\Msun}/{\yr}), the plateau is reached for
smaller baselines because the mean angular size of the disc is larger (the
hotter the disc, the larger the region of emission at a given wavelength).
However at small baselines, discs with moderate accretion present a faster
visibility drop because the resolved scattered flux is larger compared to the
still unresolved inner parts.

\begin{figure*}[p]
   \subfigure[variation of albedo]{\label{fig:alb}%
   \includegraphics[width=0.23\hsize]{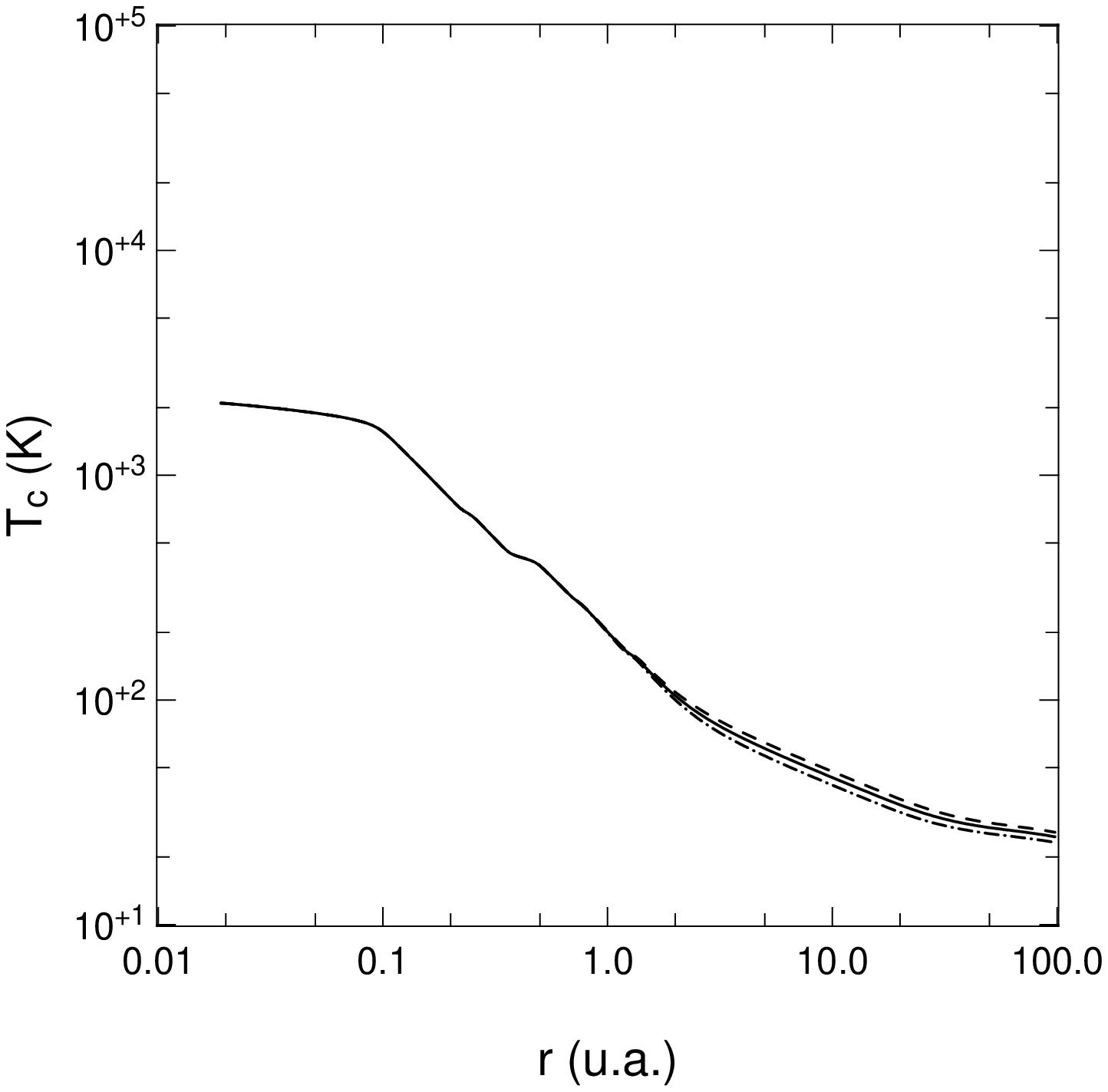}\hspace{0.02\hsize}%
   \includegraphics[width=0.23\hsize]{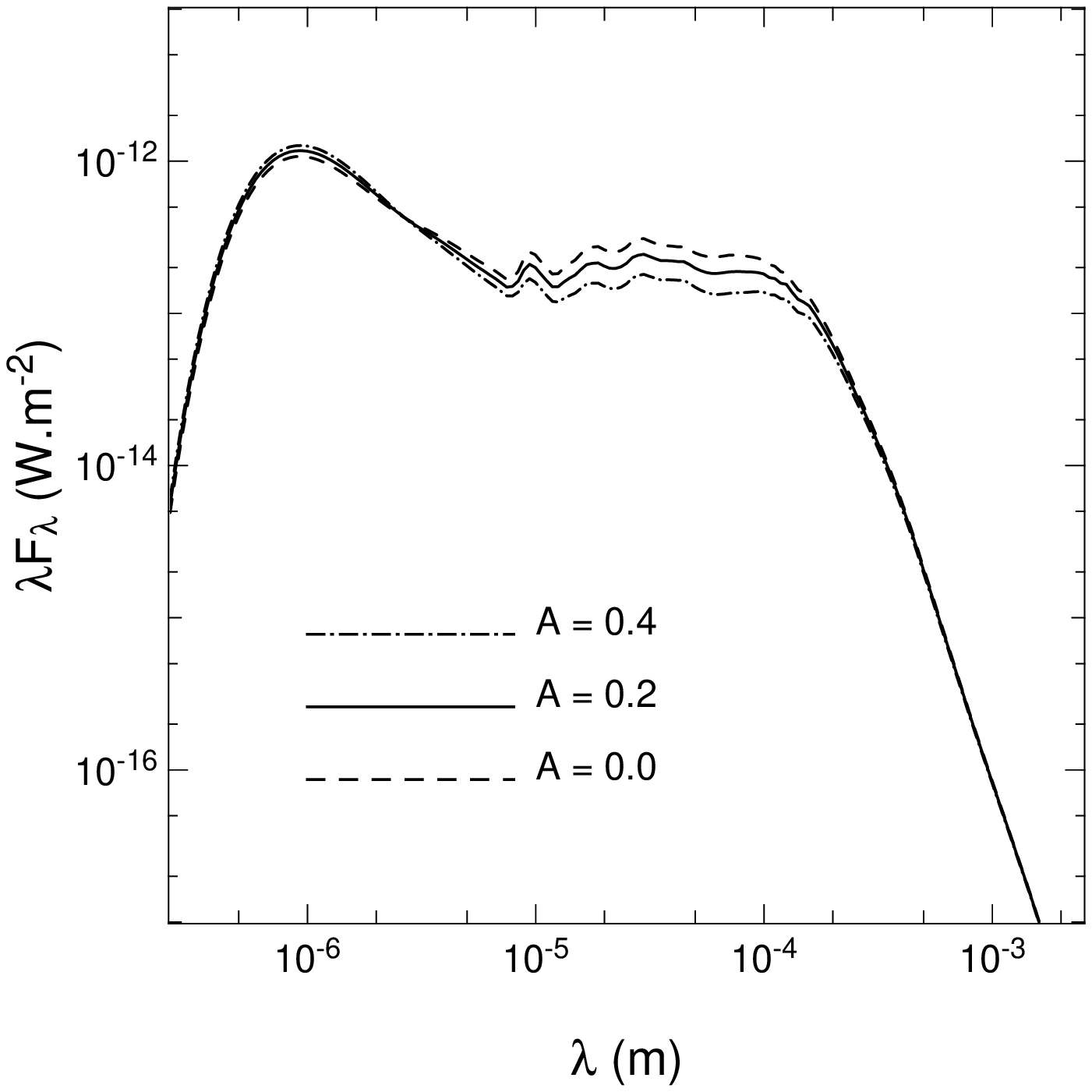}\hspace{0.02\hsize}%
   \includegraphics[width=0.23\hsize]{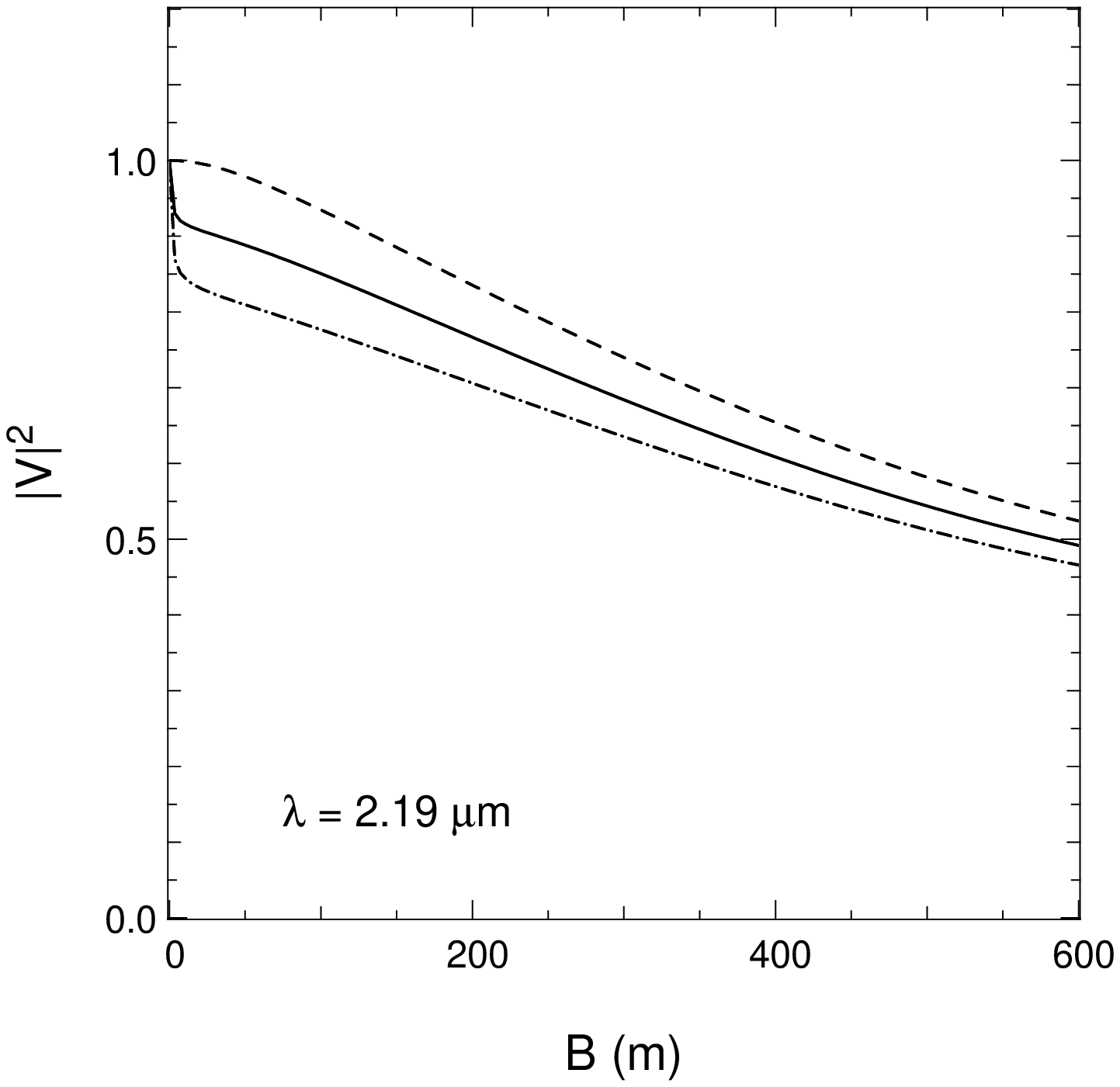}\hspace{0.02\hsize}%
   \includegraphics[width=0.23\hsize]{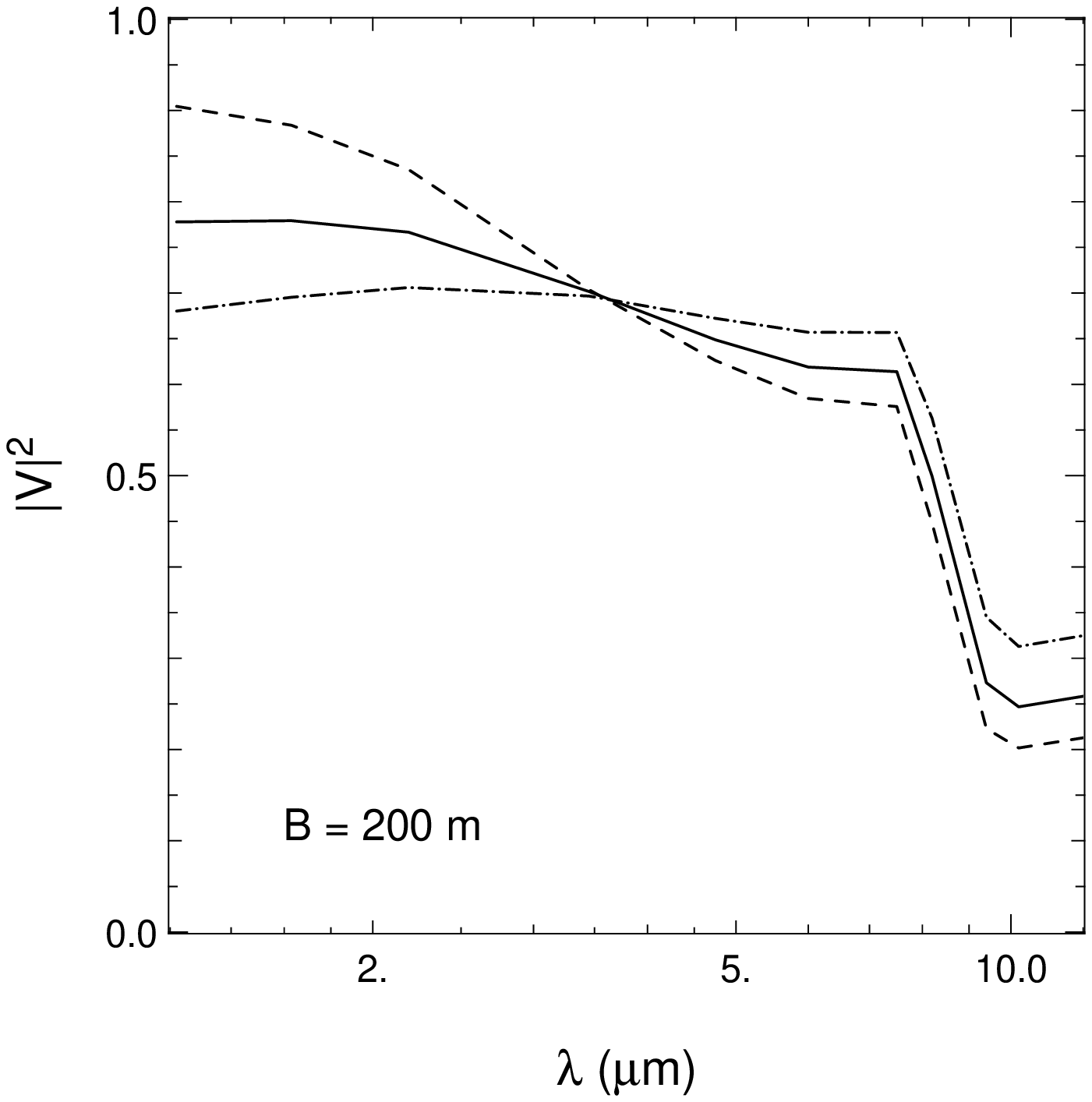}}\\[-0.5em]
   \subfigure[variation of inner radius]{\label{fig:rmin}%
   \includegraphics[width=0.23\hsize]{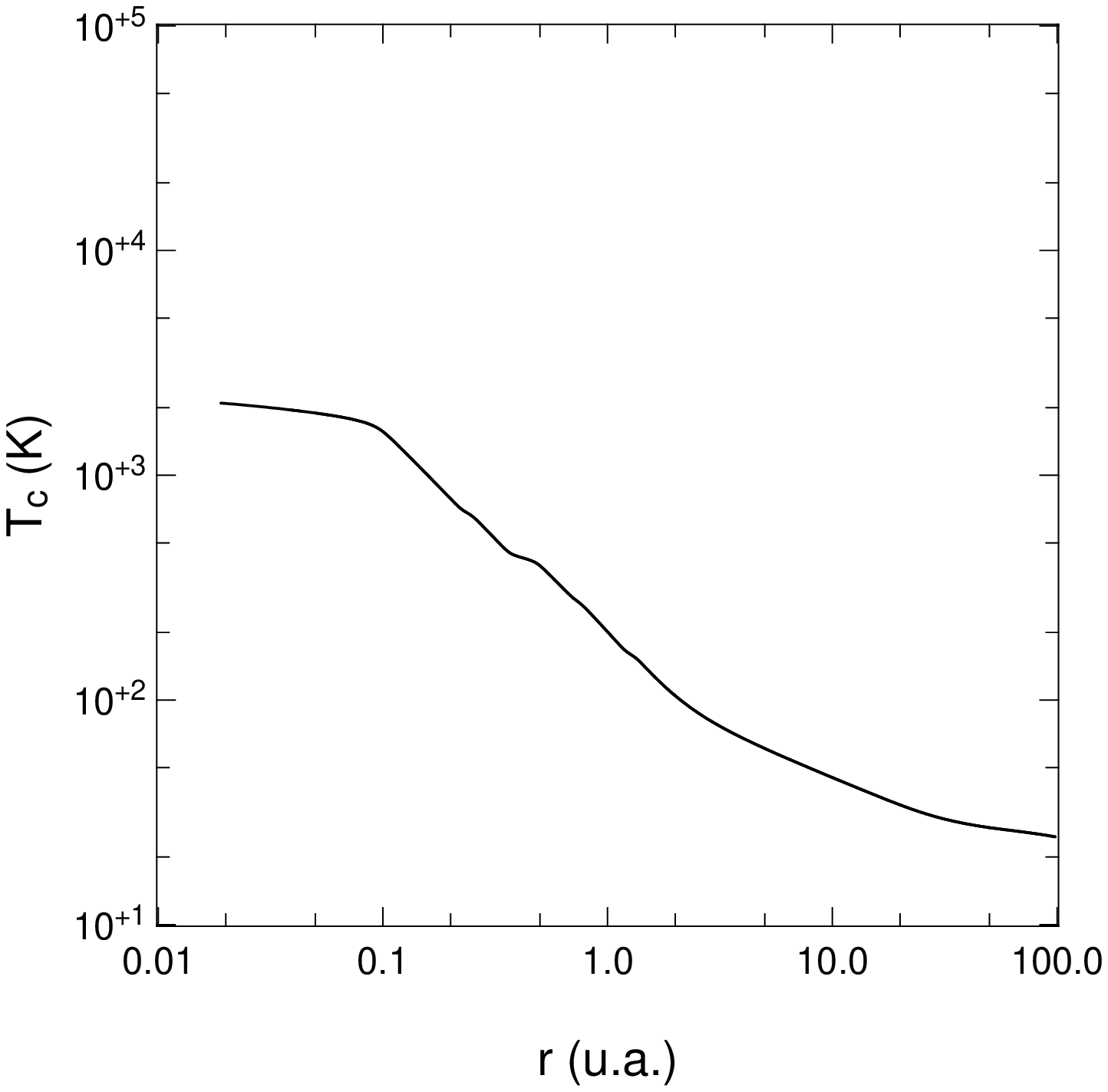}\hspace{0.02\hsize}%
   \includegraphics[width=0.23\hsize]{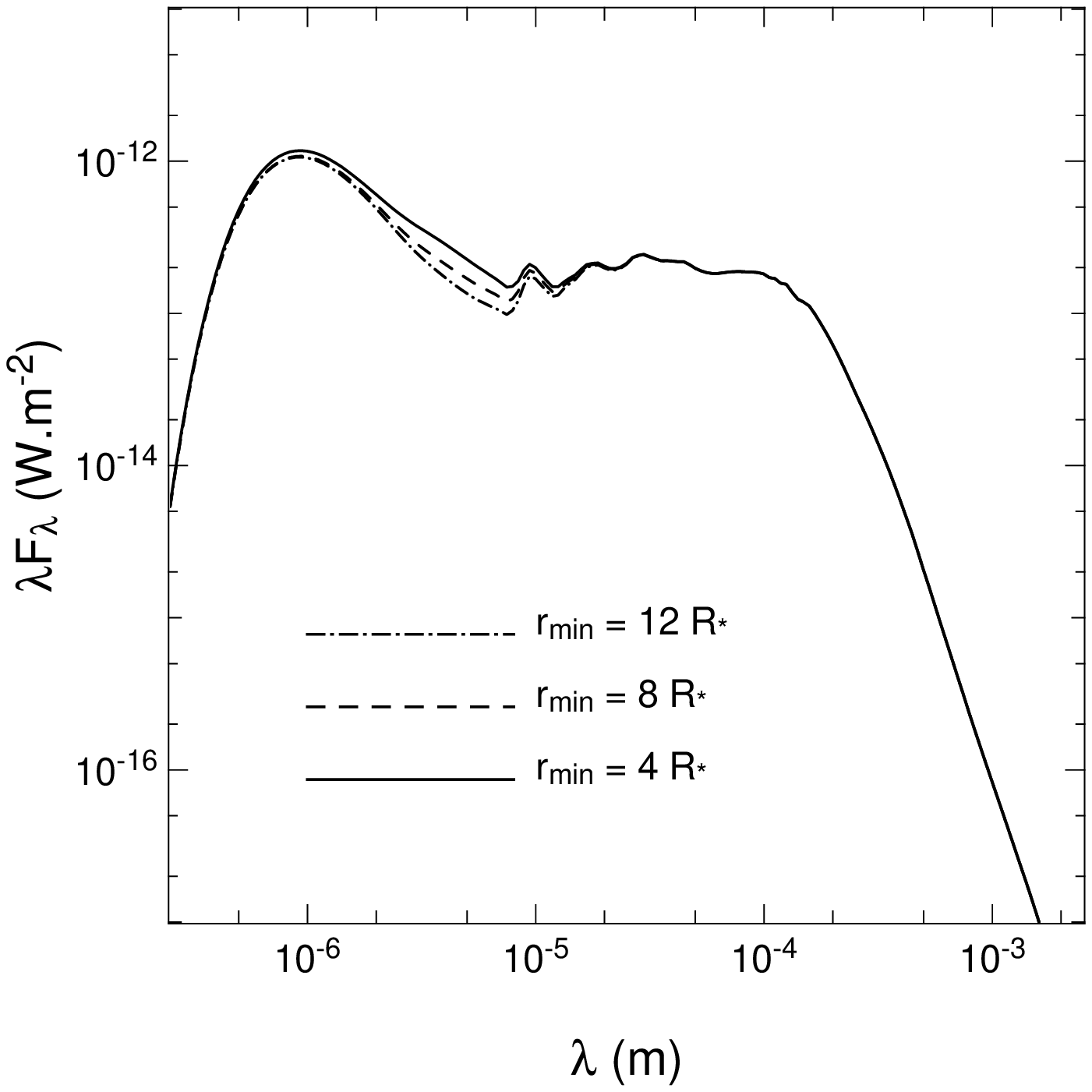}\hspace{0.02\hsize}%
   \includegraphics[width=0.23\hsize]{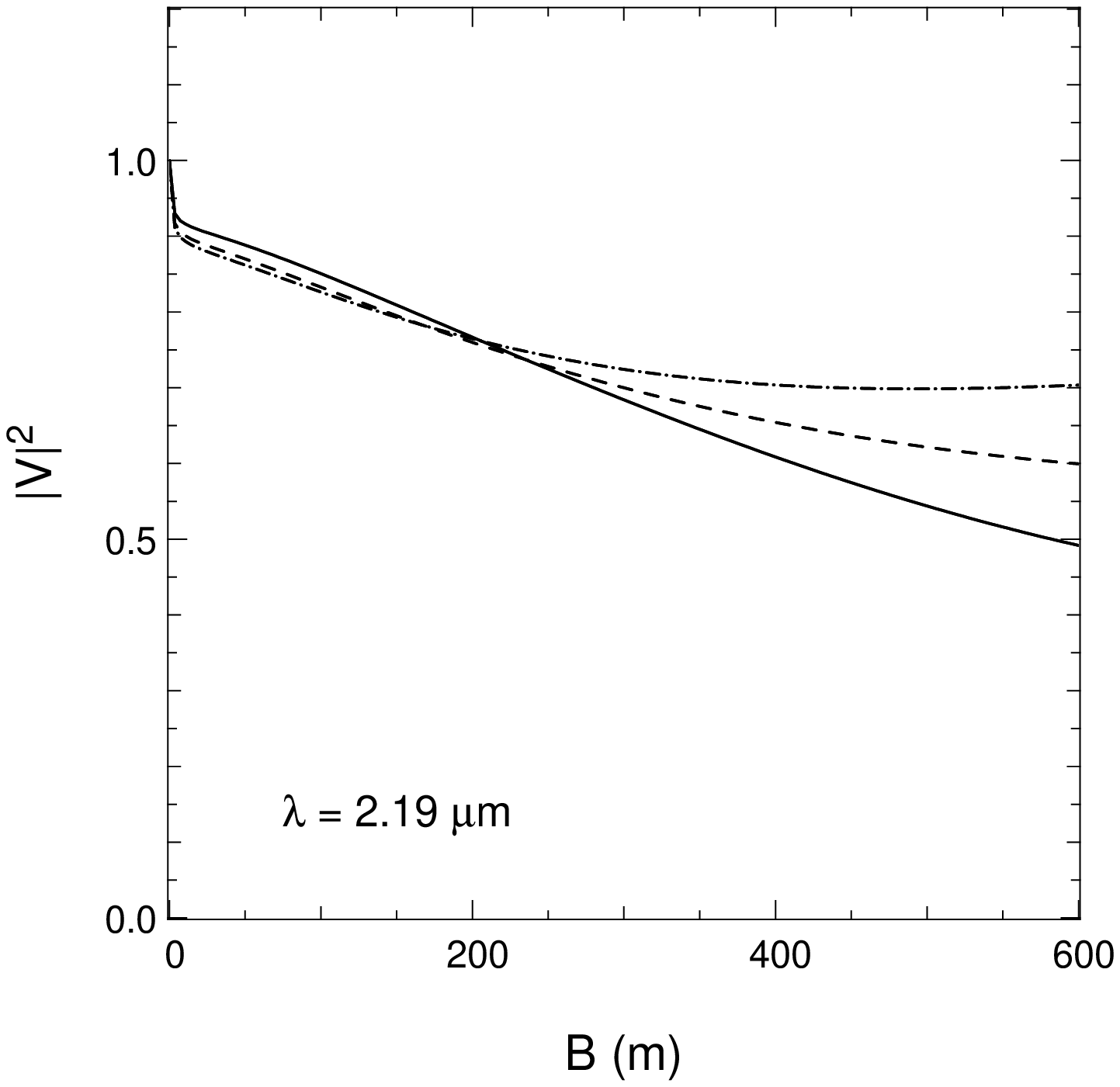}\hspace{0.02\hsize}%
   \includegraphics[width=0.23\hsize]{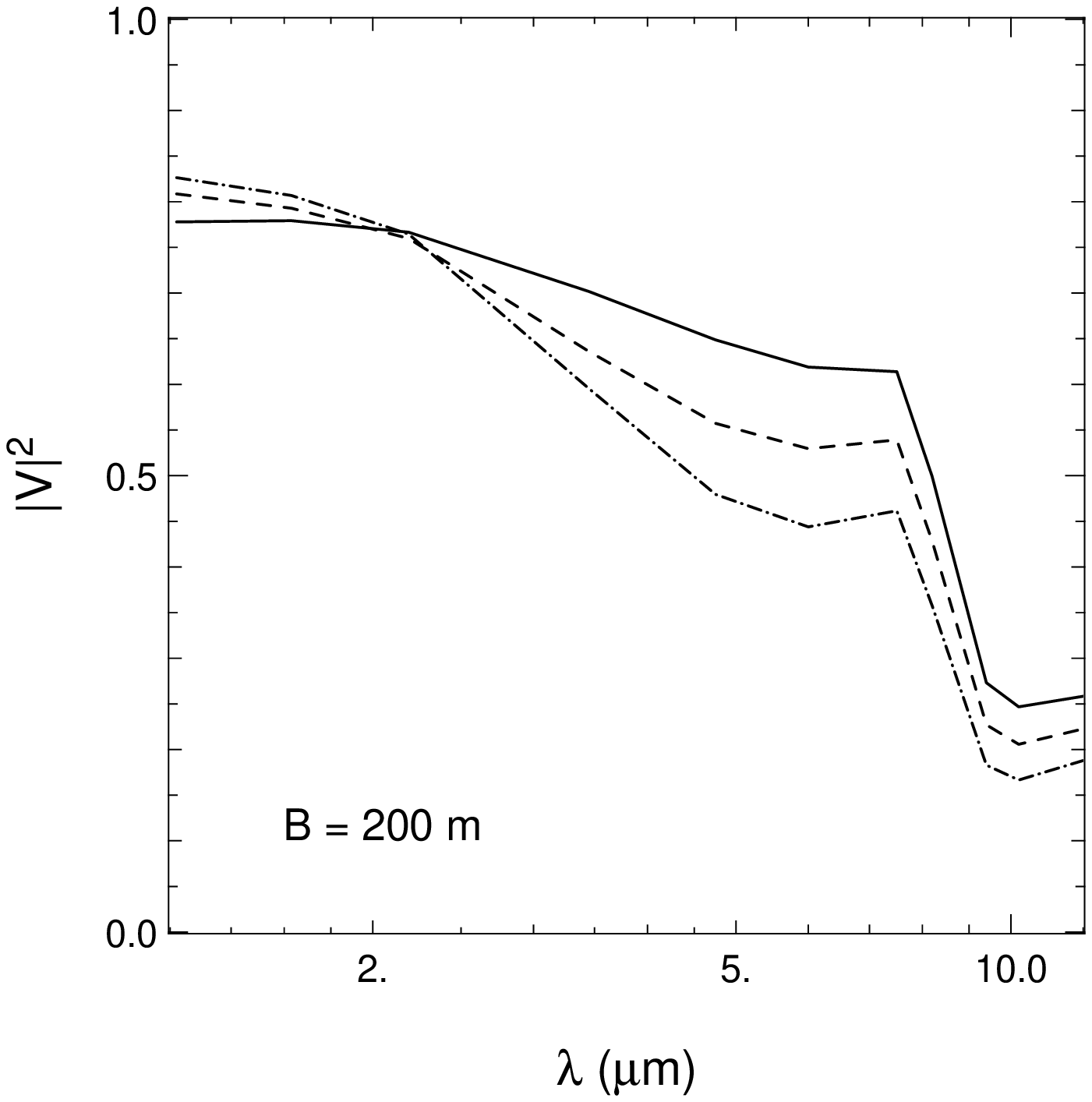}}\\[-0.5em]
   \subfigure[variation of viscosity]{\label{fig:alpha}%
   \includegraphics[width=0.23\hsize]{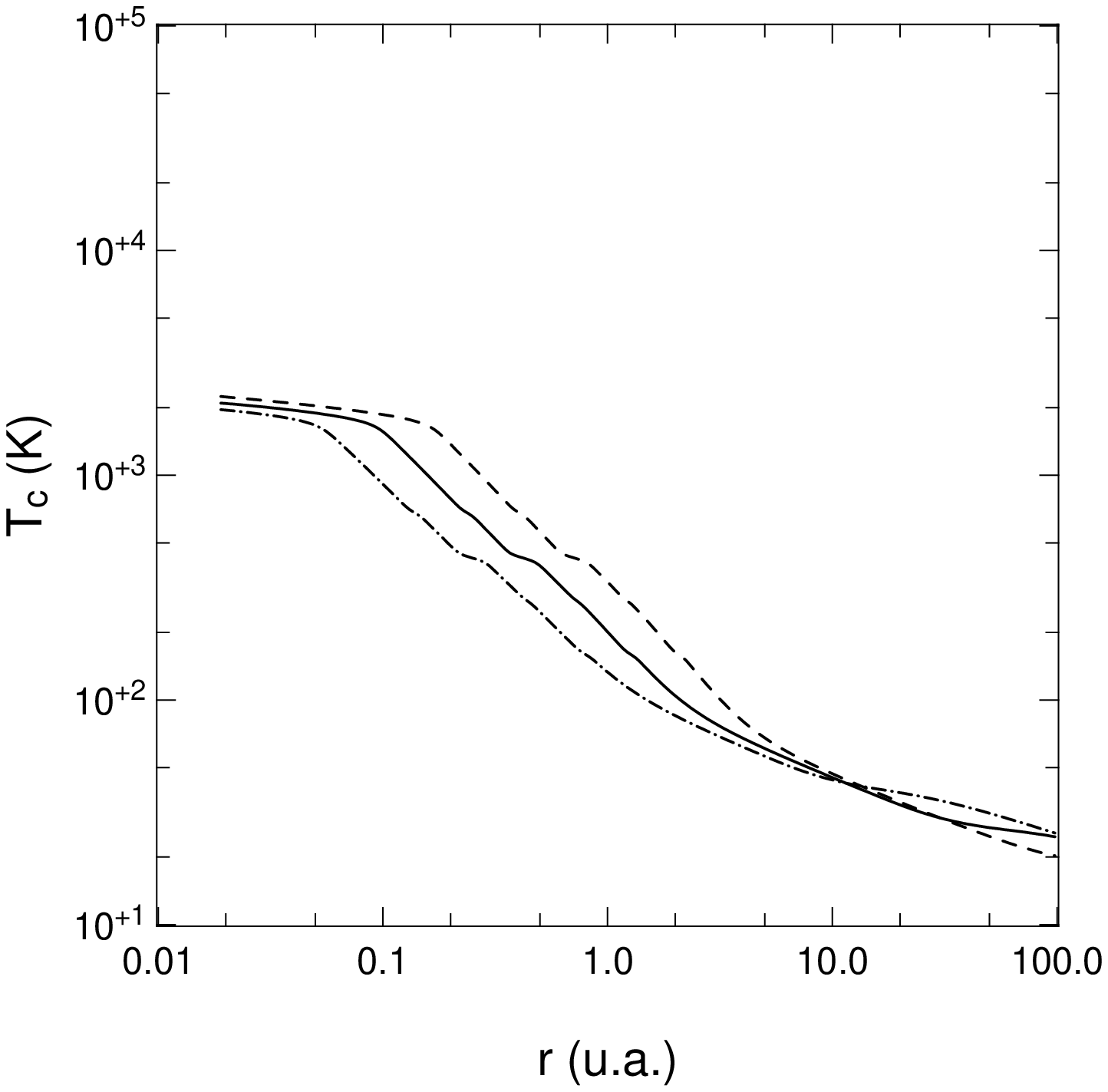}\hspace{0.02\hsize}%
   \includegraphics[width=0.23\hsize]{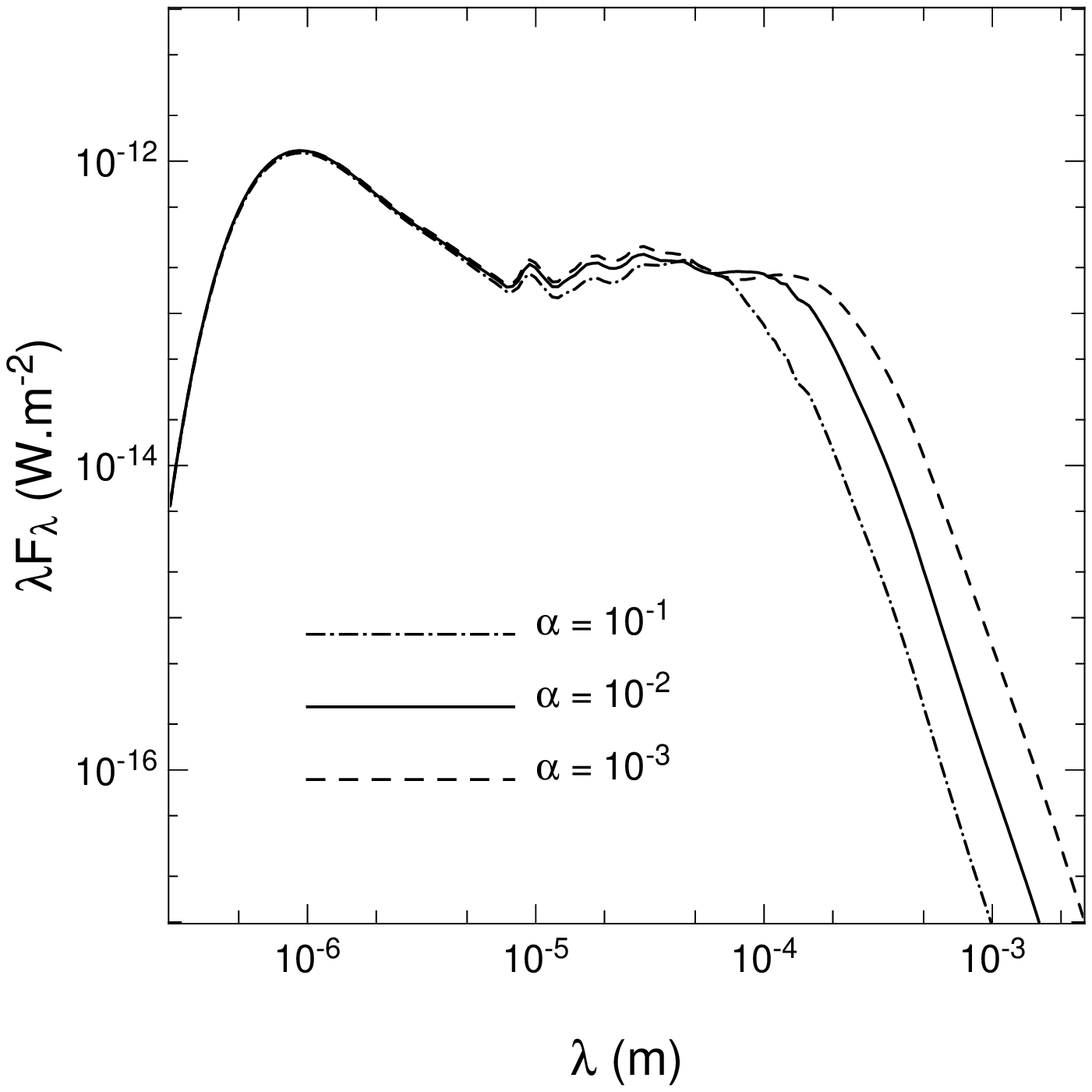}\hspace{0.02\hsize}%
   \includegraphics[width=0.23\hsize]{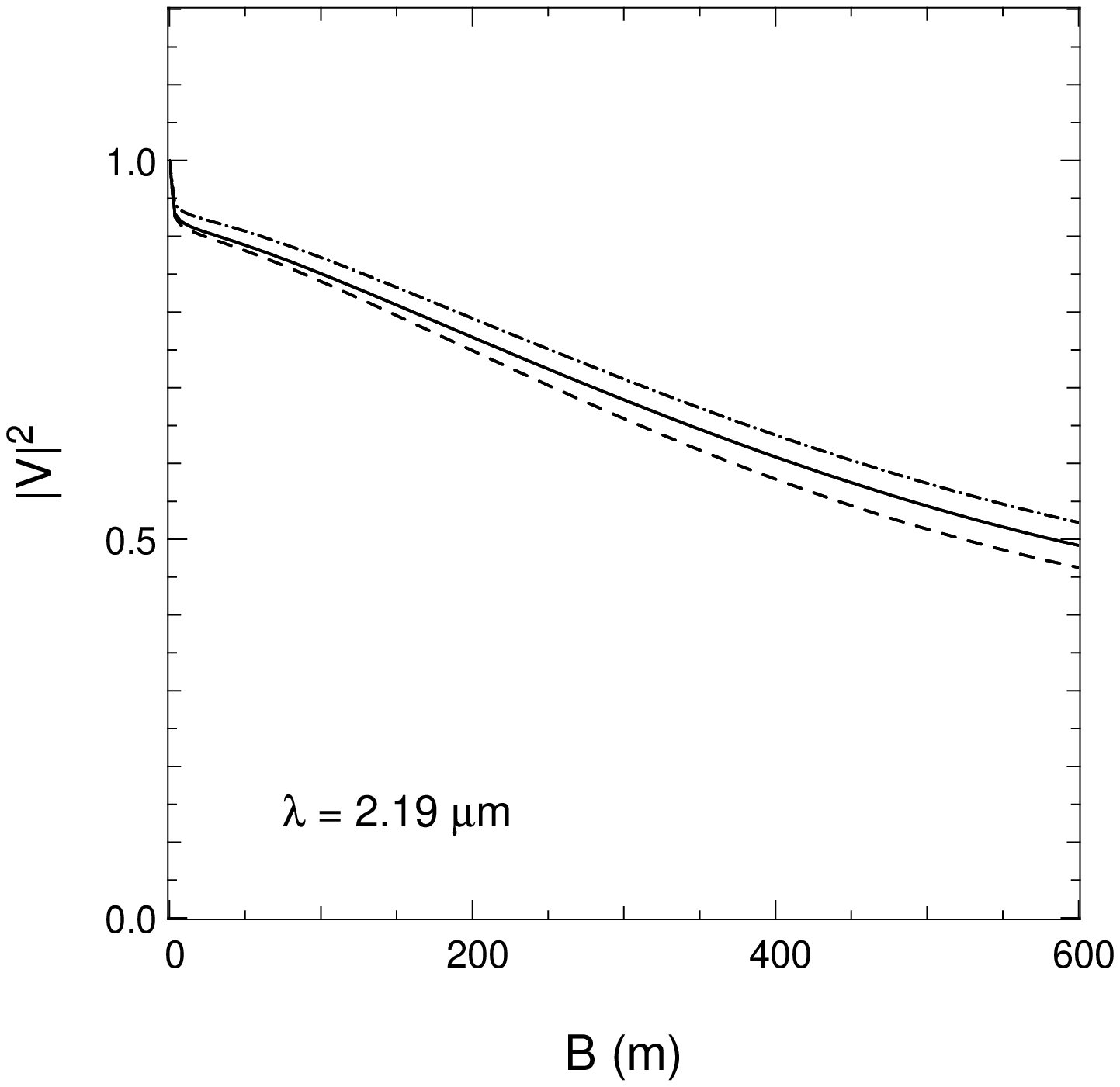}\hspace{0.02\hsize}%
   \includegraphics[width=0.23\hsize]{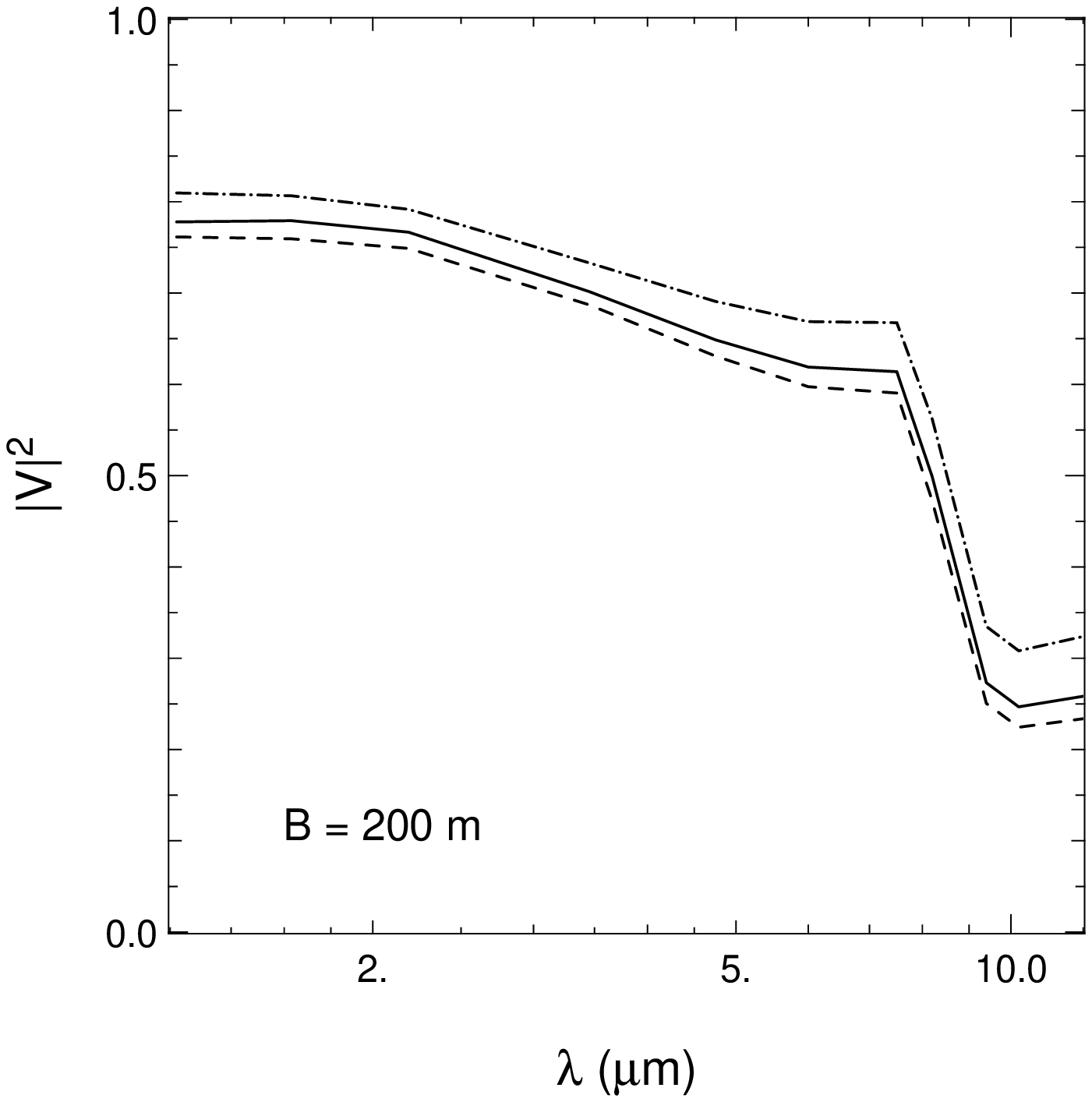}}\\[-0.5em]
   \subfigure[variation of accretion rate]{\label{fig:mdot}%
   \includegraphics[width=0.23\hsize]{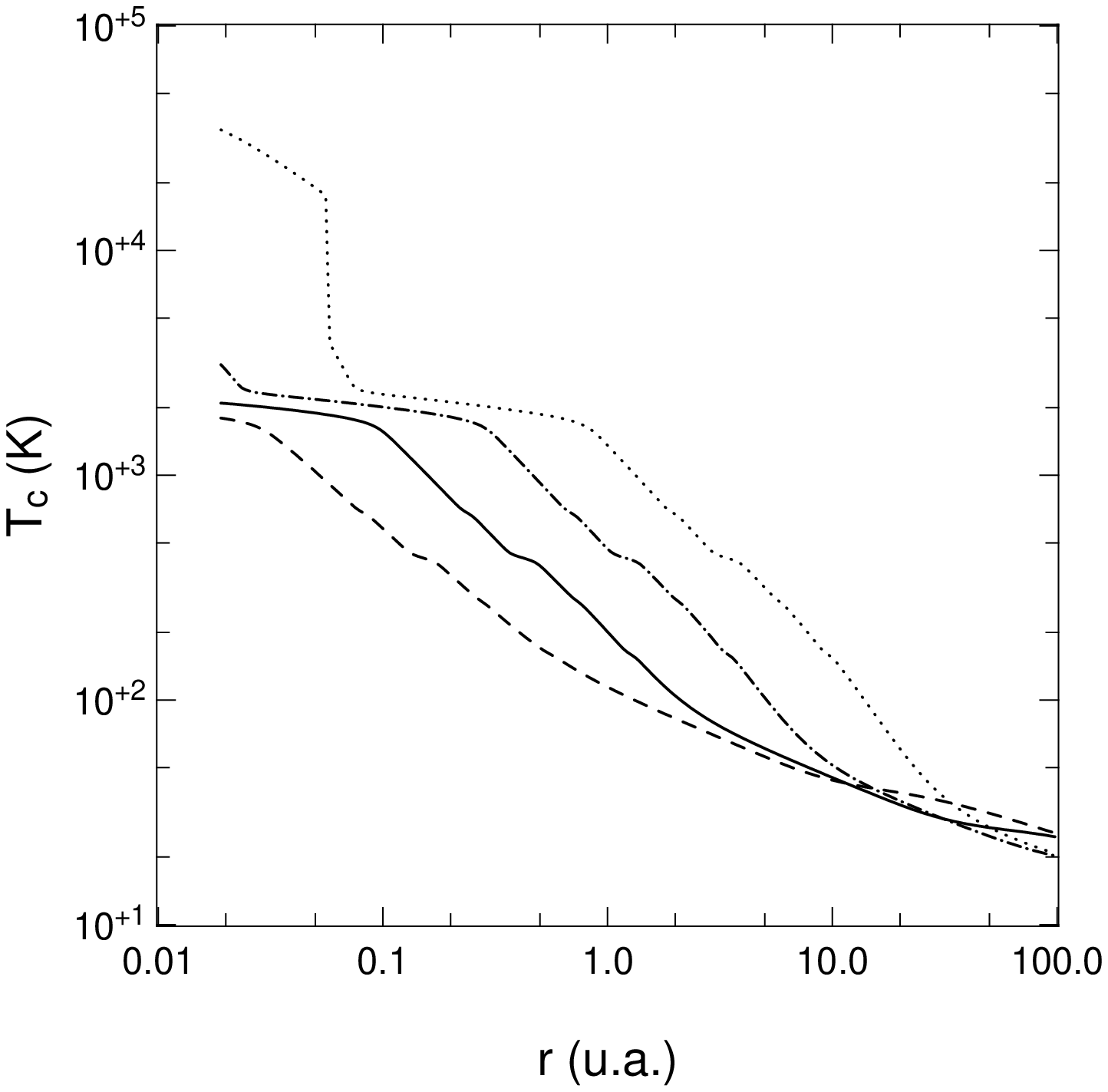}\hspace{0.02\hsize}%
   \includegraphics[width=0.23\hsize]{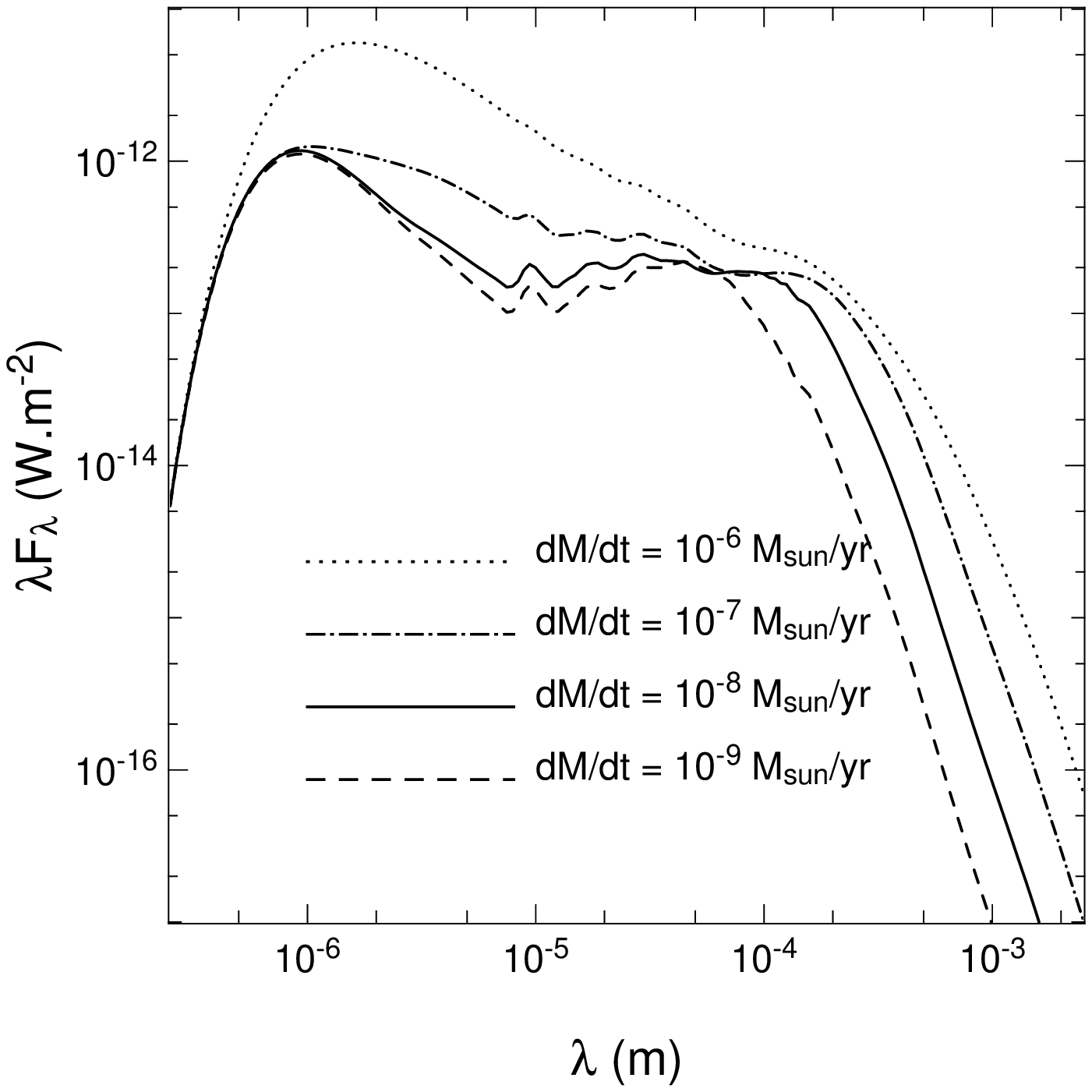}\hspace{0.02\hsize}%
   \includegraphics[width=0.23\hsize]{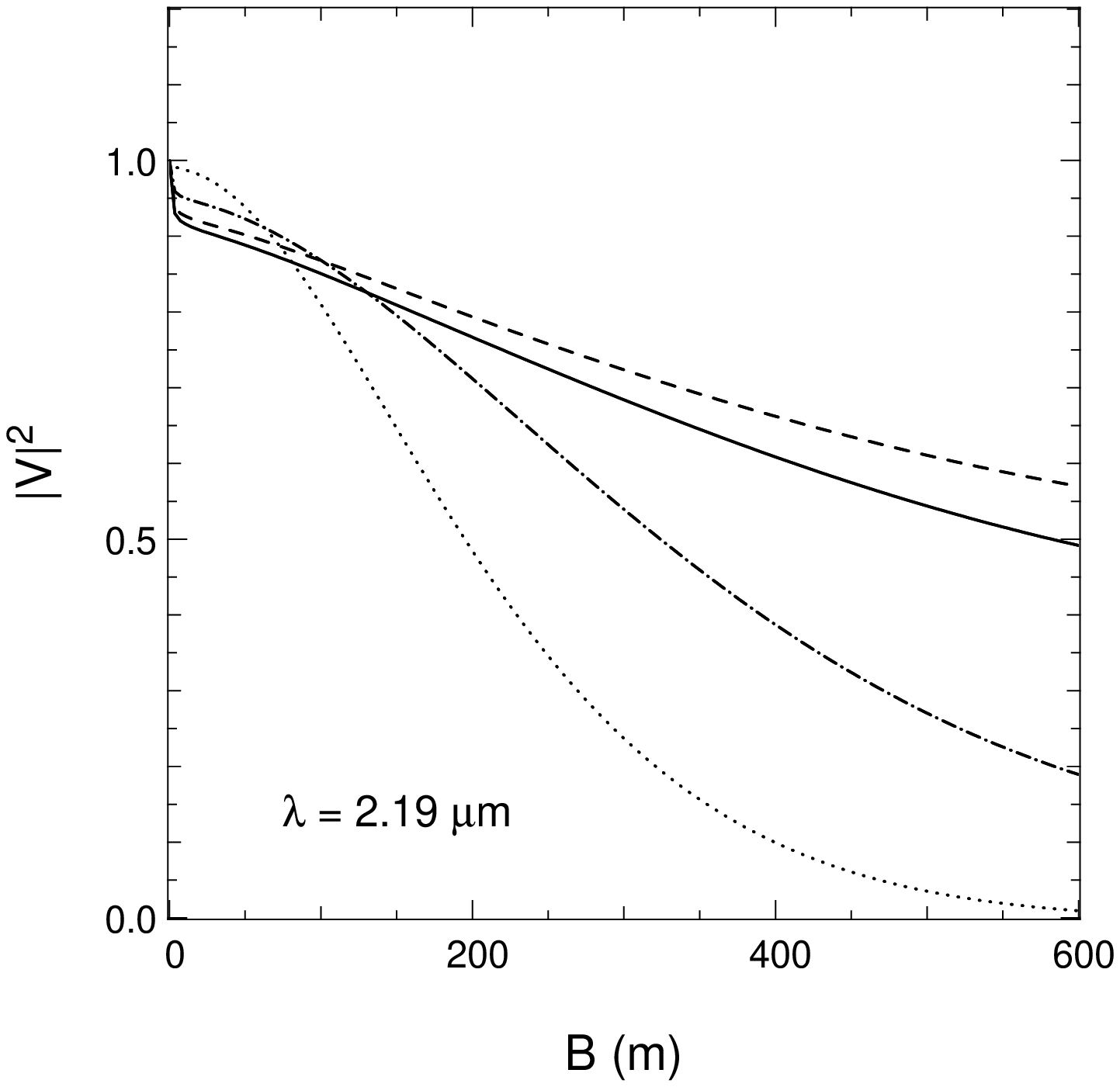}\hspace{0.02\hsize}%
   \includegraphics[width=0.23\hsize]{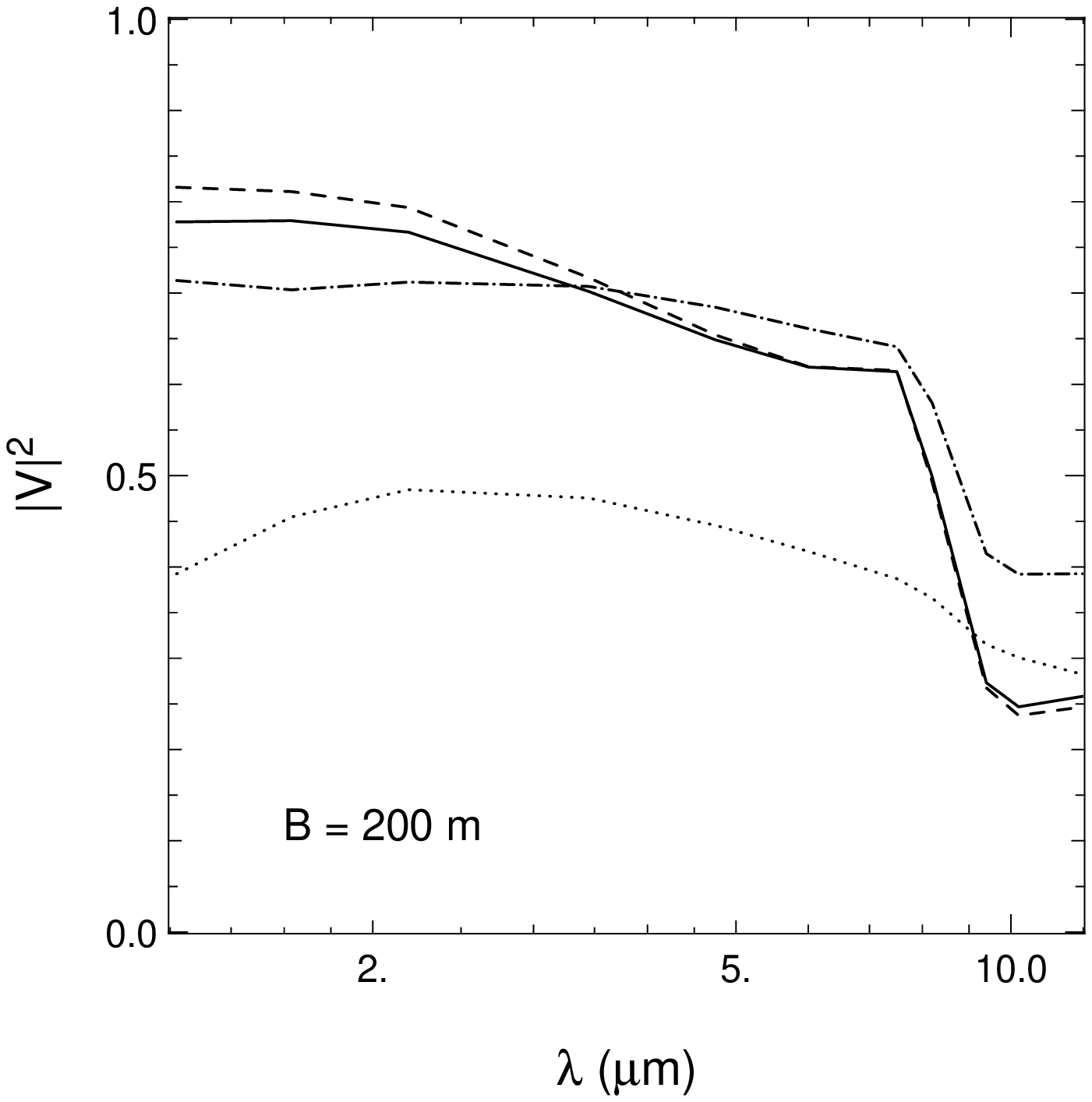}}
   \caption{%
      Variation of the structure of a disc (first column), of its SED (second
      column), of its visibility curve (third column), and of its visibility as
      a function of the wavelength (fourth column) when one of its parameters
      varies.%
   }
   \label{fig:paramsp}
\end{figure*}

\section{Comparison with observations}
\label{sec:res}

We limit our study to TTS and FU~Orionis stars for which both SEDs and optical
visibilities are available.  The sample consists of three objects: FU~Ori,
SU~Aur and T~Tau North.  Table~\ref{tab:obs} lists interferometric observations
available for these stars.  SED measurements have been taken from
\citet{Gezari99}.  Table~\ref{tab:param} lists the physical parameters of the
best fits.

\begin{table}[t]
   \caption{Interferometric observations of FU~Ori and T~Tau stars}
   \label{tab:obs}
   \begin{tabular}{llll}
      \hline\hline
      star    & band    & instrument & reference\\
      \hline
      FU~Ori  & H~\& K  & PTI/IOTA   & a\\
      T~Tau~N & K       & PTI        & b\\
      SU~Aur  & K       & PTI        & b\\
      \hline
      \multicolumn{4}{l}{a: \citet{Malbet98,Malbet01p}}\\
      \multicolumn{4}{l}{b: \citet{Akeson00,Akeson02}}\\
      \hline
   \end{tabular}
\end{table}

Model-fitting of disc visibilities has already been carried out by
\citet{Malbet01p,Malbet02p}: they consistently fitted the spectrum and K
visibilities for FU~Ori, with a self-similar flat disc model presenting an
effective temperature $\propto r^{-0.75}$ but were not able to reproduce both
H~\& K visibilities unless assuming a radial temperature law in $r^{-0.4}$ to
$r^{-0.6}$,  that most physical accretion disc models cannot reproduce: in the
case of FU~Ori, dominated by viscous dissipation, the expected temperature
exponent is $-0.75$.  \citet{Akeson02} separately fit SEDs and visibilities for
T~Tau and SU~Aur but fail to find a set of parameters consistent both with
interferometric data and SED.

\begin{figure*}[p]
   \subfigure[SU~Aur]{%
      \label{fig:suaur}
      \includegraphics[width=0.24\hsize]{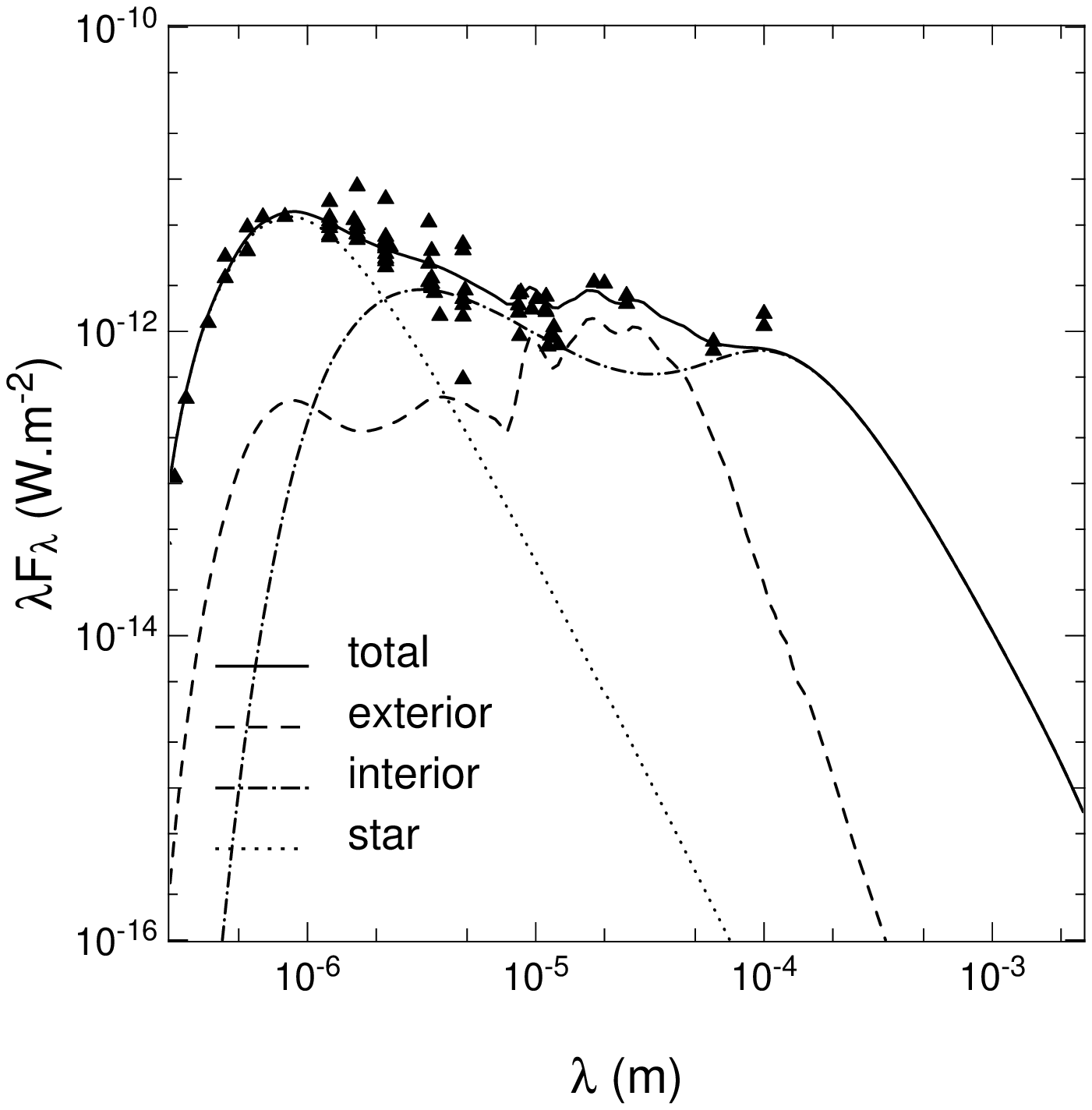}\hspace{0.01\hsize}%
      \includegraphics[width=0.24\hsize]{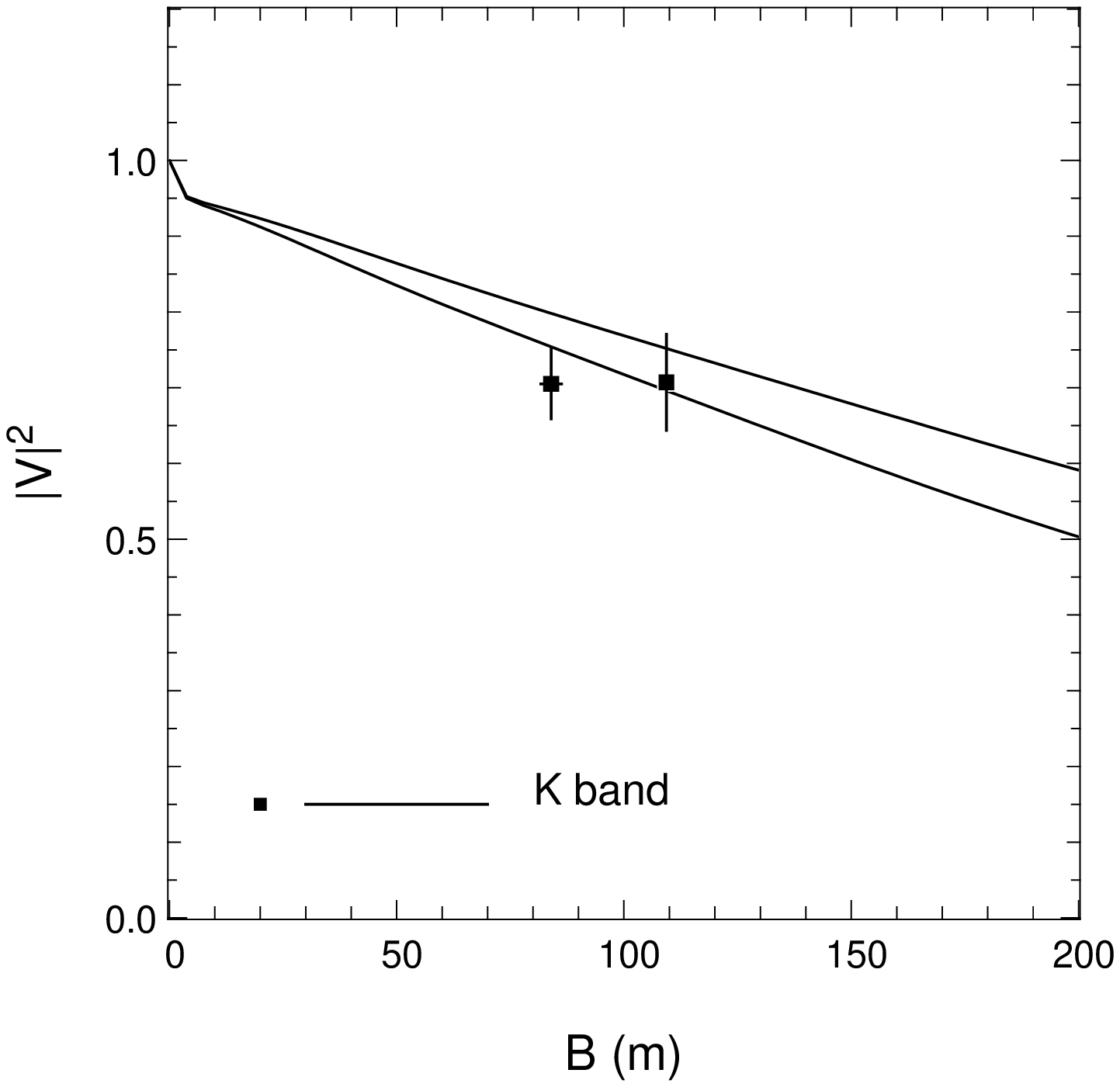}\hspace{0.01\hsize}%
      \includegraphics[width=0.24\hsize]{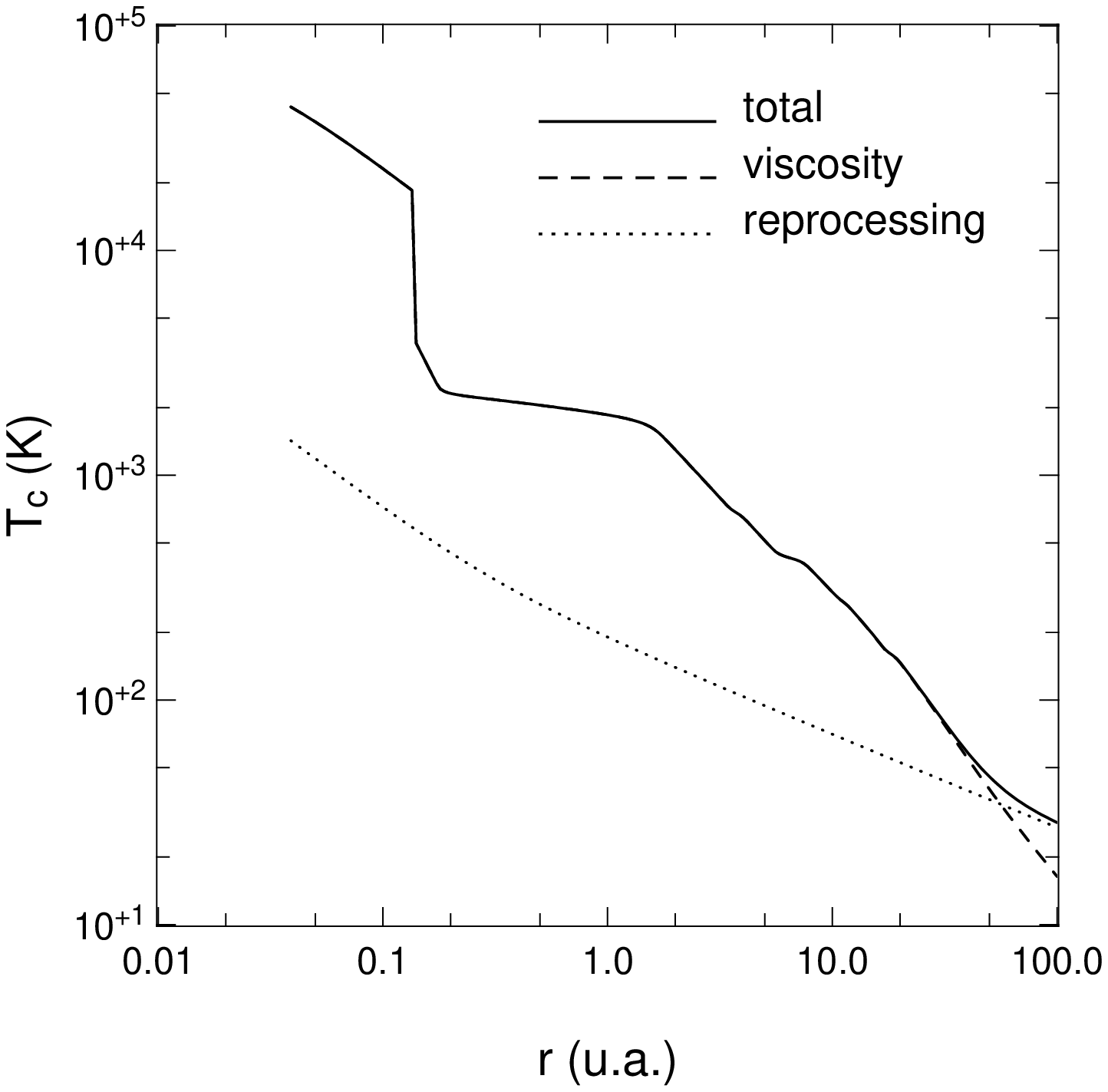}\hspace{0.01\hsize}%
      \includegraphics[width=0.24\hsize]{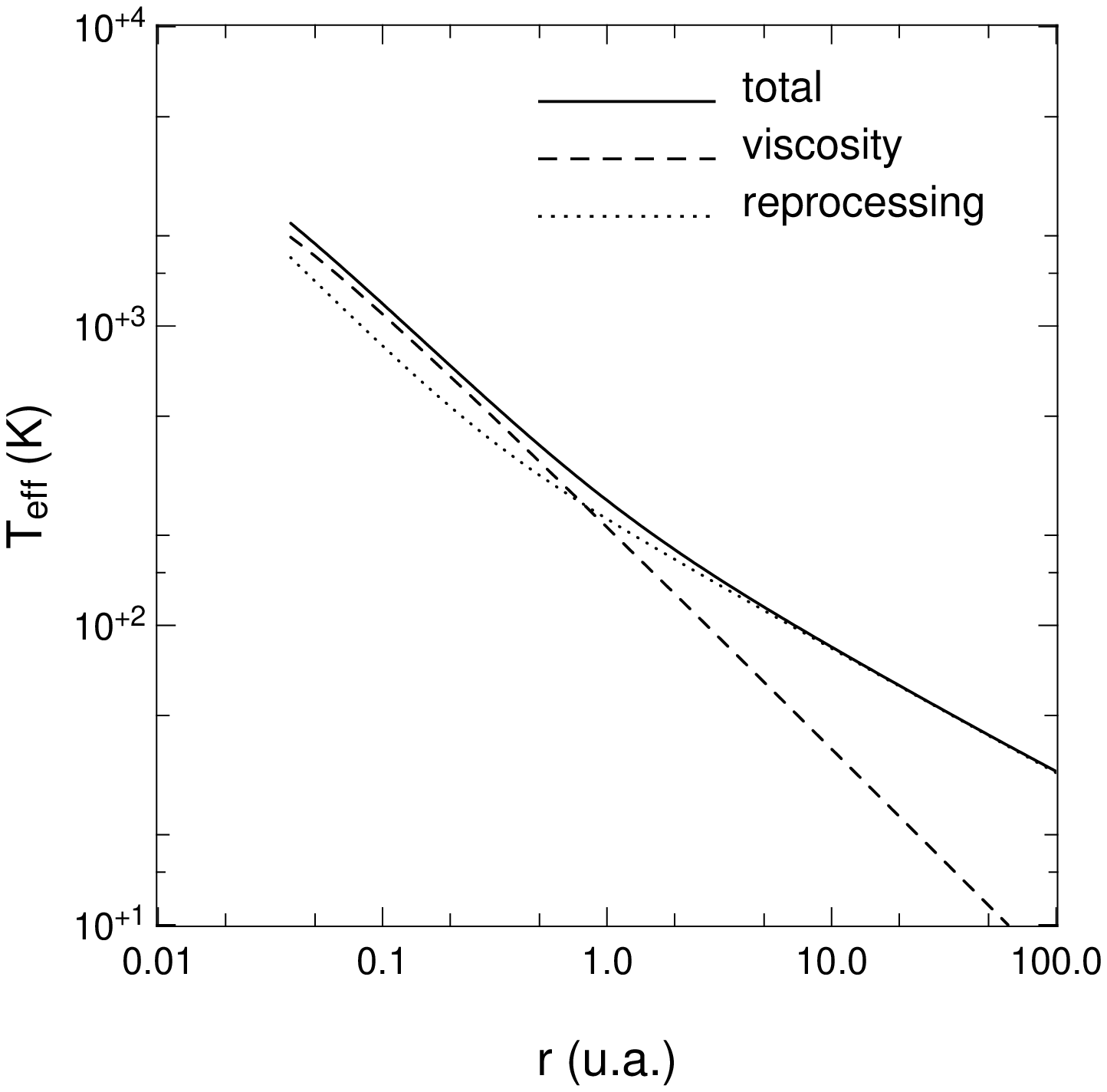}%
   }\\[-8pt]
   \subfigure[FU~Ori]{%
      \label{fig:fuori}
      \includegraphics[width=0.24\hsize]{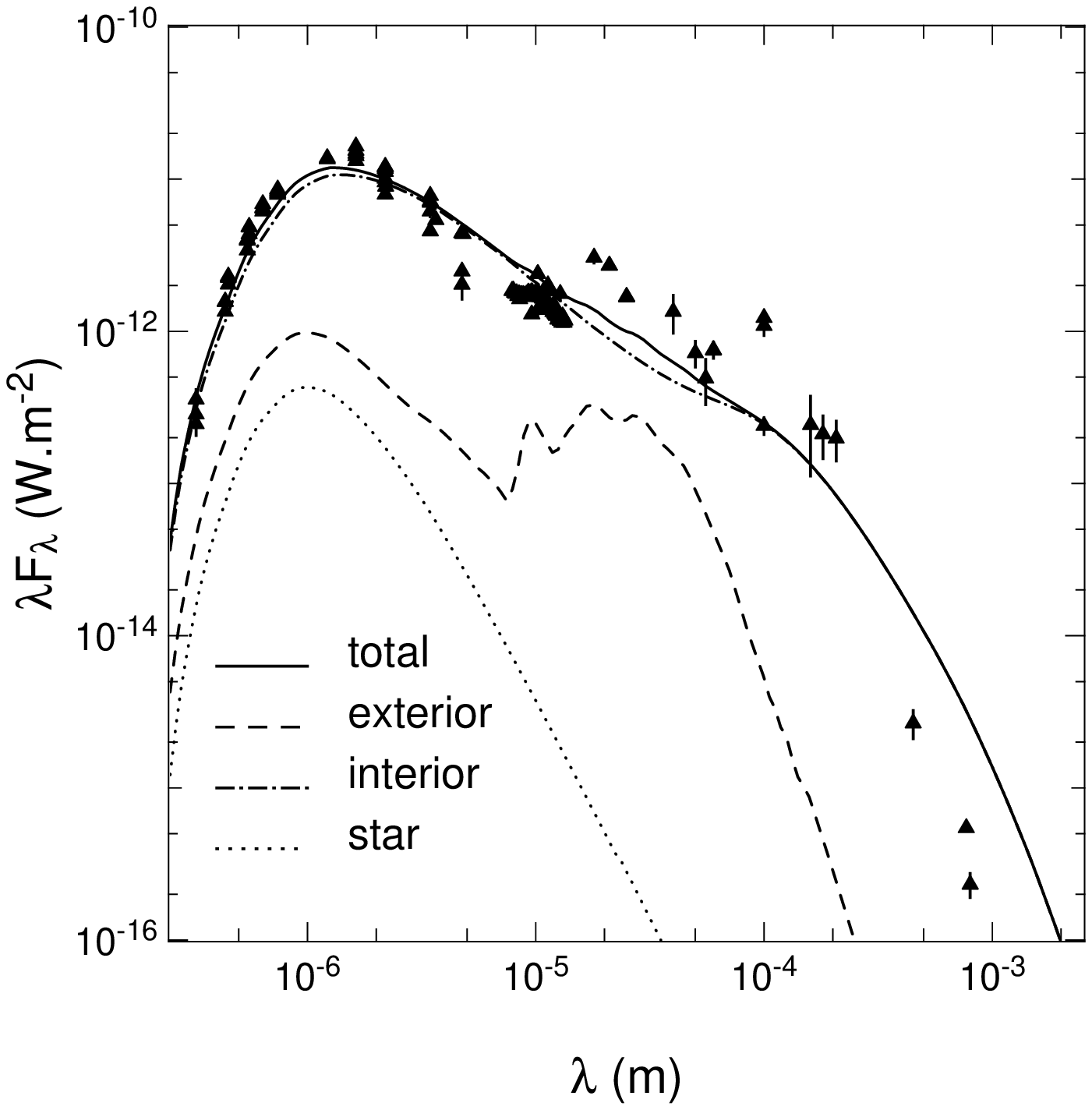}\hspace{0.01\hsize}%
      \includegraphics[width=0.24\hsize]{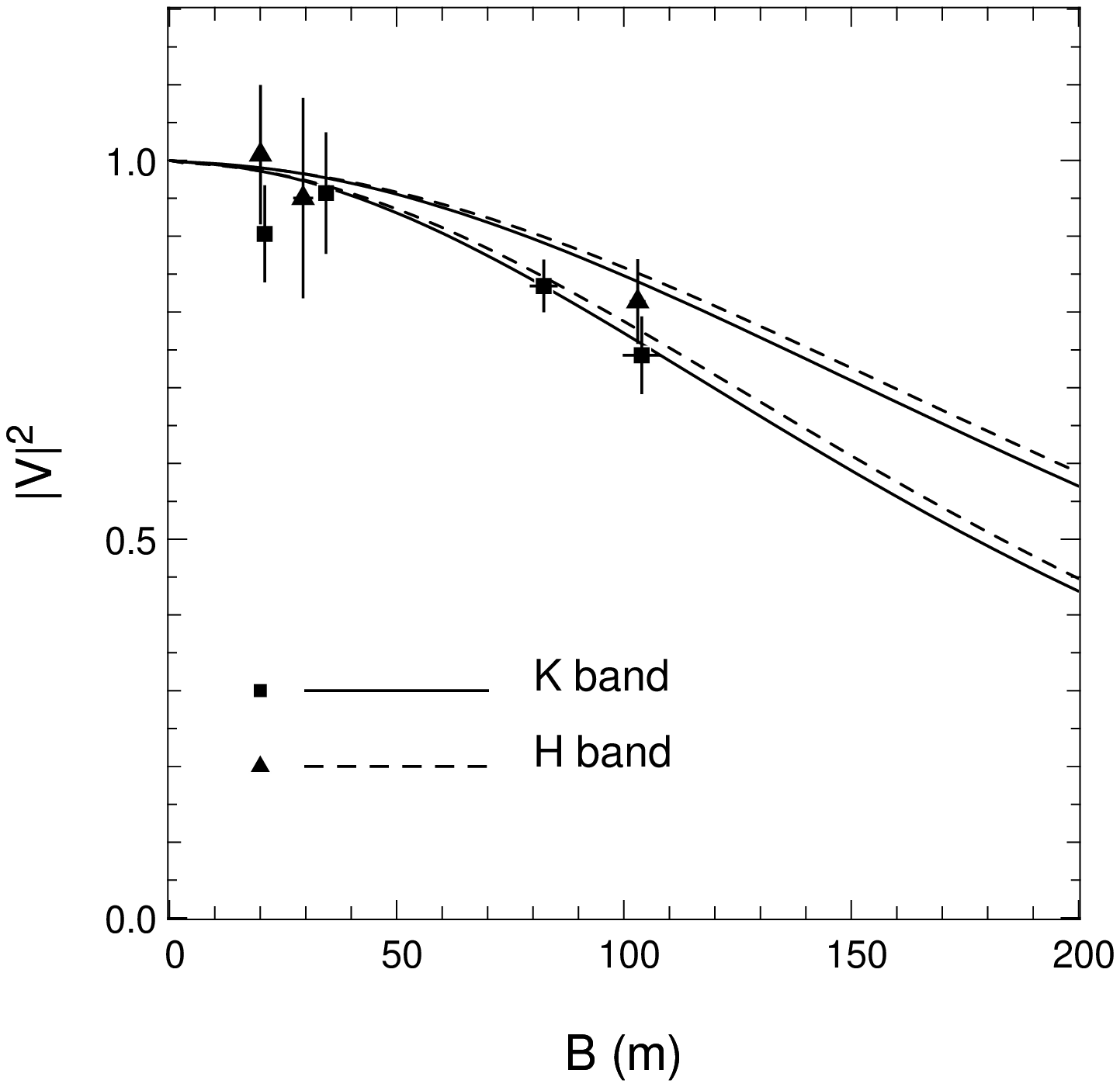}\hspace{0.01\hsize}%
      \includegraphics[width=0.24\hsize]{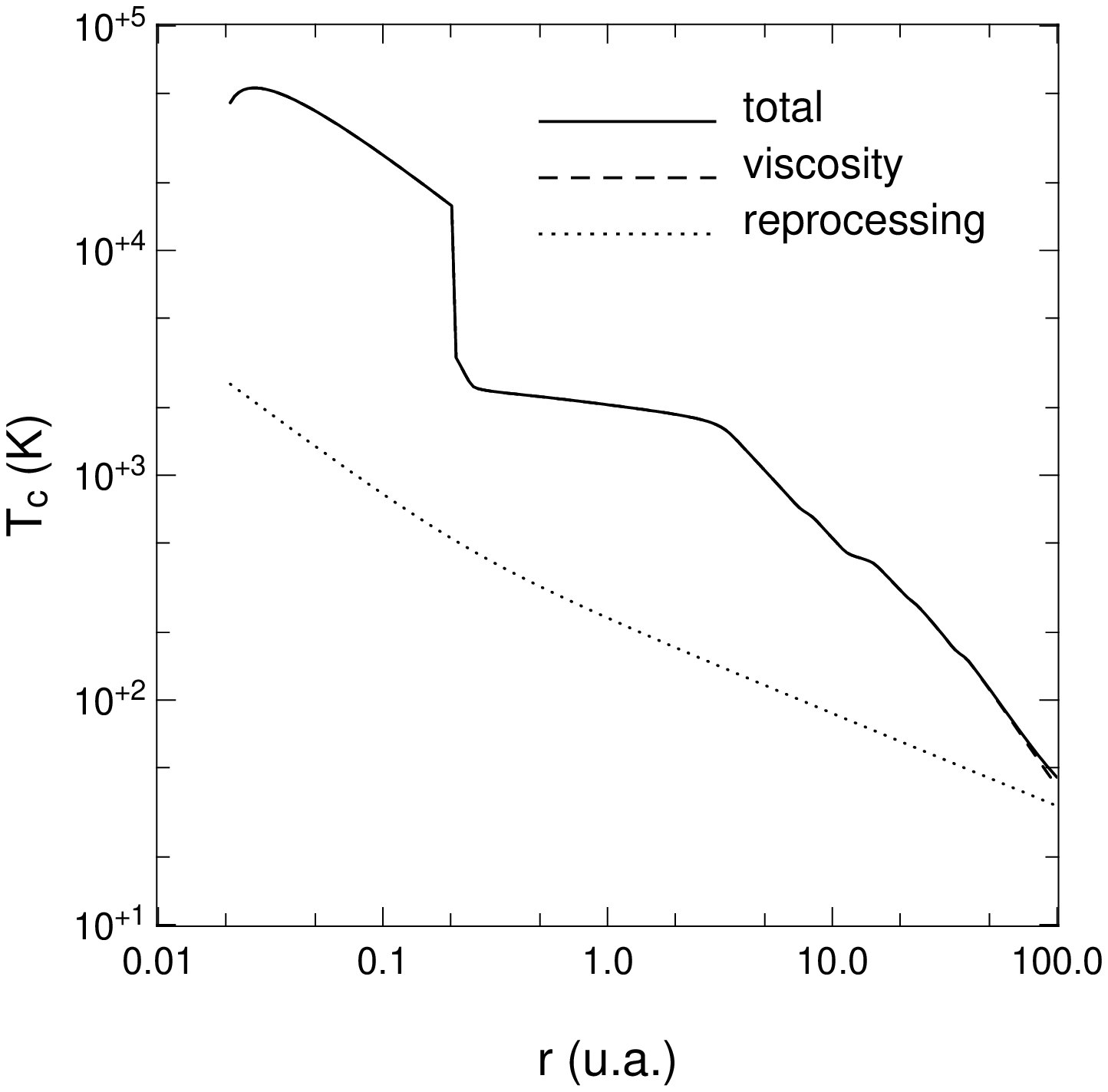}\hspace{0.01\hsize}%
      \includegraphics[width=0.24\hsize]{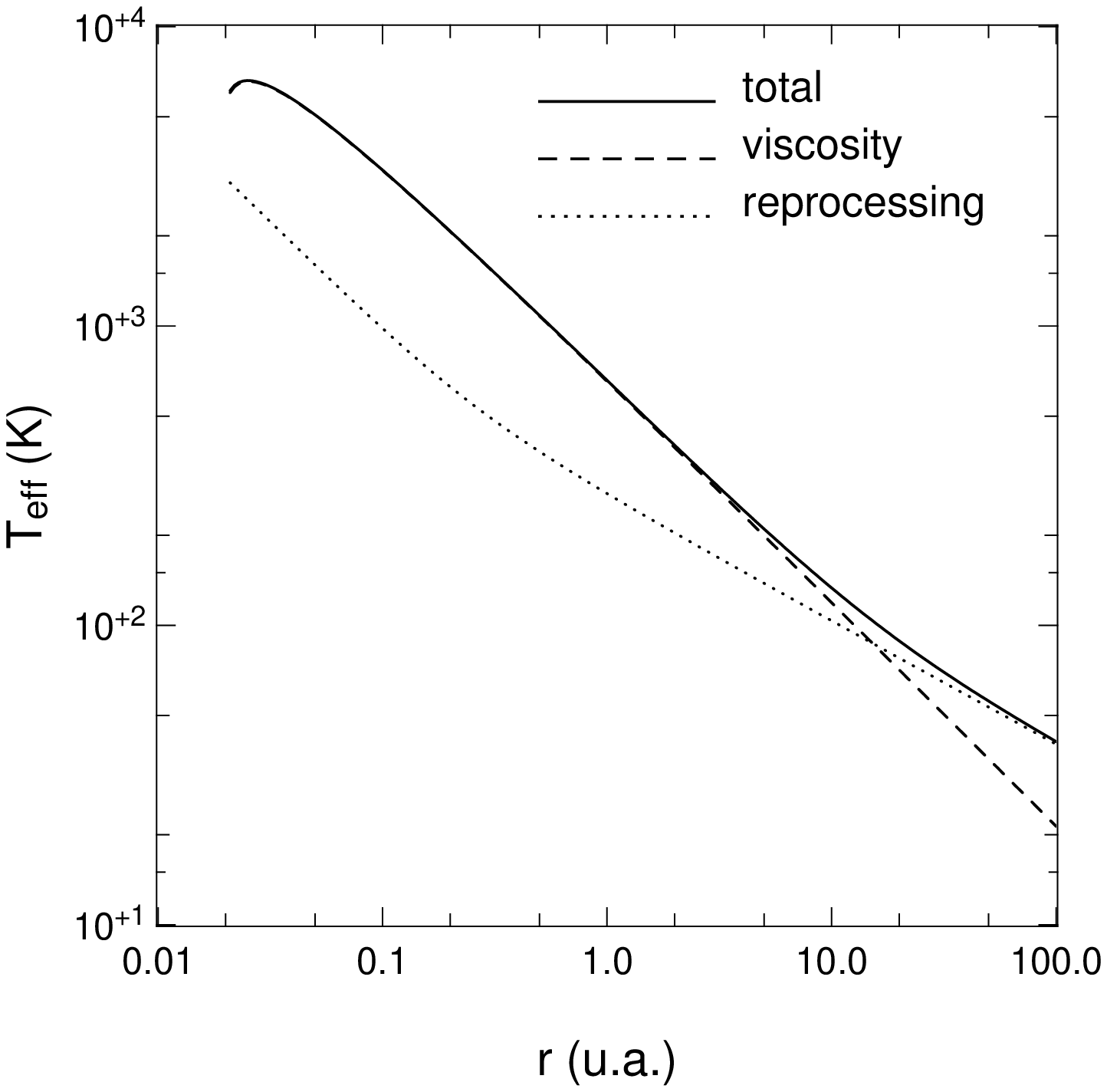}%
   }\\[-8pt]
   \subfigure[T~Tau North (SED fit)]{%
      \label{fig:ttau}
      \includegraphics[width=0.24\hsize]{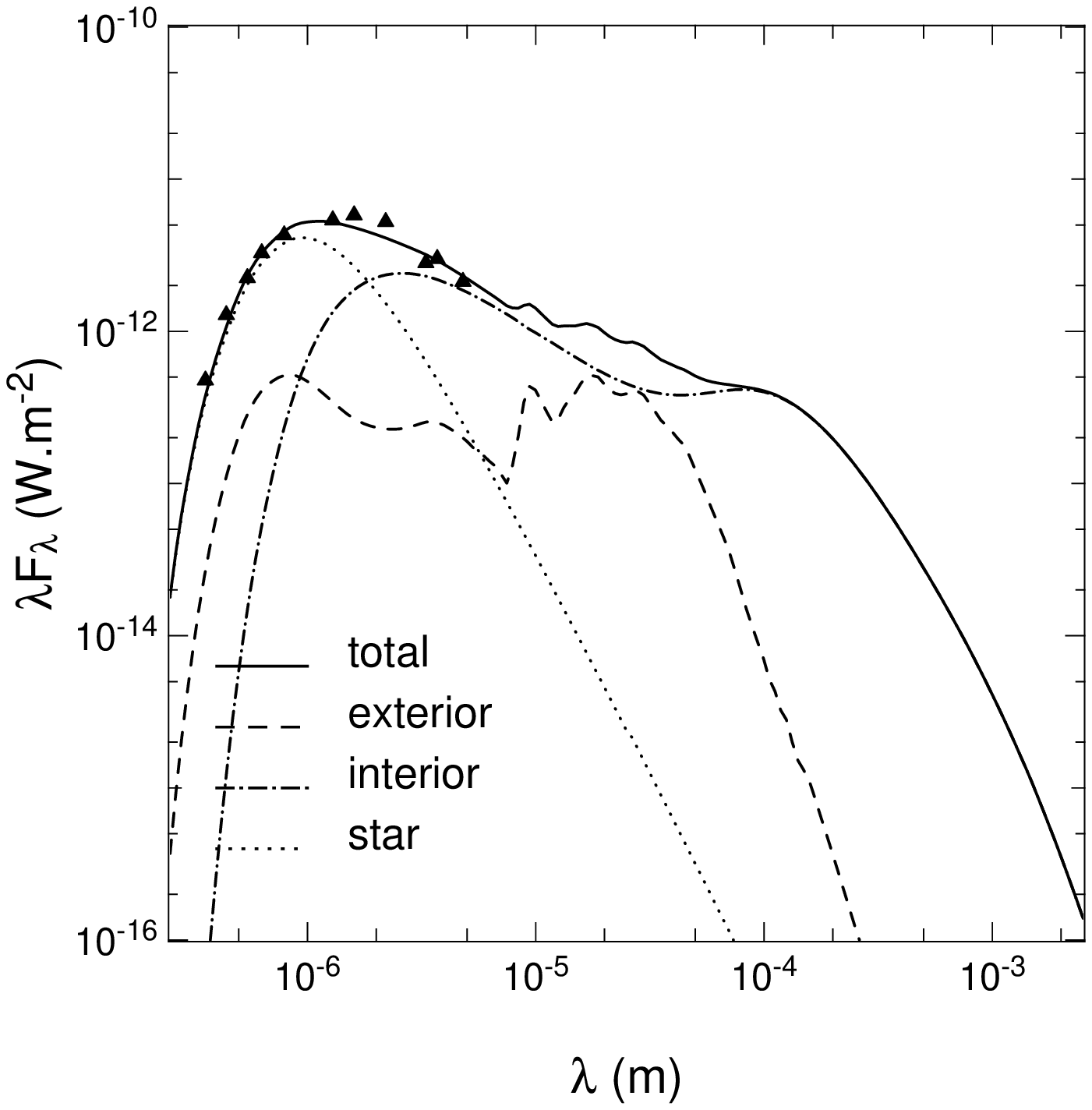}\hspace{0.01\hsize}%
      \includegraphics[width=0.24\hsize]{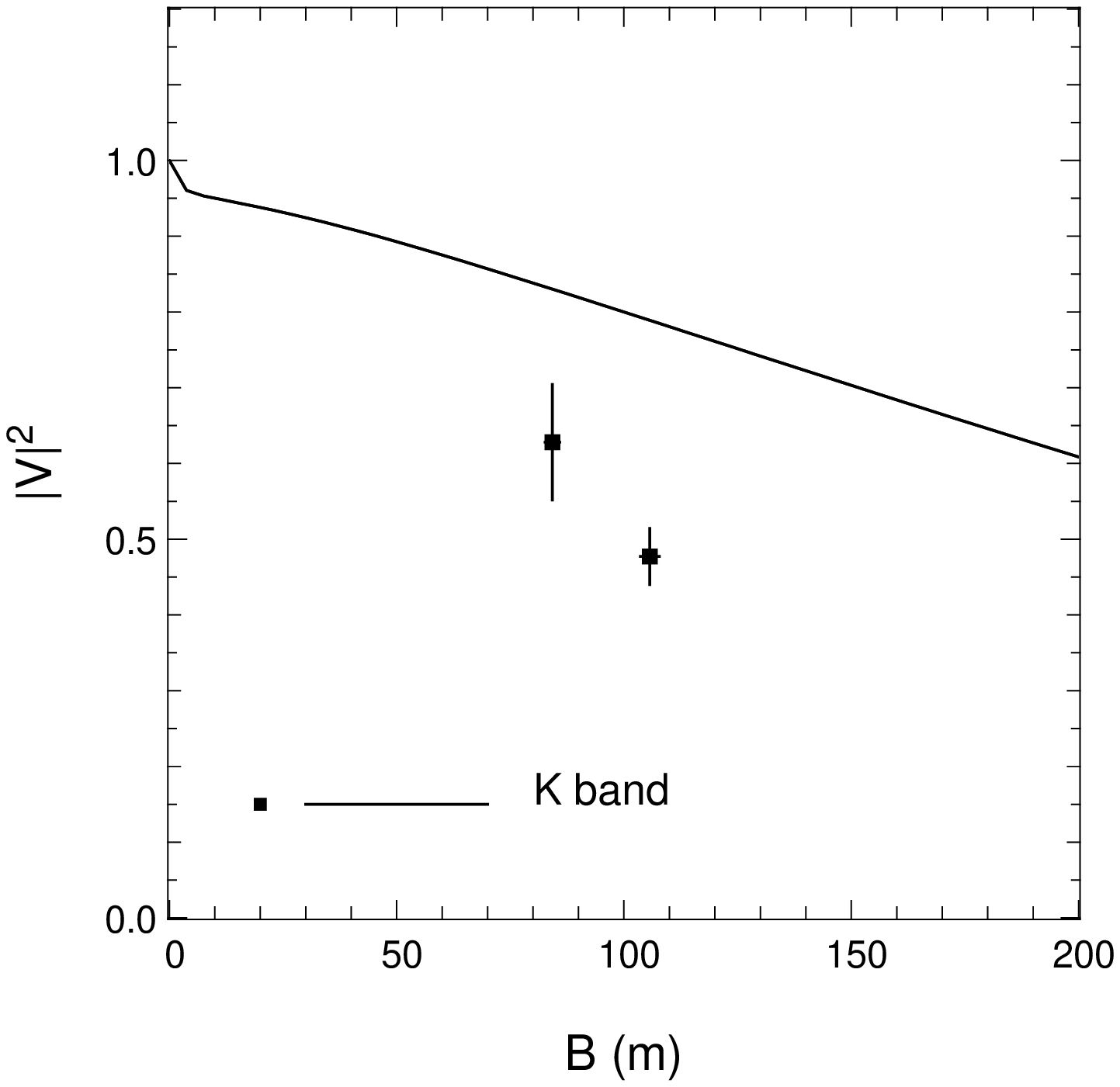}\hspace{0.01\hsize}%
      \includegraphics[width=0.24\hsize]{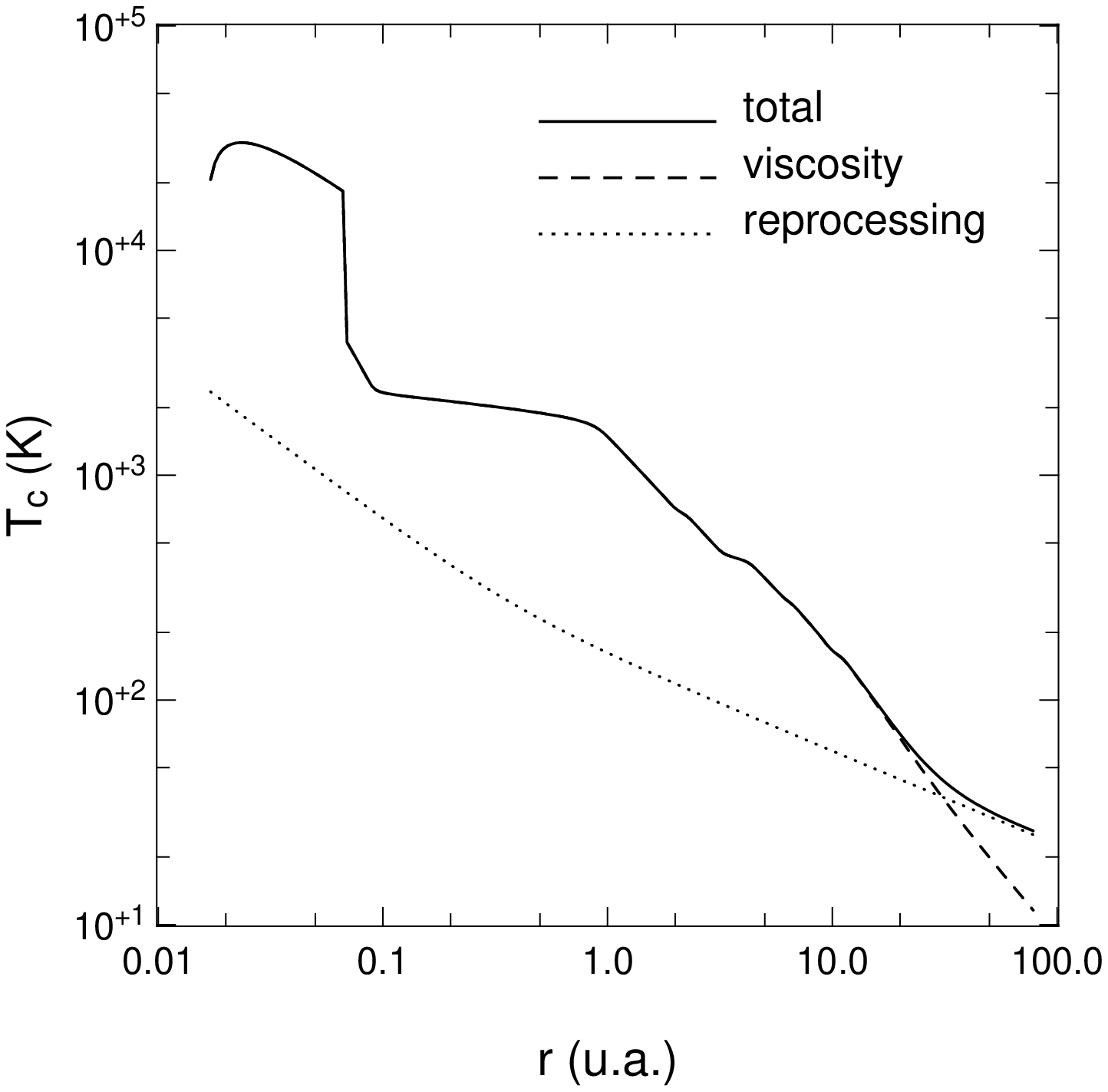}\hspace{0.01\hsize}%
      \includegraphics[width=0.24\hsize]{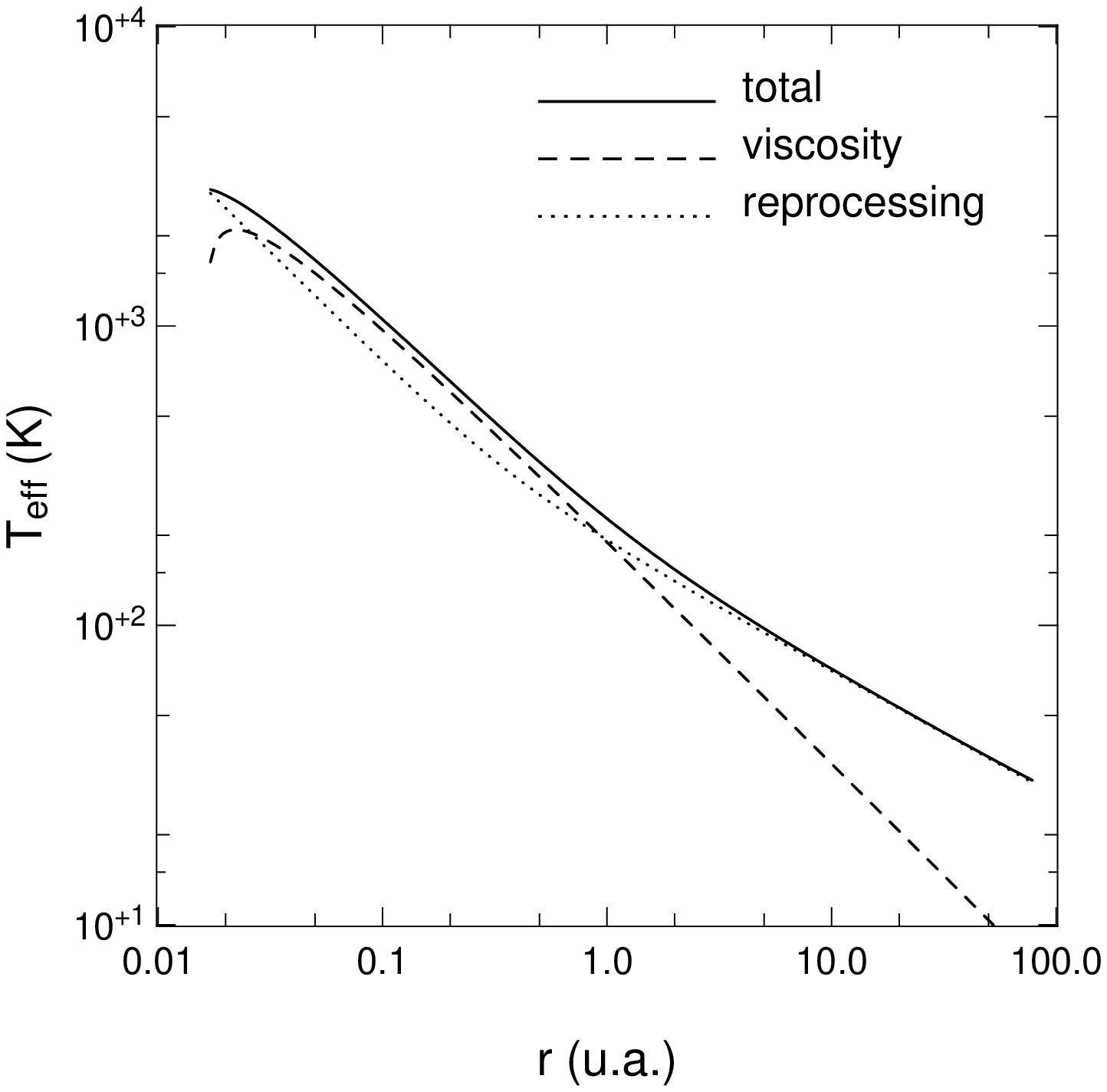}%
   }\\[-8pt]
   \subfigure[T~Tau North (visibility fit)]{%
      \label{fig:ttau2}
      \includegraphics[width=0.24\hsize]{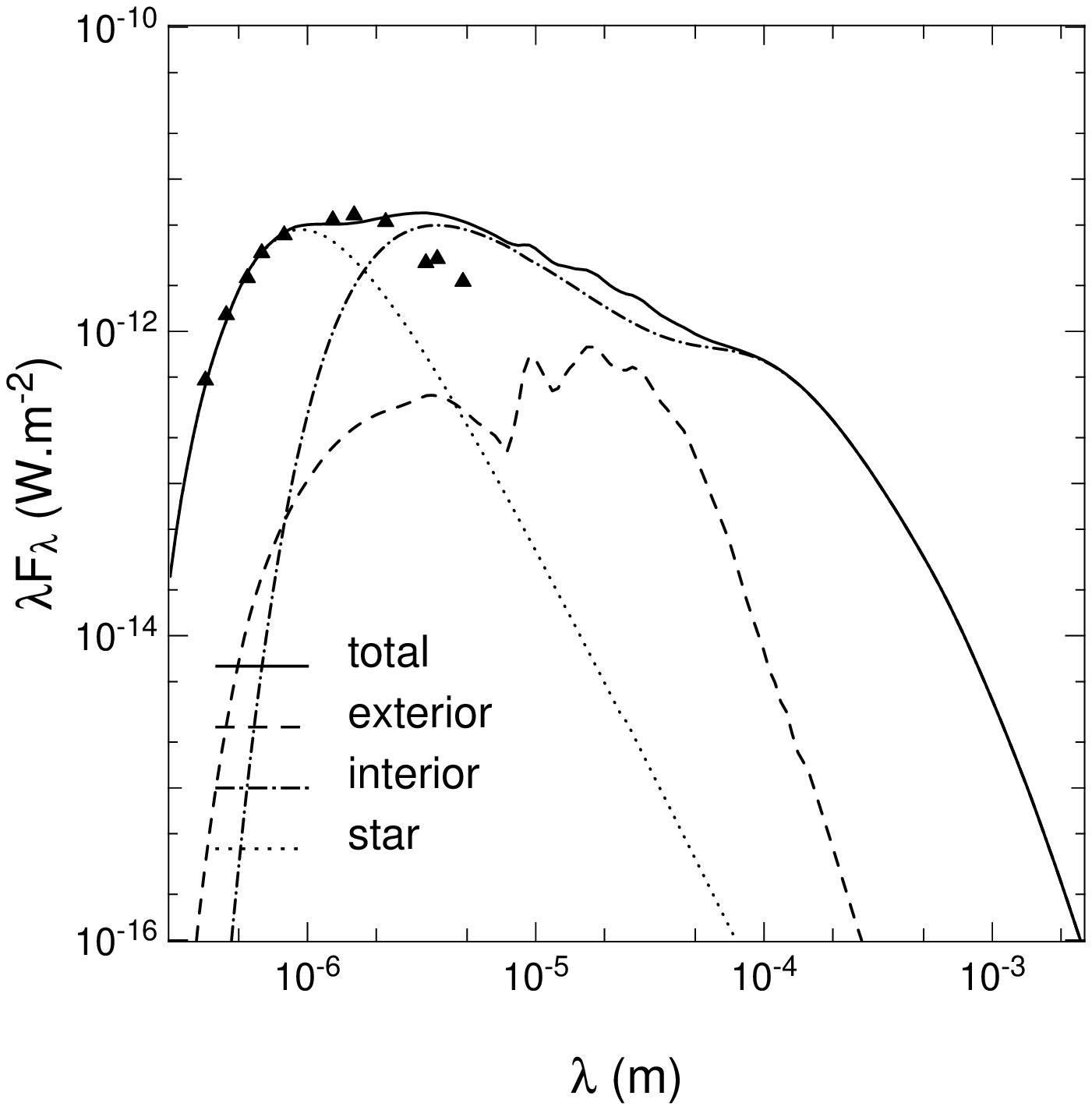}\hspace{0.01\hsize}%
      \includegraphics[width=0.24\hsize]{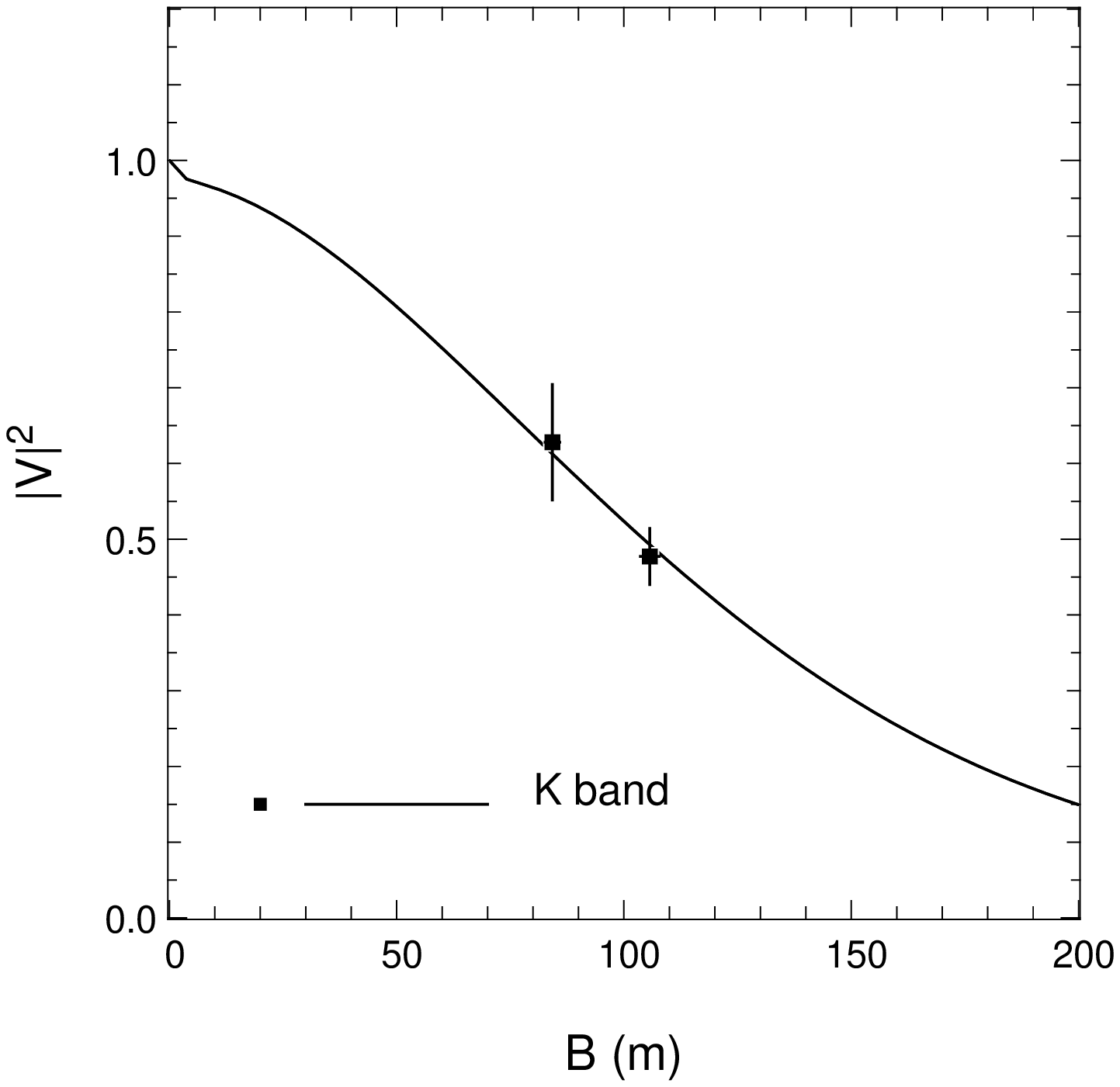}\hspace{0.01\hsize}%
      \includegraphics[width=0.24\hsize]{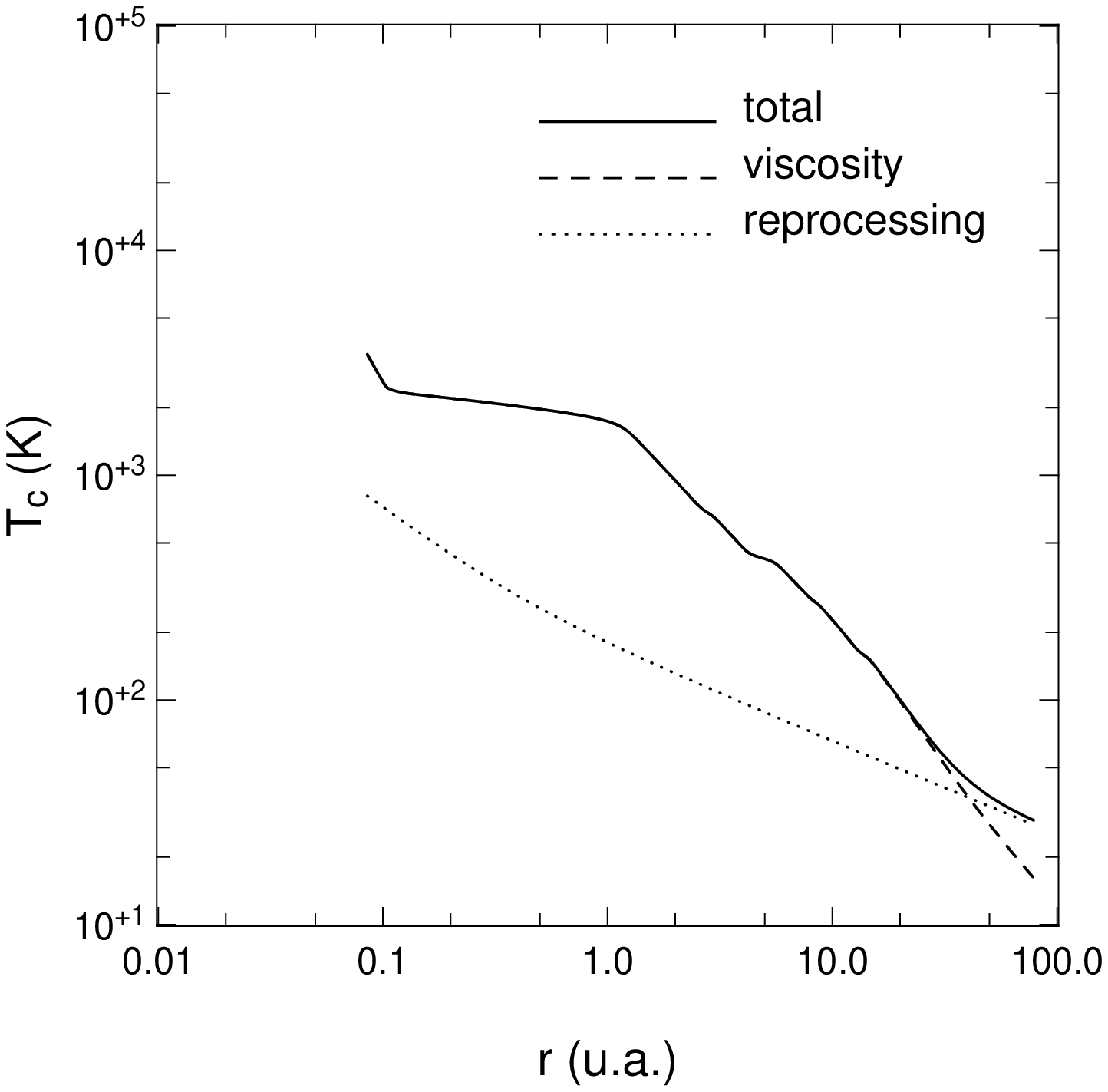}\hspace{0.01\hsize}%
      \includegraphics[width=0.24\hsize]{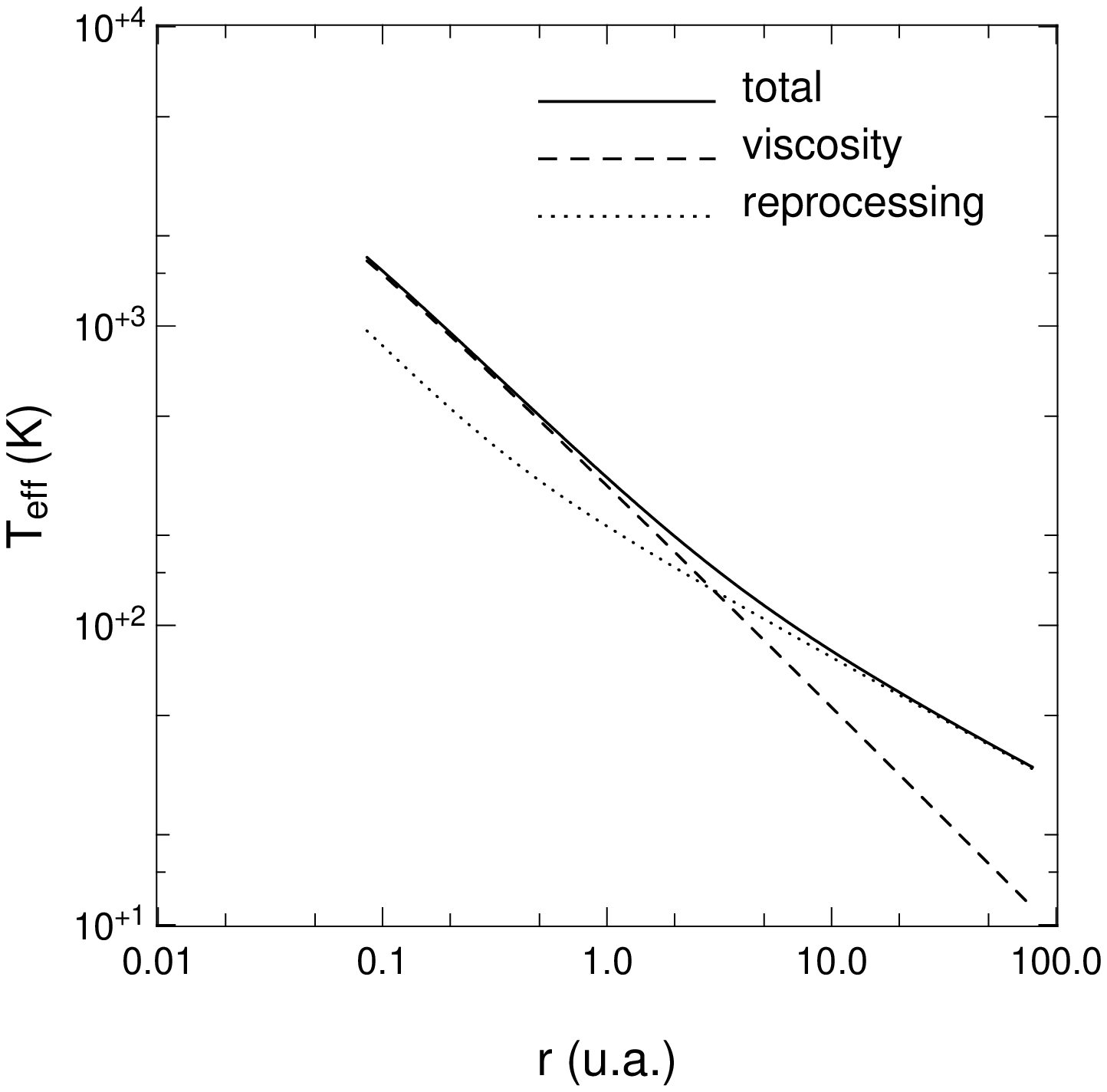}%
   }
   \caption{%
      Data and fits for YSOs (a: SU~Aur, b: FU~Ori, c,d: T~Tau North's separate
      fits for SED and visibility). Left column: SED with contributions of the
      inner and outer layers and of the star; middle-left column: visibility
      amplitude vs. baseline, for the major and minor axes of the disc image;
      middle-right column: mid-plane temperature and contributions
      of reprocessing and viscous heating vs. radius;
      right column: effective temperature and contributions of reprocessing
      and viscous heating vs. radius.%
   }
\label{fig:model-obs}
\end{figure*}

Figure~\ref{fig:model-obs} displays observational data together with our best
models fits.

\begin{table}[t]
   \caption{%
      Parameters used for the models of our best fits.  For T~Tau North the
      SED and visibility fits are displayed.
   }
   \label{tab:param}
   \setlength{\tabcolsep}{4pt}
   \def\note#1{\ensuremath{\,^{#1}}}
   \begin{tabular}{lllll}
      \hline\hline
                                 & SU~Aur              & FU~Ori               & T~Tau N              & T~Tau N           \\
                                 & ~                   & ~                    & (SED)                & ($V$)             \\
      \hline
      $r_*$ ($\Rsun$)            & $3.1$               & $4.0$\note a         & $3.5$                & $3.8$\note d      \\
      $r_{\mathrm{min}}$ (\Rsun) & $8.2$               & $4.4$                & $3.6$                & $18.0$\note d     \\
      $r_{\mathrm{max}}$ (AU)    & $200$\note a        & $150$\note a         & $80$\note a          & $80$\note a       \\
      $M_*$ (\Msun)              & $2.2$               & $1.0$                & $2.0$                & $2.0$\note d      \\
      {\Mdot} (\Msun/\yr)        & \sci{2.0}{-7}       & \sci{4.0}{-5}        & \sci{1.4}{-7}\note c & \sci{8.0}{-7}\note d\\
      $T_*$ (K)                  & $5600$              & $5000\note a$        & $4600$               & $4700$\note d     \\
      $i$ (\deg)                 & $40$                & $40$                 & $0$                  & $0$\note d        \\
      $A_V$ (mag)                & $0.85$              & $1.4$                & $0.8$                & $0.8$\note a      \\
      \alb                       & $0.3$               & $0.2$                & $0.3$\note c         & $0.2$\note d      \\
      $\alpha$                   & \sci{2}{-4}\note b  & \sci{1}{-1}\note b   & \sci{2}{-3}\note a   & \sci{1}{-2}\note d\\
      \hline
      \multicolumn{5}{l}{a: parameter not constrained}\\
      \multicolumn{5}{l}{b: parameter loosely constrained}\\
      \multicolumn{5}{l}{c: $(1-\alb) \Mdot$ constrained}\\
      \multicolumn{5}{l}{d: parameters not separately constrained}\\
      \hline
   \end{tabular}
\end{table}

\subsection{Fitting strategy}

We have identified various resulting characteristics of our modelled SEDs and
visibilities that relate more specifically to physical parameters of the model.
We describe hereafter how we use them to adjust our fits. 

\begin{itemize}
   \item The level of the near-IR and mid-IR SED are very sensitive to the
   accretion rate (see Fig.~\ref{fig:mdot}), and more precisely to $M_* \dot M$
   (used to fit $M_* \dot M$). 
   \item The shape of near-IR and mid-IR SED is sensitive to the inner disc
   truncation (see Fig.~\ref{fig:rmin}) (used to fit $r_\mathrm{min}$).
   \item The shape of the optical SED is sensitive to the stellar temperature,
   the stellar radius and the extinction (used to fit $A_V$, $T_*$ and $r_*$).
   However, in massive discs, the shape of the optical SED is governed by the
   disc only and the values of $r_*$ and $T_*$ have little impact on the
   observables.
   \item The shape of the far-IR SED is sensitive to the stellar temperature,
   the albedo and the parameter $\omega$ (used to fit $\omega$ and $\alb$).
   \item The inclination of the disc implies a spread of visibility
   measurements at a given baseline modulus because the visibilities along
   major and minor axes then differ.  We  fit the inclination $i$ using the
   spread in visibility.
\end{itemize}

Visibilities have not been used to fit parameters directly but were added
as a constraint.  The reason is that many parameters have an influence
on visibilities, as shown in Fig.~\ref{fig:paramsp}.

We find values close to 0.2 for $\omega$ \citep[0.1 in][]{Chiang97}, and albedo
values close to 0.2. These parameters will no longer be free when we have
included wavelength-dependent opacities.

\subsection{SU~Aur}

We were able to find a set of parameters to adjust simultaneously the
visibilities and the spectrum of this object (see
Table~\ref{tab:param}).

As expected with an accretion rate of $2 \times 10^{-7}\Msun/\yr$, the
structure of SU~Aur is dominated by viscous heating within 10~AU from the star.
(See the mid-plane temperature curves on the right panel of
Fig.~\ref{fig:suaur})  On the contrary, the effective temperature of the disc
is never dominated by viscous dissipation, as shown by effective temperature
curves on the same figure.  Therefore, the inner parts of the disc can be
considered as an active disc in terms of structure but not in terms of
observables.  The reason for this contradictory behaviour is the following:
$\TV \approx \TR$, but since $\Ti \approx \TV \taui + \TR/2$ with $\taui \gg
1$, we indeed have an inner temperature $\Ti \approx \TV \taui$, that is
dominated by viscous heating.

\subsection{FU~Ori}

The visibility measurements present a very large scatter. It may be produced
by: (i) a large inclination of the disc (ii) a fully resolved structure.  The
first hypothesis can be discarded with scattered light images that do not show
any jet, suggesting that the disc is seen almost pole on.  Moreover the low
value for a baseline of 30~m also backs the second hypothesis.
\citet{Malbet01p} derived that the visibility measurements are compatible with
a standard disc if a faint punctual source is located at 35~mas from the star.
It might be interpreted as a hot spot in the disc.

We are able to reproduce the SED until 20~$\mu$m and the average of H and K
measurements.  The excess in the far IR is a well-know feature likely due to
some phenomenon occurring in the outer parts of the discs. For instance,
\citet{Lodato01} explain this excess with self-gravitation.

Our model does not explain why the visibility value in the H-band at 100~m is
much higher than in the K-band while the visibility curve in H is only slightly
above the K curve.  However our fit is consistent because of large
uncertainties on visibility values. We indeed need more accurate measurements
to disentangle between uncertainties and model inaccuracy.
\citet{Malbet01p,Malbet02p} find that this higher value in H is consistent with
an effective temperature distribution $\TV \sim r^{-0.4\ \mathrm{to}\ -0.6}$,
that our model cannot reproduce.  Figure~\ref{fig:fuori} displays the effective
temperature of the disc: its law is $\TV \sim r^{-0.75}$, in agreement with the
active disc model.

As expected, the structure of FU~Ori (see mid-plane temperature curves
Fig.~\ref{fig:fuori}, right panel) is dominated by viscous dissipation.  Its
effective temperature is also dominated by this process within a few tens of
AUs from the star.

\subsection{T~Tau North}

The available interferometric data for T~Tau North include the T~Tau South
component.  We have used the \citet{Akeson02} corrected data to adjust our
model but we were unable to derive a unique set of physical parameters to
reproduce simultaneously the visibility and the SED of T~Tau North.  Therefore
we present separate fits for the SED and the visibility, respectively displayed
on Figs.~\ref{fig:ttau}~\& \ref{fig:ttau2}  The best SED fit is obtained
with a small inner disc truncation and a moderate accretion rate of
$\sci{2.0}{-7}\Msun/\yr$.  This fit overestimates the visibility by at least
$2\sigma$.  Visibility fits imply a large inner disc truncation and a
larger accretion rate, inconsistent with the spectral data. 

We cannot exclude caveats in the visibility correction mentioned above.  Hence
we conclude that the physical parameters derived from the SED fit are the
actual T~Tau North's ones.  Future VLTI or Keck Interferometer (KI)
observations with a field of view of 50~mas should remove the T~Tau N/S
ambiguity.

Like SU~Aur, T~Tau disc structure is dominated by viscosity within 10~AU from
the star (see Fig.~\ref{fig:ttau}), but in terms of effective temperature both
viscosity and reprocessing must be taken into account.


\section{Conclusion}
\label{sec:ccl}

We have developed an analytical model for T~Tauri accretion discs based on a
two-layer approximation, and including the main heating processes: viscous
dissipation, reprocessing, and thermalization with the surrounding medium.  The
outer layer is directly heated by visible stellar light and the thermal flux
from the inner layer, whereas the inner layer is heated by viscous dissipation
and light reprocessed by the outer layer.  The strength of the model is an
analytical prediction of the mid-plane temperature and, yet with less accuracy,
of the flaring of the disc; it is a suitable tool in grasping the physical
conditions taking place in these discs: it allows to predict easily the
relative importance of heating processes, the contribution of scattering,  the
influence of the viscosity prescription, etc.

Despite of simplifications made in order to keep the model analytical, it
compares well with other disc models available in the literature.  Its
predictions in terms of mid-plane temperature, density scale height or disc
mass are consistent with those of numerical models by \citet{DAlessio98}, or by
\citet{Bell97} in the inner regions where this model is valid.  

For the first time, we are able to consistently fit the infrared spectra and
the optical visibility of two young stars, SU~Aur and FU~Ori.  A third object
has been considered (T~Tau North), but we could not find a set of parameters
that would fit simultaneously the SED and the visibility.  This might result
from the peculiar structure of this object which is a triple system, and
demonstrates that even a single interferometric measurement at one infrared
wavelength can be a very strong constraint to interpret disc models.  We
therefore expect a breakthrough in disc physics understanding when new
generation interferometers, like the VLTI or the KI, are able to observe
hundreds of young stellar objects.  Providing an analytical description and a
fast computation, while including essential physical phenomena taking place in
discs, our model will be a useful tool to interpret these forthcoming
observations.

Concerning the influence of the viscosity on the model output, our results show
that current instruments cannot significantly probe the influence of the
viscosity law, because it has no direct impact on the flux emerging from
optically thick regions.  The dependency on viscosity observed in our model
remains small and results from the variation of the disc geometrical
thickness when the amount of material changes.  It is only at larger
wavelengths, where the central regions are optically thin, that the mid-plane
temperature can be probed, thus indirectly the viscosity.  The future Atacama
Large Millimetre Array (ALMA) should be able to measure the column density and
the mid-plane temperature of discs within 10 AUs from the star, and probe the
influence of the viscosity.

In forthcoming work, we will address the following points.
\begin{itemize}
   \item Including direct heating of the inner rim of the disc by the star will
   allow us to make predictions for more massive stars (Herbig Ae/Be systems)
   as shown by \citet{Dullemond01}.
   \item The treatment of self-shadowing can also be improved: our model~1
   assumes that the regions where the disc flares inwards are simply not
   illuminated at all, while vertical structure models tend to prove that
   there is still enough material at large heights to catch stellar
   radiation.  Model~2 is neither satisfactory because the flaring index
   is taken \latin{ad hoc}, though it gives quite faithful predictions.
   \item A precise determination of the chemistry species in the outer parts of
   the disc implies to model the temperature profile in the outer layer
   instead of assuming an isothermal one.  \citet{Aikawa02} showed that the
   model by \citet{Chiang97} cannot reproduce observed element abundances in
   the outer part of the disc whereas the vertical structure by
   \citet{DAlessio98} can.  
\end{itemize}

\begin{acknowledgements}
We thank the referee Dr. O.~Regev for a quick response and a report
that helped to clarify the paper.  This research has made use of NASA's
Astrophysics Data System Bibliographic Services, of CDS's Service for
Astronomical Catalogues, and of the free software Yorick by D.~Munro.
\end{acknowledgements}


\appendix

\section{General $\alpha$-disc solution}
\label{sec:thin}

In simplified disc models, dealing with optical thickness is always cumbersome,
for five different mean optical thicknesses of the inner layer have to be
considered: Rosseland optical thickness {\tauR}, Planck thickness {\tauP},
Planck-like thickness for the radiation of the outer layer {\tauPe} and the
Planck-like thickness for the radiation of the star {\tauPstar}.  It leads to
considering five different regimes summarised in Table~\ref{tab:opacreg}.
\begin{table}[t]
   \caption{Five optical thickness regimes of the disc.}
   \label{tab:opacreg}
   \begin{tabular}{ccccc}
      \hline\hline
      regime &  \tauR    & \tauP    & \tauPe    & \tauPstar\\
      \hline
      1      &  $\gg 1$  & $\gg 1$  &  $\gg 1$  & $\gg \cos \phi$\\
      2      &  $\lt 1$  & $\gg 1$  &  $\gg 1$  & $\ll \cos \phi$\\
      3      &  $\lt 1$  & $\ll 1$  &  $\gg 1$  & $\ll \cos \phi$\\
      4      &  $\lt 1$  & $\ll 1$  &  $\ll 1$  & $\ll \cos \phi$\\
      5      &  $\lt 1$  & $\ll 1$  &  $\ll 1$  & $\ll \cos \phi$\\
      \hline
   \end{tabular}
\end{table}
Most T~Tauri discs present the regimes 1--4, while the fifth one is seldom
encountered.  For each of these regimes the mid-plane temperature follows
different laws, expressed below:
\begin{subequations}
\begin{align}
     \Ti^4 &\approx \frac{1}{2} \TR^4 + \TV^4 \xi \tauR + \TISM^4,\\
     \Ti^4 &\approx \frac{1}{2} \TR^4 + \TV^4 + \TISM^4,\\
     \Ti^4 &\approx \frac{1}{2} \frac{1     }{2\tauP} \TR^4 + \frac{\TV^4}{2\tauP} + \TISM^4,\\
     \Ti^4 &\approx \frac{1}{2} \frac{\tauPe}{\tauP}  \TR^4 + \frac{\TV^4}{2\tauP} + \TISM^4,\\
     \Ti^4 &\approx \frac{1}{2} \frac{\tauPe}{\tauP} \frac{2\tauPstar}{\cos \phi} \TR^4 + \frac{\TV^4}{2\tauP} + \TISM^4,
\end{align}
\end{subequations}
where $\xi$ is a constant of the order of a few tenths related to the
opacity local density exponent $l$ (see Eq.~\ref{eq:kappa}), as defined in
Eq.~(\ref{eq:xi}).

Some factors 2 in these expressions arise from our choice to deal with
half-optical thickness (from mid-plane to outer layer).  Instead of using four
developments for the first four domains, we use an expression that follows the
same asymptotic behaviours:
\begin{equation}
     \Ti^4 = \frac{\TV^4}{F\left(\invtauR\right)F(2\tauP)} + \frac{\TR^4 F(2\tauPe)}{2 F(2\tauP)} + \TISM^4
     \label{eq:tiadhoc}
\end{equation}
with
\begin{equation}
     F(x) = 1-\exp(-x).
\end{equation}
In Eq.~(\ref{eq:tiadhoc}), all F-terms can be interpreted using simple 
radiative transfer arguments, except for $\invtauR$ that has been taken 
\latin{ad hoc}.

The different optical thicknesses of Eq.~(\ref{eq:tiadhoc}) are determined from 
monochromatic opacities $\kappa_\lambda(\rho, T)$ by
\begin{align}
   \tauR  &= \Sigma \kappar(\rho, T, \Te)\\
   \tauP  &= \Sigma \kappap(\rho, T, T) \\
   \tauPe &= \Sigma \kappap(\rho, T, \Te)
\end{align}
where the generalised Planck and Rosseland opacities of the material at 
$(\rho, T)$ for black-body radiation at temperature $\Trad$ are given by:
\begin{align}
   \kappar &= \frac{\lambdaint{\pderiv{B_\lambda}{\Trad}(\Trad)}}{\lambdaint{\frac{1}{\kappa_\lambda(\rho, T)}\pderiv{B_\lambda}{\Trad}(\Trad)}}\\
   \kappap &= \frac{\lambdaint{\kappa_\lambda(\rho, T) B_\lambda(\Trad)}}{\lambdaint{B_\lambda(\Trad)}}
\end{align}

The opacity ratio in the outer layer (thermal radiation/stellar radiation) is 
then described as
\begin{equation}
   \omega = \frac{\kappap(\rho, \Te, \Te)}{\kappap(\rho, \Te, \Tstar)}.
\end{equation}

\section{Flaring index}
\label{sec:gamma}

The mid-plane temperature is given by
\begin{equation}
   \Ti^4   = \sum_k t_k(r, \Ti)^4. \label{eq:ap:Ti}
\end{equation}
Using Eqs.~(\ref{eq:hi}, \ref{eq:Q}) one derives
\begin{align}
   \Ti = \zeta \epsilon(r)^2  r,
\end{align}
where $\zeta$ is a constant and $\epsilon(r) = \Hdisk/r$.  Substituting this 
equation in to Eq.~(\ref{eq:ap:Ti}) we have
\begin{equation}
   \epsilon(r)^8 = (\zeta r)^{-4} \sum_k t_k(r, \zeta r \epsilon(r)^2).
   \label{eq:ap:epsilon}
\end{equation}
   
The flaring is given by the variation of the relative geometrical thickness
$\epsilon(r)$ (Eq.~\ref{eq:gamma}):
\begin{equation}
   \gamma = \frac{r}{\epsilon} \deriv{\epsilon}{r}.
\end{equation}
Derivating Eq.~(\ref{eq:ap:epsilon}) in respect to $r$ and introducing
\begin{subequations}
   \label{eq:ap:akbk}
   \begin{align}
      a_k &= \deriv{\log{t_k^4}}{\log r},\\
      b_k &= \deriv{\log{t_k^4}}{\log \Ti}
   \end{align}
\end{subequations}
we derive
\begin{equation}
   \gamma = \frac{\displaystyle \sum_k \gamma_k  t_k(r, \Ti)^4}{\displaystyle \sum_k           t_k(r, \Ti)^4},
\end{equation}
with
\begin{equation}
   \gamma_k = \frac{4+a_k-b_k}{8-2b_k}.
\end{equation}

The flaring index of an optically thick $\alpha$-disc is given in the
main matter, by substituting the $t_k$'s in Eq.~\ref{eq:ap:akbk}.  In the
following we give the formulae in the case of an arbitrary optical thickness 
and of a $\beta$-disc.

\subsection{$\alpha$-disc of arbitrary optical thickness}
\label{sec:alphathin}

The expression for the temperature contributions $t_k$ are in the present
case quite tangled (see Appendix~\ref{sec:thin}), though straightforward to 
derive, so we only give the results for the flaring index. We note
\begin{equation}
     \eta(x) = \frac{ x \exp(-x)  } { 1 - \exp(-x) }
\end{equation}
the logarithmic derivative of $F$.  The flaring associated to each
heating process are:
\begin{subequations}
   \begin{align}
     \gamma_1 &=
        \frac{\plm\left(\dtaua\right)+(2+g(r)) }{\qlm\left(\dtaua\right)+16},\\
     \gamma_2 &=
        \frac{\plm\left(\dtaub\right)+4        }{\qlm\left(\dtaub\right)+14},\\
     \gamma_3 &=
        \frac{\plm\left(\dtaub\right)+2        }{\qlm\left(\dtaub\right)+16},\\
     \gamma_4 &= \frac12
   \end{align}
\end{subequations}
with
\begin{subequations}
   \begin{align}
     \plm     &= 2m+3l+1-(1+l)g(r),\\
     \qlm     &= 4m-6l-4.
   \end{align}
\end{subequations}

\subsection{Optically thick $\beta$-disc}
\label{sec:beta}

The switch to $\beta$-prescription changes terms that involve column density:
\begin{align}
   \Sigma_0 &= \frac{f(r)\Mdot}{3\pi\beta\Omega r^2},\\
   \rho_0   &= \frac{\Sigma_0}{r} \sqrt{\frac{2}{\pi}}.
\end{align}

Then, only the temperature contribution and flaring index related to
viscous dissipation are changed:
\begin{align}
   t_1(r)^4 &=       \frac34 k\mu^{l/2}\Tgrav^{l/2}\rho_0^l
                          \Sigma_0\TV^4 \Ti^{m-l/2},\\
   t_1(r)^4 &\propto f(r)^{l+2} r^{-2l-7/2} \Ti^{m-l/2},\\
   \gamma_1 &=       \frac{1-3l-2m}{16+2l-4m} + \frac{2+l}{16+2l-4m} g(r).
\end{align}

\subsection{$\beta$-disc of arbitrary optical thickness}
\label{sec:betathin}

The flaring indices are given by the same set of equations as in 
Appendix~\ref{sec:alphathin} except for a change of $\qlm$:
\begin{subequations}
   \begin{align}
      \plm     &= 2m+3l+1-(1+l)g(r),\\
      \qlm     &= 4m-2l.
   \end{align}
\end{subequations}

\section{Disc thickness on the outer edge}
\label{sec:Qedge}

In the main matter, we used an approximate $g_z \propto z$ in the determination
of the structure of the inner layer, because most of the matter is concentrated
within one or two scale heights, with $\hi \ll r$.  In determining the
vertical location of the disc surface with the same approximation, we noticed a
large underestimation of {\Hdisk}.  The reason is that $\Hdisk \approx
4\mbox{--}6 \hi$ is not so small compared to $r$, so that the departure of
$g_z$ from the linear law cannot be ignored.  So, we are bound to considering
the exact expression for the gravity field.

In the outer layer, the Planck optical thickness $\taue$, the surface
density $\Sigmae$, and the opacity $\kappae$ are linked by
\begin{equation}
   \taue = \kappae \Sigmae.
\end{equation}
If we use an isothermal layer, $\Sigmae$ can be substituted by $\rhoe\he$
where $\he$ is the scale height of the outer layer and $\rhoe$ the density
at its base.  $\taue$ can be expressed as $\phif\omega$, because we expect
the incidence $\phi$ of stellar radiation to be dominated by the flaring 
term.  Therefore
\begin{equation}
   \kappae\rhoe\he = \phi\omega
\end{equation}
To determine $\rhoe$, we solve the vertical hydrostatic equilibrium of
the inner layer in the gravity field of the central star
\begin{equation}
   g_z (z) = - \frac{z\ci^2}{2\hi^2} \left[ 1 + \left(\frac zr\right)^2 \right]^{-3/2},
\end{equation}
which leads to
\begin{align}
   \rho(z) &= \rhoi \exp\left( -\left(\frac rh\right)^2 \left\{ 1 - \left[ 1 + \left(\frac zr\right)^2\right]^{-1/2} \right\} \right),\\
\intertext{with a scale height}
   h(z)    &\propto \sqrt{T(z) \left[ 1 + \left( \frac{Q\hi}{r}\right)^2\right]^{-3/2}}.
\end{align}
This allows us to write
\begin{align}
   \rhoe  &= \rhoi\exp\left(- \frac{Q^2}{2} \delta_\rho\right),\\
   \he    &= \hi  \sqrt{\frac\Te\Ti \delta_h},
\end{align}
$\delta_\rho$ and $\delta_h$ being corrective terms due to the
non-linearity of $g(z)$. They are given in Sect.~\ref{sec:Q}
(Eqs.~\ref{eq:deltarho} \& \ref{eq:deltah}).  Therefore
\begin{equation}
   Q = \sqrt{ \frac{2}{\delta_\rho} \log \left( \frac{\kappae\rhoi\hi}{\phi\omega} \sqrt{\frac\Te\Ti \delta_h}\right) }.
\end{equation}
Equation~(\ref{eq:scQ}) in Sect.~\ref{sec:Q} is then obtained after the
substitution $\phi = \gamma Q\hi/r$.


\bibliographystyle{aa}
\bibliography{M2735}

\end{document}